\documentclass[aps,superscriptaddress,nofootinbib,floatfix,notitlepage]{revtex4-1}
\usepackage[utf8x]{inputenc}
\pdfoutput=1
\usepackage{graphicx}
\usepackage{hyperref}
\usepackage{bm}
\usepackage{slashed}

\newcommand{\be}{\begin{equation}}
\newcommand{\ee}{\end{equation}}
\newcommand{\beq}{\begin{equation}}
\newcommand{\eeq}{\end{equation}}
\newcommand{\bea}{\begin{eqnarray}}
\newcommand{\eea}{\end{eqnarray}}
\newcommand{\msbar}{\overline{\textrm{MS}}}

\newcommand{\nn}{\nonumber}

\newcommand{\msb}{\overline{\rm{MS}}}

\newcommand{\smomq}{RI-SMOM$(\slashed{q}, \slashed{q})$ }
\newcommand{\smomgamma}{RI-SMOM$(\gamma^\mu, \gamma^\mu)$ }

\newcommand{\beqn}{\begin{eqnarray}}   
\newcommand{\eeqn}{\end{eqnarray}}

\def\utfit{{\bf{U}}\kern-.20em{\bf{T}}\kern-.15em{\it{fit}}\@}  
\def\butfit{{\bf{U}}\kern-.18em{\bf{T}}\kern-.10em{\it{fit}}\@}

\begin{document}
\vskip 0.5 cm 
\title{New \utfit\   Analysis of the  Unitarity Triangle  \\ 
in  the  Cabibbo-Kobayashi-Maskawa  scheme}

\author{Marcella \surname{Bona}}
\affiliation{School of Physics and Astronomy, Queen Mary University of London, London; United Kingdom}

\author{Marco \surname{Ciuchini}}
\affiliation{INFN, Sezione di Roma Tre, Via della Vasca Navale 84, I-00146 Rome, Italy}

\author{Denis \surname{Derkach}}
\affiliation{HSE University, 20 Myasnitskaya Ulitsa, Moscow 101000, Russia}

\author{Fabio \surname{Ferrari}}
\affiliation{INFN, Sezione di Bologna, Via Irnerio 46, I-40126 Bologna, Italy}
\affiliation{Dipartimento di Fisica e Astronomia, Universit\`a di Bologna, Viale Berti Pichat 6/2, 40127 Bologna, Italy}

\author{Enrico \surname{Franco}}
\affiliation{INFN, Sezione di Roma, P.le A. Moro 2, I-00185 Roma, Italy}

\author{Vittorio \surname{Lubicz}}
\affiliation{INFN, Sezione di Roma Tre, Via della Vasca Navale 84, I-00146 Rome, Italy}
\affiliation{Dipartimento di Matematica e Fisica,  Universit\`a  Roma Tre, Via della Vasca Navale 84, I-00146 Rome, Italy}

\author{Guido \surname{Martinelli}}
\affiliation{INFN, Sezione di Roma, P.le A. Moro 2, I-00185 Roma, Italy}
\affiliation{Dipartimento di Fisica, Universit\`a di Roma La Sapienza, P.le A. Moro 2, I-00185 Roma, Italy}

\author{Davide \surname{Morgante}}
\affiliation{INFN, Sezione di Milano, Via Celoria 16, I-20133 Milano, Italy}
\affiliation{Dipartimento di Fisica, Universit\`a  degli Studi di Milano, Via Celoria 16, I-20133 Milano, Italy}

\author{Maurizio \surname{Pierini}}
\affiliation{CERN CH-1211 Geneve 23, Switzerland}

\author{Luca \surname{Silvestrini}}
\affiliation{INFN, Sezione di Roma, P.le A. Moro 2, I-00185 Roma, Italy}

\author{Silvano \surname{Simula}}
\affiliation{INFN, Sezione di Roma Tre, Via della Vasca Navale 84, I-00146 Rome, Italy}

\author{Achille  \surname{Stocchi}}
\affiliation{Universit\'e Paris Saclay, CNRS/IN2P3, IJCLab, 91405 Orsay, France}

\author{Cecilia \surname{Tarantino}}
\affiliation{INFN, Sezione di Roma Tre, Via della Vasca Navale 84, I-00146 Rome, Italy}
\affiliation{Dipartimento di Matematica e Fisica,  Universit\`a  Roma Tre, Via della Vasca Navale 84, I-00146 Rome, Italy}

\author{Vincenzo \surname{Vagnoni}}
\affiliation{INFN, Sezione di Bologna, Via Irnerio 46, I-40126 Bologna, Italy}

\author{Mauro \surname{Valli}}
\affiliation{C.N. Yang Institute for Theoretical Physics, Stony Brook University, Stony Brook, NY 11794, USA}

\author{Ludovico \surname{Vittorio}}
\affiliation{LAPTh, Universit\'e  Savoie Mont-Blanc and CNRS, Annecy, France}

\collaboration{The \utfit\ Collaboration}
\vskip 0.5 cm  
 {\it In honour of Nicola Cabibbo,  father of  flavour physics. \\
He introduced the idea of universality between the leptonic current and a single 
 hadronic current,  combination of the $SU(3$)  currents allowing  $\Delta S =
0$ and $\Delta S = 1$  transitions. It was thus  Nicola Cabibbo who reconciled strange particle decays
with the the  universality of weak interactions  paving the way to the  modern electroweak
unification within the Standard Model.}

\begin{abstract}
Flavour mixing and CP violation as measured in weak decays and mixing of neutral mesons are a fundamental tool to test the  Standard Model (SM)  and 
to search for new physics.   New analyses  performed at the LHC experiment   open an  unprecedented insight  into the Cabibbo-Kobayashi-Maskawa  (CKM) metrology  and new evidence for rare decays.  Important progress has also been achieved in theoretical calculations of several hadronic quantities with a 
remarkable reduction of  the uncertainties. This improvement is essential since previous studies of the Unitarity Triangle did show that possible contributions from new physics, if any, must be tiny and could easily be hidden by   theoretical and experimental errors. 
 Thanks to the experimental and theoretical advances, the CKM picture  provides very precise SM predictions
through global analyses. We present here the results of   the  latest global SM analysis performed by
the UTfit collaboration including all the most updated inputs from experiments, lattice QCD and
phenomenological calculations.  
 
 \end{abstract}

\maketitle

\section{Introduction}
\label{sec:intro}
In this paper we enlarge and update the Unitarity Triangle  Analysis  within  the Standard Model (SM)
using  the most recent  progress of the theoretical inputs and the latest measurements of
the experimental observables. A more general analysis  beyond the SM, with constraints on new physics contributions, will be presented in a subsequent publication. 

The \utfit\ collaboration has been routinely updating the Unitarity Triangle  Analysis (UTA) within and beyond the SM   via
its online results (http://utfit.org/), regular presentations to topical conferences and  continuous collaboration with the flavour community, see \,\cite{Ciuchini:2000de, Alpigiani:2017lpj, Bona:2022zhn, Bona:2022xnf} and references therein. 
Within the SM,  the UTA is a fundamental  tool to determine precisely the SM   parameters of the flavour sector, to test the compatibility of the experimental results with the theoretical calculations  and to predict yet unmeasured   flavour SM observables.
Beyond the SM  a quantitative  evaluation of the degree of discrepancy between measurements and theoretical predictions offers the possibility 
of discovering  New Physics (NP) effects due to the presence of new particles or interactions  at still unexplored energy scales.
In this respect,  the UTA is complementary to the search of new particles at colliders working at multi-TeV  energies. 
For a recent discussion about predictions of  rare decays in the SM  see\,\cite{Buras:2022qip}.

Within the SM,  flavour mixing and weak CP violation  are  described by several free parameters, namely the  quark masses and the CKM matrix elements\, \cite{Cabibbo:1963yz,Kobayashi:1973fv}.  Indeed these parameters can be reduced to  only ten independent, physically determinable quantities,  that we choose to be  the quark masses, $m_q$, defined in a suitable scheme,  and  the value of the Wolfenstein parameters  $\lambda, {\cal A}, \bar \rho, \bar \eta$\,\cite{Wolfenstein:1983yz,Buras:1994ec}.  In addition,  the SM is characterised by  two important properties: the absence of tree-level Flavour Changing Neutral Currents (FCNC)  and the GIM suppression mechanism\,\cite{Glashow:1970gm}. The latter manifests itself either as {\it mild GIM} suppression,  proportional to $\log m_q^2/M_W^2$, for QCD and radiative penguin operators/amplitudes, or as {\it hard GIM} suppression, proportional to $m_q^2/M_W^2$,  for $\Delta F=2$ transitions. 
Thus, beyond the SM,  CKM and/or  GIM suppressed processes   are the most interesting  quantities to study since they are highly sensitive to NP  contributions.  Among them, $\Delta F=2$ transitions are the best cases since they  are both CKM and {\it hard GIM} suppressed. This is the reason why accurate SM estimates of CP violation in neutral meson oscillations  is a crucial ingredient   of the UTA analysis that we will discuss in the following. 

In the recent past,  there has been a lot of excitement about apparent violations of Lepton Flavour Universality (LFU).   On the one hand,  in the case of semileptonic decays,   we have the so-called $\vert V_{cb}\vert$  puzzle, i.e. the  tension between the inclusive\,\cite{Gambino:2013rza,Alberti:2014yda,Gambino:2016jkc} and the exclusive determinations of the CKM matrix  element $\vert V_{cb}\vert$\,\cite{BaBar:2007nwi, BaBar:2007cke, BaBar:2008zui, BaBar:2009zxk, Belle:2010qug, Belle:2015pkj, Belle:2017rcc, Belle:2018ezy}. On the other hand, a discrepancy exists between the theoretical   expectation value and the measurements of $R(D^{(*)})$\,\cite{HFLAV:2019otj}, defined as the ratios of the   branching fractions of  $B \to D^{(*)}\tau \nu_\tau$ over $B \to   D^{(*)}\ell \nu_\ell $ decays, $\ell  = e, \mu$, performed by Belle, BaBar
and LHCb\,\cite{BaBar:2012obs, BaBar:2013mob, LHCb:2015gmp, Belle:2015qfa, Belle:2016ure, Belle:2016dyj, LHCb:2017smo, Belle:2017ilt, LHCb:2017rln}.  For $\vert V_{cb}\vert$ we included in our analysis the latest experimental measurements and  the results of a new approach to evaluate the form factors based on unitarity and analyticity.  Besides  semileptonic charged-current $B$ decays,  LFU seems to be violated by  the ratios  $R_{K^{(*)}}= BR(B\to K^{(*)}  \mu^+\mu^-)/BR(B\to K^{(*)}  e^+ e^-)$ which are sensibly lower than the value close to one expected if LFU holds\,\cite{LHCb:2021trn, LHCb:2021lvy, LHCb:2017avl, Belle:2019oag}. Although the UTA is not able to shed any light on the $R_{K^{(*)}}$ (and $R(D^{(*)})$)   problem, it is very useful to clarify the issues connected to the $\vert V_{cb}\vert$  puzzle, since a determination of this CKM matrix element can be derived indirectly from the UTA  when omitting  the exclusive and inclusive semileptonic $B$ decays in the  inputs.  
In addition to the new values for $\vert V_{cb}\vert$ from exclusive decays, our updated   analysis is also based on  the latest determinations of other  relevant  theoretical inputs and  recent  measurements of  the experimental flavour observables. The basic constraints used in the global fit and contributing to the
sensitivity of the CKM matrix elements are: $\vert V_{ub}\vert$ and $\vert V_{cb}\vert$  from semileptonic $B$ decays, $\Delta M_d$ and $\Delta M_s$
from $B^0_{d,s}$  oscillations, $\varepsilon$ from neutral $K^0$ mixing,  the unitarity triangle angles $\alpha$  from charmless hadronic $B$ decays, 
 $\gamma$ from charm hadronic $B$ decays and  $\sin 2\beta$ from $B^0 \to J/\psi K^0$ decays. To these {\it classical} quantities,  we added $\mathrm{BR}(B^+ \to \tau^+ \nu_\tau)$, $\mathrm{BR}(B_s \to \mu^+\mu^-)$ and  the ratio
$\varepsilon^\prime/\varepsilon$ for which a solid theoretical  prediction now exists\,\cite{RBC:2020kdj}. Although the present precision is not enough to further constrain the parameters of the SM, we believe that is important to include this quantity in view of further theoretical improvements and also because it can help to constrain the contribution of some operators present  in models of NP.  In the present work we also included a new evaluation of long distance charm contributions to $\varepsilon$\,\cite{Ciuchini:2021zgf}. 

The values of most experimental inputs are taken from the Heavy Flavour Averaging Group  (HFLAV)\,\cite{HFLAV:2019otj} and from the  online update.   When  updated individual results are available,  however,  the UTfit collaboration performs its own averages.  We also use  the updates of the Particle Data Group 2022\,\cite{ParticleDataGroup:2022pth}. On the theoretical side, the non-perturbative QCD parameters are taken from the
most recent Lattice QCD (LQCD)  determinations: as a general prescription, we average the   $N_f =2+ 1 + 1$  and $N_f = 2 + 1$ FLAG numbers\,\cite{Aoki:2021kgd}.
The continuously updated set of numerical values used as inputs can be found at the URL http://www.utfit.org/.

As for the output results of the UTA and their uncertainties,   for all fits the  quoted numbers  correspond to the highest probability intervals containing at least $68\%$ and $95\%$ of the sample. The $68\%$ probability intervals are then presented as $V (E_V)$,  where $V$  is the center and $E_V$ the half width of the interval.

 The main results of  this work are the following: i)  there is a general consistency, at the percent level,   between the  SM predictions and the experimental measurements. Thus in order to discover new physics effects a further effort in theoretical and experimental accuracy is required;
ii)  although the tension between  exclusive and  inclusive determination of $\vert V_{cb}\vert$ requires further investigation,   in particular of the form factors relevant in semi-leptonic $B \to   D^{(*)}\ell \nu_\ell $ decays and of the fitting procedures of these processes,  we find that the UT analysis strongly favours  a larger value of  $\vert V_{cb}\vert$, close to its inclusive determination,  and a smaller value of  $\vert V_{ub}\vert$, close to the exclusive value;  iii)  the value of $\varepsilon^\prime/\varepsilon$ as predicted by using the weak Hamiltonian operator  matrix elements computed in
 ref.\,\cite{RBC:2020kdj} and the Wilson coefficients computed within the general UT analysis are in very good agreement  with the experimental value whereas   the \utfit\ prediction  for $\varepsilon$  is  lower  than its experimental value and a   further  improvement in  the accuracy of  the theoretical calculation is welcome. The results  of this work, written for  the {\it Rendiconti Lincei. Scienze Fisiche e Naturali}, will be combined with an \utfit\ analysis of 
$D^0$-$\bar D^0$ mixing and of searches for physics beyond the Standard Model in a paper to be submitted to a physics journal. 

The paper is organized as follows: in sec.\,\ref{sec:inputs} we describe the theoretical and experimental inputs used in our analysis; in sec.\,\ref{sec:eps} we discuss the recent progress in the calculation of the CP violating  amplitudes related to $\varepsilon$ and $\varepsilon^\prime/\varepsilon$, which is included in our analysis for the first time; in  sec.\,\ref{sec:SMUTA} we present the results of our updated analysis in the Standard Model. Finally, in sec.\,\ref{sec:concl} we present our conclusion and an outlook for future developments.

\section{Update of theoretical and experimental inputs}
\label{sec:inputs}
\begin{table}[ht]
\centering
\begin{tabular}{|c |c |c| c|c|}
\hline
Input &Reference & Measurement  & \utfit\  Prediction   & Pull \\ [0.5ex]
\hline
$\sin 2\beta$ & \cite{HFLAV:2019otj}, \utfit\ & $0.688(20)$ & $0.736(28)$ & $- 1.4$ \\
$\gamma$      & \cite{HFLAV:2019otj}          & $66.1(3.5)$   & $64.9(1.4)$ & $+0.29$ \\
$\alpha$      & \utfit\                       & $94.9(4.7)$   & $92.2(1.6)$ & $+0.6$ \\
$\varepsilon \cdot 10^3$ & \cite{ParticleDataGroup:2022pth} & $2.228(1)$ & $2.00(15)$ & $+1.56$ \\
\hline
$|V_{ud}|$    & \utfit\                       & $0.97433(19)$ & $0.9738(11)$ & $+0.03$ \\
\hline
$|V_{ub}| \cdot 10^3$   $^\bullet$        & \utfit\     & $3.77(24)$ & $3.70 (11)$ & $+0.25$ \\ 
~~~$|V_{ub}| \cdot 10^3$ (excl) &\cite{Aoki:2021kgd}& $3.74(17)$ & &  \\
~~~$|V_{ub}| \cdot 10^3$ (incl) &\cite{HFLAV:2019otj}& $4.32(29)$ & &  \\
\hline
$|V_{cb}| \cdot 10^3$  $^\bullet$  & \utfit\            & $41.25(95)$ & $42.22(51) $ & $-0.59$ \\
~~~$|V_{cb}| \cdot 10^3$ (excl) & \utfit\     & $39.44(63)$  & & \\
~~~$|V_{cb}| \cdot 10^3$ (incl) & \cite{Bordone:2021oof}& $42.16(50) $  &&  \\
\hline
~~~$\vert V_{ub}\vert /\vert V_{cb}\vert$ &  \cite{Aoki:2021kgd} &0.0844(56)&& \\ \hline 
$\Delta M_d \times 10^{12}\, {\rm s}^{-1}$ &  \cite{ParticleDataGroup:2022pth}& $ 0.5065(19)  $   & $ 0.519(23)$& $ -0.49 $ \\ 
$\Delta M_s \times 10^{12}\, {\rm s}^{-1}$ &  \cite{ParticleDataGroup:2022pth}& $ 17.741(20)  $  & $ 17.94(69) $& $ -0.30 $ \\ 
BR$(B_s\to\mu\mu)\times 10^{9}$ & \cite{ParticleDataGroup:2022pth}& $ 3.41(29) $ & $ 3.47(14)  $  &  $ -0.14 $ \\ 
BR$(B \to \tau\nu)\times10^4$ & \cite{ParticleDataGroup:2022pth} & $1.06(19)$ & $0.869(47)$ & $+0.96$ \\
\hline
$ \mathrm{Re}\left(\varepsilon^\prime/\varepsilon\right) \times10^4$&\cite{ParticleDataGroup:2022pth} &$16.6 (3.3) $  & $15.2(4.7)$ & $+0.27$ \\
$\phi_{\varepsilon}$  & \cite{ParticleDataGroup:2022pth} & $0.7596 \,\mathrm{rad}$ &   &  \\
$\omega$              & \cite{ParticleDataGroup:2022pth}& $0.04454(12)$   &  &  \\
$\delta_0(s)$ $^{*}$  & \cite{ParticleDataGroup:2022pth}&$32.3(2.1)^\circ$& $32.3 (1.7)^\circ$ at $s=471.0$\,MeV  \cite{Blum:2015ywa}&\\
$\delta_2(s)$ $^{*}$  & \cite{ParticleDataGroup:2022pth} &$-11.6(2.8)^\circ$& $-7.96(37)^\circ$ at $s=479.1$\,MeV   \cite{Blum:2015ywa}&\\
$ \mathrm{Re}\left(A_{2}\right)$  & \cite{ParticleDataGroup:2022pth}$^{**}$  & $1.479(4) \times 10^{-8}\, \mathrm{GeV}$ &$1.50(10) \times 10^{-8}\, \mathrm{GeV}$ & \\
$ \mathrm{Re}\left(A_{0}\right)$  & \cite{ParticleDataGroup:2022pth}$^{***}$ & $3.3201(18) \times 10^{-7}\, \mathrm{GeV}$& $3.01(41) \times 10^{-7}\, \mathrm{GeV}$ &  \\
$ \mathrm{Im}\left(A_{0}\right)$  &\utfit\ & &$-6.75 (86) \times 10^{-11}$\,GeV  &   \\
$ \mathrm{Im}\left(A_{2 }\right)$ &\utfit\ & &$ -8.4 (1.2) \times 10^{-13}$\,GeV  & \\
  \hline
\end{tabular}
\caption{{\it  Full SM inputs with their predictions from the SM global fit. When the averages are  made by us, using values and errors from other calculations,  the reference is  denoted by \utfit\  and the procedure used to obtain the final value and  uncertainty is explained in the text. $^\bullet$  These values have been obtained by combining the exclusive and inclusive  values  of $\vert V_{ub}\vert$, $\vert V_{cb}\vert$ and $\vert V_{ub}\vert /\vert V_{cb}\vert$ reported in this table as explained in sec.\,\ref{sec:CKM}.  $^{*}$ $s$ denotes  the  c.o.m. energy of the two pions; $^{**}$ $ \mathrm{Re}\left(A_{2}\right)$ is extracted from the $K^+ \to \pi^+ \pi^0$ decay widths;  $^{***}$ $\mathrm{Re}\left(A_{0}\right)$ is computed  using the $K^0 \to \pi^+ \pi^- $ and $K^0 \to \pi^0 \pi^0$  decay widths as explained in the text. The theoretical  values of $\delta_{0,2}(s)$ in the fourth column of the table  have  been taken  from ref.\,\cite{Blum:2015ywa}.}}
\label{tab:fullSM}
\end{table}
\begin{table}[ht]
\centering
\begin{tabular}{|c |c |c| }
\hline
Input & Reference & Measurement  \\ [0.5ex]
\hline
$\tau_{D^0 }\cdot 10^{13} s$                 & \cite{ParticleDataGroup:2022pth}   & $0.4101 \pm 0.0015$    \\
$\tau_{B^0} \cdot 10^{12} s$             & \cite{HFLAV:2019otj} & $1.519 (4)$           \\
$\tau_{B^+} \cdot 10^{12} s$             & \cite{HFLAV:2019otj} & $1.638 (4)$            \\
$\tau_{B_s} \cdot 10^{12} s$             & \cite{HFLAV:2019otj} & $1.516 (6)$          \\
$\Delta \Gamma_s/\Gamma_s$ & \cite{HFLAV:2019otj} & $0.112 (10)$         \\
\hline
$\alpha_s(M_Z)$ & \cite{deBlas:2022hdk} & 0.11792(94) \\ 
$m_t^{\overline{\mathrm{MS}}}(m_t^{\overline{\mathrm{MS}}})$ \,(GeV)                 &\cite{deBlas:2022hdk}            &  $163.44 (43)$       \\ 
 \hline 
\end{tabular}
\caption{\it Extra inputs used in our UTA analysis. The quoted value of $m_t^{\overline{\mathrm{MS}}}(m_t^{\overline{\mathrm{MS}}})$ corresponds to $m_t^\mathrm{pole} = (171.79 \pm 0.38)$ GeV [40].}
\label{tab:extraSM}
\end{table}

\begin{table}[ht]
\centering
\begin{tabular}{|c |c|c|}
\hline
Input & Lattice \\ [0.5ex]
\hline
$\hat B_K$ & $0.756(16)$   \\ 
$f_{B_s}$ & $230.1(1.2)$\,MeV \\
$f_{B_s}/f_{B}$ & $1.208(5)$  \\
$\hat B_{B_s}$ & $1.284(59)$  \\
$\hat B_{B_s}/\hat B_{B}$ & $1.015(21)$  \\ 
$m^{\msb}_{ud}(2$\,GeV$)$ & $3.394(29) $\,MeV  \\
$m^{\msb}_{s}(2$\,GeV$)$ & $93.11(52) $\,MeV \\
$m^{\msb}_{c}(3$\,GeV$)$ & $991(5) $\,MeV  \\
$m^{\msb}_{c}(m^{\msb}_{c})$  &    $1290(7)     $ \,MeV \\
$m^{\msb}_{b}(m^{\msb}_{b})$  &   $ 4196(14)     $ \,MeV  \\
\hline 
\end{tabular}
\caption{\it  Full lattice  inputs. The values of the different quantities  have been obtained by taking the weighted average of  the   $N_f =2+ 1 $  and $N_f = 2 + 1+1$  FLAG numbers\,\cite{Aoki:2021kgd}.}
\label{tab:fullLattice}
\end{table}
In this section we  discuss the main experimental and theoretical updated  inputs that have been used in our 
UTA analysis.   A list of the experimental inputs can  be found in tables\,\ref{tab:fullSM} and \ref{tab:extraSM}.
Most of the experimental inputs, as the values of the CKM matrix elements $\vert V_{ij}\vert$, 
are extracted from  data using some theoretical calculation, e.g. the form factors in the case of semi-leptonic decays. We will also present a  discussion  of the decay constants, form factors and $B$-parameters used in this work with a comparison
of  the values used  by other groups. Most of the  inputs for these non-perturbative QCD parameters   are taken from the most recent lattice determinations and given in table\,\ref{tab:fullLattice}.   As already mentioned,  our  general prescription  is to  take the weighted average of the $N_f = 2 + 1$  and $N_f = 2 + 1+1$ FLAG numbers\,\cite{Aoki:2021kgd}.  We will discuss in this paper only  those  cases where we followed a different procedure and explain the reasons for the different choice. 

In table\,\ref{tab:fullSM}, the values denoted as \utfit\  prediction  in the fourth column are obtained from  the UT analysis by excluding the quantity under consideration. Thus, for example,  $|V_{cb}| \cdot 10^3=42.22(51) $  and $|V_{ub}| \cdot 10^3=3.70(11) $ have been obtained by excluding from the UT analysis the input values of these CKM matrix elements.
\subsection{The CKM matrix elements}
\label{sec:CKM}
In this subsection we discuss the updated   absolute values of several elements of  the Cabibbo-Kobayashi-Maskawa (CKM) 
matrix\,\cite{Cabibbo:1963yz,Kobayashi:1973fv}.
\subsubsection{$\vert V_{ud}\vert$   and  $\vert V_{us}\vert$}
$\vert V_{ud}\vert$   and $\vert V_{us}\vert$ are determined from super allowed $0^+ \to 0^+$ nuclear $\beta$ decays 
and from a combined analysis of $K_{\mu 2}$, $K_{\ell 3}$ and $\pi_{\mu 2}$ decays.

One  possibility is the  extraction of  $\vert V_{ud}\vert$   and $\vert V_{us}\vert$ from the lattice determination of $f^+(0)$  and $f_K/f_\pi$ using the relations\,\cite{Moulson:2017ive,ParticleDataGroup:2022pth}
\begin{equation}  \vert V_{us}\vert  f^+(0) = 0.2165(4)  \, , \qquad   \frac{\vert V_{us}\vert f_K}{\vert V_{ud}\vert  f_\pi} =0.2760(4)\, . \end{equation}
The lattice  results for  $f^+(0)$  or  $f_K/f_\pi$  have been obtained  averaging the results of lattice  calculations performed either within the four flavour, $N_f = 2 + 1+1$, or within the three flavour,  $N_f = 2 + 1$, theory\,\cite{Aoki:2021kgd}, the corresponding results for  $\vert V_{ud}\vert$  and $\vert V_{us}\vert$  are given in table\,\ref{table:vudvus}.   
The weighted average for $\vert V_{ud}\vert$  is 
\beq \vert V_{ud}\vert=0.974387(98)\,  . \label{eq:vudl}\eeq

In the past the super allowed nuclear $\beta$   transitions provided the most precise determination of $\vert V_{ud}\vert$.  Its accuracy is limited by the hadronic uncertainties of the  electroweak  radiative corrections, some of which  are universal for all the nuclei whereas others  depend on the specific nucleus structure.  A new critical survey which takes into account  the most recent experimental results and  the new theoretical calculations of the radiative corrections\,\cite{Marciano:2005ec, Seng:2018yzq, Seng:2018qru, Czarnecki:2019mwq}  has been presented in ref.\,\cite{Hardy:2020qwl}.    The average of the data, including  radiative and isospin-symmetry-breaking corrections,  yields the  CKM matrix    element
\beq \vert V_{ud}\vert = 0.97373(31) \, , \label{eq:vudb}\eeq
the uncertainty of which is larger than the value quoted in Eq.\,(\ref{eq:vudl}).
 The value in Eq.\,(\ref{eq:vudb}) is  lower   than the previous  2015 result by one standard deviation and its uncertainty is  increased by $50\%$. The variation   is a consequence of new calculations for  the radiative corrections  and of the  spread between different estimates of these corrections.  Hopefully,  at least for the universal corrections,  a new more accurate determination will come from lattice calculations.   Indeed,   only the easiest cases, namely the radiative corrections to $K_{\mu 2}$  and $\pi_{\mu 2}$ decays,  have been computed  so far on the lattice from first principles and without any model assumption/approximation\,\cite{Giusti:2017dwk,DiCarlo:2019thl}. 

By combining with the PDG method~\cite{ParticleDataGroup:2022pth} (see also ref.\,\cite{DAgostini:1999niu})  the most precise results   from   refs.\, \cite{Aoki:2021kgd} and \cite{Hardy:2020qwl},   eqs.\,(\ref{eq:vudl}) and (\ref{eq:vudb}),  we quote the final results
\bea \vert V_{ud}\vert= 0.97433 (19)\, ,    \qquad \qquad     \vert V_{us}\vert=  0.2251 (8)\, , \eea
where $\vert V_{us}\vert$ is indeed obtained from the unitarity of the CKM matrix in the SM and it is therefore not used in the UTA. 
\begin{table}[]
\begin{tabular}{|c|c|c|c|}
\hline
      & Reference      & $\vert V_{ud}\vert$       & $\vert V_{us}\vert$              \\ \hline
$\beta$  decay & \cite{Hardy:2020qwl}  & 0.97373 (31)&   \\ \hline
$N_f=2+1$ & \cite{Aoki:2021kgd}  & 0.97438 (12)    & 0.2249 (5)  \\ \hline
$N_f=2+1+1$ &\cite{Aoki:2021kgd}  &0.97440 (17)   & 0.2248 (7)  \\ \hline
$\tau$  decay & \cite{HFLAV:2019otj}  & 0.97561 (40)  & 0.2195 (19)  \\ \hline
$\tau$  decay & \cite{Hudspith:2017vew, Maltman:2019xeh}  & 0.97461 (43)  & 0.2240 (18)  \\ \hline
\end{tabular}
\caption{{\it Values of $\vert V_{ud}\vert$  and $\vert V_{us}\vert$  from different physical processes. For completenesss  we also give the values obtained from $\tau$ decays although they are not used in the present paper.}}
	\label{table:vudvus}
\end{table}
For a recent reappraisal of the determination of  $\vert V_{ud}\vert$ and $\vert V_{us}\vert$ and of the problem of the unitarity of the first  CKM  matrix row see also  ref.\,\cite{Cirigliano:2022yyo}.
\subsubsection{$\vert V_{cb}\vert$ }
Semileptonic $B\to  D^{(*)}\ell \nu_\ell$  decays and their inclusive counter part  are very  important
processes in the phenomenology of flavor physics.  From their measurement and the corresponding theoretical predictions depends the value of  $\vert V_{cb}\vert$ which plays a fundamental role in the UTA analyses\,\cite{Alpigiani:2017lpj,Blanke:2018cya,Buras:2021nns}.  For years we had to live with the apparent strong tension between the inclusive\,\cite{Gambino:2013rza,Alberti:2014yda,Gambino:2016jkc} and the exclusive determination of this CKM matrix  element \cite{BaBar:2007nwi, BaBar:2007cke, BaBar:2008zui, BaBar:2009zxk, Belle:2010qug, Belle:2015pkj, Belle:2017rcc, Belle:2018ezy}.  Some important novelties have, however, recently  changed the previous situation: on  the one hand the inclusive predictions were recently reconsidered and the uncertainties of the calculation performed in the Heavy Quark Effective Theory were reevaluated\,\cite{Bordone:2021oof,Bernlochner:2022ucr}. On the other hand,  new lattice calculations of the relevant form factors in the small recoil region\,\cite{FermilabLattice:2021cdg}, new  approaches to their  determination in the full  kinematical  range\,\cite{DiCarlo:2021dzg, Martinelli:2021frl, Martinelli:2022adr, Martinelli:2022xir, Martinelli:2021myh} and measurements of the exclusive differential decay rates were presented. We think that it is possible to argue that  for $\vert V_{cb}\vert$, although some difference remains,  the tension is finally resolved, see the recent average from ref.\,\cite{Martinelli:2022xir} given in eq.\.(\ref{eq:vcbDMex}) below.   A set of values from different estimates of $\vert V_{cb}\vert$ from inclusive and exclusive decays are given in table\,\ref{table:vcb}.

\begin{table}[]
\begin{tabular}{|c|c|c|c|}
\hline
   &  Process & Reference      & $\vert V_{cb}\vert \cdot 10^3$                \\ \hline
1& $b\to c $ inclusive  & \cite{Bordone:2021oof}  &42.16  (50)   \\ \hline
2&$B \to D $ &\cite{Martinelli:2022adr}  DM &41.0  (1.2)    \\ \hline
3&$B \to D $  $N_f=2+1$ &\cite{Aoki:2021kgd} &   40.0 (1.0) \\ \hline
4&$B_s \to D_s $  $N_f=2+1$ &\cite{Martinelli:2022xir} DM &   41.7 (1.9) \\ \hline
5&$B \to D^*$ & \cite{Martinelli:2021myh}  DM & 41.3  (1.7)      \\ \hline
6&$B \to D^*$ & \cite{Gambino:2019sif} & 39.6  (1.1)      \\ \hline
7&$B \to D^*$ &\cite{Jaiswal:2020wer}& 39.6  (1.1)      \\ \hline
8&$B \to D^*$  $N_f=2+1$ &\cite{Aoki:2021kgd} & 38.9  (0.9)      \\ \hline
9&$B \to D^*$  $N_f=2+1+1 $ &\cite{Aoki:2021kgd} & 39.3  (1.4)$^a$      \\ \hline
10&$B \to D^*$ and  $B \to D $  $N_f=2+1$ &\cite{Aoki:2021kgd} & 39.4  (0.7)      \\ \hline
11&$B_s \to D^*_s $  $N_f=2+1$ &\cite{Martinelli:2022xir}  DM &  40.7 (2.4) \\ \hline
\end{tabular}
\caption{{\it Values of $\vert V_{cb}\vert$ from inclusive or exclusive determinations. $^a$ This value of $\vert V_{cb}\vert \times 10^3$ has been derived using the value of the form factor at zero recoil given in Eq.\,(267) of ref.\cite{Aoki:2021kgd}. DM in the rows 2-4-5-11 denotes the values obtained by using the Dispersive Matrix approach mentioned in the text.  For completenesss  we also give some determinations  which are not used in the present paper.  
More recent determinations of  $\vert V_{cb}\vert$ from $B\to D^{(*)}$ semileptonic decays by Belle II have been presented at the 2022 ICHEP Conference in Bologna by T. Koga (KEK).  Since, however,   LQCD form factors and experimental  data  were simultaneously used to extract the value of  $\vert V_{cb}\vert$, we decided not to use these results for the time being.}}
	\label{table:vcb}
\end{table}

For the exclusive determination  of $\vert V_{cb}\vert$ we proceed as follows:
\begin{itemize} 
\item For  $ B \to D^*$ semileptonic decays, rather than making an average of the values of $\vert V_{cb}\vert $ given in the rows 8-9 of table\,\ref{table:vcb}, following the procedure adopted for other quantities in this paper,  we average the form factor $F(1)$ obtained from  $N_f=2+1$, $F(1)=0.906(13)$, and  $N_f=2+1+1$, $F(1)=0.895(10)(24)$\,\cite{Aoki:2021kgd}, obtaining  $F(1)=0.904(11)$;
\item Then, using the formula derived from  the rate,  $ F(1)\, \eta_{EW} \, \vert V_{cb}\vert = 35.44 (64) 10^{-3}$\,\cite{Aoki:2021kgd},   and $ \eta_{EW} = 1.00662$\,\cite{Sirlin:1981ie}  we get $\vert V_{cb}\vert = 38.95 (86) 10^{-3}$; 
\item For $B\to D$, following\,\cite{Aoki:2021kgd},    we quote     $\vert V_{cb}\vert= 40.0(1.0) 10^{-3}$;
\item Averaging   the above  values  of $\vert V_{cb}\vert$ from   $ B \to D^*$ and $ B \to D$    we obtain
\beq \vert V_{cb}\vert \cdot 10^3 \, \, \, {\rm (excl.)} =  39.44 (65)\, . \label{eq:eq3}\eeq
\end{itemize} 
This procedure uses all the available information from $ B \to D^*$ but neglects the correlation  of the lattice determination of form factors for $ B \to D^*$ and $ B \to D$   decays obtained using the same gauge field configurations.  Alternatively,  we combined   the $N_f=2+1$ value of  $\vert V_{cb}\vert$  by FLAG, obtained by averaging  $ B \to D^*$ and $ B \to D$   decays and taking into account the correlation  of the lattice determination of form factors for these   decays obtained using the same gauge field configurations,  $\vert V_{cb}\vert=39.48 (68)  10^{-3} $,  with the value that we can obtain from 
the form factor $F(1)$ with $N_f=2+1+1$,   $\vert V_{cb}\vert=39.34 (1.35) 10^{-3}$.  We obtain
\beq \vert V_{cb}\vert \cdot 10^3 \, \, \, {\rm (excl.)} =  39.45 (61)\, ,  \eeq
from which, after combining  using the PDG method~\cite{ParticleDataGroup:2022pth} with the result in eq.\,(\ref{eq:eq3}),  we get  our final result 
\beq \vert V_{cb}\vert \cdot 10^3 \, \, \, {\rm (excl.)} =  39.44 (63)\, ,  \label{eq:vcbex}\eeq
which differs  by   3.4  $\sigma$  from the inclusive value in table\,\ref{table:vcb}.
We may combine the inclusive value of  $\vert V_{cb}\vert $ in table\,\ref{table:vcb} with the result  in Eq.\,(\ref{eq:vcbex}) obtaining
\beq \vert V_{cb}\vert \cdot 10^3 \, \, \,  =  41.1 (1.3)   \, \, \, {\rm (incl.+excl.)}\, . \label{eq:vcbtot}\eeq

We anticipate that a more accurate determination of $\vert V_{cb}\vert$,  that is the one used in this \utfit \, analysis and quoted in Table\,\ref{tab:fullSM}, will be obtained by combining Eq.\,(\ref{eq:vcbtot}) with the determination of $\vert V_{ub}\vert$  and of the ratio $\vert V_{ub}\vert/\vert V_{cb}\vert$ from $B_s\to (K^-,D_s^-)\mu^+\nu_\mu$ decays  in Eq.\,(\ref{eq:raubcb}).
 
An  alternative determination of the exclusive value of $\vert V_{cb}\vert$ can be obtained by using the values obtained using the Dispersive Matrix (DM) approach of Ref.\,\cite{DiCarlo:2021dzg}  and given in Table\,\ref{table:vcb}, Refs.\,\cite{Martinelli:2022adr, Martinelli:2022xir, Martinelli:2021myh}.
 By combining these results, which include $B_s \to D^{(*)}_s$ decays, we obtain
\beq \vert V_{cb}\vert \cdot 10^3 \, \, \, {\rm (DM \,\,\,  excl.)} =  41.2 (8)\, ,  \label{eq:vcbDMex}\eeq
namely a value much closer and compatible at  the 1 $\sigma$ level with  the inclusive one,  with an uncertainty comparable to the uncertainty quoted in Eq.\,(\ref{eq:vcbex}) . By combining the inclusive value of Table\,\ref{table:vcb} with the DM result in Eq.\,(\ref{eq:vcbDMex})
we obtain the   result
\beq \vert V_{cb}\vert \cdot 10^3 \, \, \,  =  41.9 (4)  {\rm  (incl.+ DM \,\,\,  excl.)}\, , \label{eq:vcbtotDM}\eeq
more precise than the result of Eq.\,(\ref{eq:vcbtot}).

\subsubsection{$\vert V_{ub}\vert$ }

The  matrix element $V_{ub}$ is determined from the measurements of the
branching ratios of  leptonic $B \to \tau \nu_\tau$ decays and from {\it exclusive} and {\it inclusive} semileptonic
$b \to u$ decays. Theoretically, its precision is limited by the uncertainty of the
 calculations of the $B$ meson decay constant and of the relevant form factors, for leptonic and exclusive semileptonic decays, and of the
matrix elements of the operators appearing in the HQET expansion of the inclusive rate.   For   $B \to \tau \nu_\tau$, which is very interesting because it is particularly sensitive to physics beyond the SM,   the main  source of  uncertainty comes for the large  error in the experimental measurement of the  rate.   Although the determinations from inclusive semileptonic decays are  systematically higher than  the exclusive ones, the two values are  compatible, once the spread of the inclusive determinations using   different theoretical models is considered. 

For the leptonic  decays  we average  the results given in Eqs.\,(289) and (290) of ref.\,\cite{Aoki:2021kgd} obtaining the  value  $a)$ of eq.\,(\ref{eq:eq9}). For the exclusive semileptonic decays  we take the number of table\, 57  of  the same reference, quoted in  $b)$.   We  did  not use the recent value of  $\vert V_{ub}\vert$ from the Belle II  Collaboration\,\cite{Belle-II:2022imn}  since it is rather preliminary and based only on the form factors of the FNAL/MILC Collaboration\,\cite{FermilabLattice:2015mwy}.  We plan to include  it in the future  by using in the determination  of  $\vert V_{ub}\vert$, besides the form factors  computed by FNAL/MILC,   also the recently computed form factors   by RBC/UKQCD\,\cite{Flynn:2015mha} and  JLQCD\,\cite{Colquhoun:2022atw}.  Finally,    for inclusive semi-leptonic  decays  we use the  value from ref.\,\cite{HFLAV:2019otj}.   For completeness, we give the average of  $a)$ and   $b)$ in $d)$;  the average of  $c)$ and $d)$ in $e)$
and the average of $b)$ and $c)$ in $f)$.   We note that, in  the case of the value  of $\vert V_{ub}\vert$,  a difference  between the  inclusive and exclusive determinations at  the 1.7 $\sigma$ level still persists,  although with large relative errors.

We observe  that the effect of including $B \to \tau \nu_\tau$ is almost invisible and, for reasons explained below,  adopt in the following the average in $f)$. 
\begin{eqnarray}
 V_{ub}^{B\to \tau} &=&   4.05 (64)\cdot 10^{-3}  \quad   a) \nonumber \\
V_{ub}^{B\to \pi}&=&  3.74 (17) \cdot 10^{-3}  \quad   b) \nonumber \\
 V_{ub}^{\rm incl.\,\,}&=&  4.32 (29) \cdot 10^{-3}  \quad   c)  \label{eq:eq9}\\
 V_{ub}^{a+b\,\,} &=&  3.76 (16) \cdot 10^{-3}  \quad   d) \nonumber \\
  V_{ub}^{c+d\,\,}&=&  3.89 (24) \cdot 10^{-3}  \quad   e)\nonumber \\ 
  V_{ub}^{b+c\,\,}&=&  3.89 (25) \cdot 10^{-3}  \quad   f)\nonumber 
 \end{eqnarray}
A  percent precision is expected to be reached by LQCD using
Exaflops CPUs for $f_B$ and for the form factors entering the exclusive
determination of $\vert V_{ub}\vert$.  A higher  precision will require the non-perturbative calculation of the radiative 
corrections to the decay rates\,\cite{DiCarlo:2019thl}.  Considering
how challenging the measurement of $BR(B \to \tau \nu)$ in a hadronic
environment is, it is difficult to imagine a similar improvement in
precision of the experimental measurement for this channel for which a higher theoretical accuracy can be reached. 
On the other hand, it was pointed out in ref.~\cite{UTfit:2006vpt}  that the indirect determination of $\vert V_{ub}\vert$ from the 
fit in the SM is presently more accurate than the measurements, yielding a   central value close to the {\it exclusive} determination.  Therefore the most
 precise prediction of $BR(B \to \tau \nu)$ in the SM can be   obtained by combining the {\it indirect} knowledge of $\vert V_{ub}\vert$ from the rest of   the UT fit,  fourth column in table\,\ref{tab:fullSM},    combined  with   $f_B$ derived  from $f_{B_s}$ and $f_B/f_{B_s}$ in  table\,\ref{tab:fullLattice}
\beq  BR(B \to \tau \nu)_{\rm {\bf UT}fit }  = 0.882 (45)   \, .  \eeq

The progress of lattice calculations allow us to use in the analysis also the constraint coming from the ratio $\vert V_{ub}\vert/\vert V_{cb}\vert$
determined either from $\Lambda_b \to (p,\Lambda_c) \mu^-\bar\nu_\mu$ or $B_s\to (K^-,D_s^-)\mu^+\nu_\mu$ decays.  We use only the latter  decays since the lattice form factors relevant for  $\Lambda_b$ decays do not satisfy the quality criteria of  FLAG\,\cite{Aoki:2021kgd}. Following\, \cite{Aoki:2021kgd}  we quote
\beq \frac{\vert V_{ub}\vert}{\vert V_{cb}\vert} =0.0844(56) \, .\label{eq:raubcb} \eeq 

We have  combined the information from exclusive  and inclusive decays,  eqs.\,(\ref {eq:vcbex}) and (\ref{eq:eq9})-b) and table\,\ref{table:vcb}
and  eq.\,(\ref{eq:eq9})-c) respectively,   and the ratio $\vert V_{ub}\vert/\vert V_{cb}\vert$ in eq.\,(\ref{eq:raubcb}) using the average procedure of refs.~\cite{DAgostini:1999niu} obtaining the following input values that we have  used in the present UT analysis
\beq \vert V_{cb}\vert \cdot 10^3 = 41.25 (95)   \qquad \qquad \vert V_{ub}\vert \cdot 10^3 = 3.77 (24)  \, ,   \qquad \qquad  \rho_c=0.11\label{eq:usedVubcb} \eeq
where $\rho_c$ is the correlation  matrix element; had we used the results from the DM analysis, eq.\,(\ref{eq:vcbDMex}), we would have obtained 
\beq \vert V_{cb}\vert \cdot 10^3 = 41.94 (80)   \qquad \qquad \vert V_{ub}\vert \cdot 10^3 = 3.79 (24) \, ,   \qquad \qquad  \rho_c=0.09 \, . \label{eq:usedVubcbb} \eeq
Had  we  used the most recent PDG value  of  $\vert V_{ub}\vert$    from   inclusive decays\,\cite{ParticleDataGroup:2022pth},   $\vert V_{ub}\vert \cdot 10^3=4.13(26) $, we  would have obtained 
$\vert V_{cb}\vert \cdot 10^3 = 41.24 (95)$  and    $\vert V_{ub}\vert \cdot 10^3 = 3.76 (23)$ with $ \rho_c=0.11$,  substantially identical to the values given in Eq.\,(\ref{eq:usedVubcb}),  or, in the DM case,   $\vert V_{cb}\vert \cdot 10^3 = 41.94 (80)$ and  $ \vert V_{ub}\vert \cdot 10^3 = 3.77 (23)$ with   $\rho_c=0.09$,   substantially identical to the numbers of  Eq.\,(\ref{eq:usedVubcbb}).

\begin{figure}[!t]
\hspace*{-0.6cm}
  \centering
    \begin{tabular}{ccc}
    \includegraphics[width=0.35\linewidth]{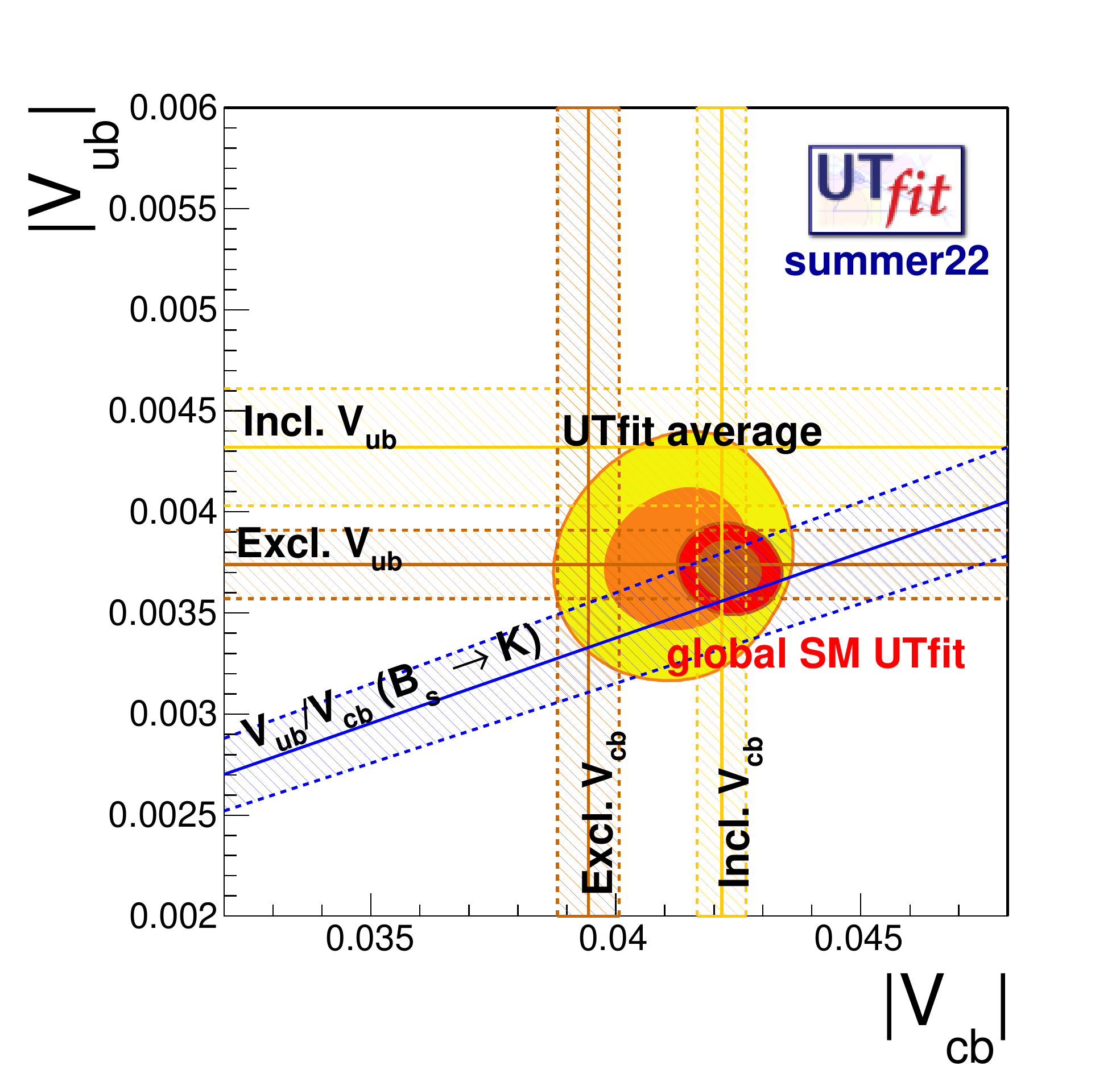} &
    \hspace*{-0.6cm}
    \includegraphics[width=0.39\linewidth,trim=0 -1.1cm 0 0]{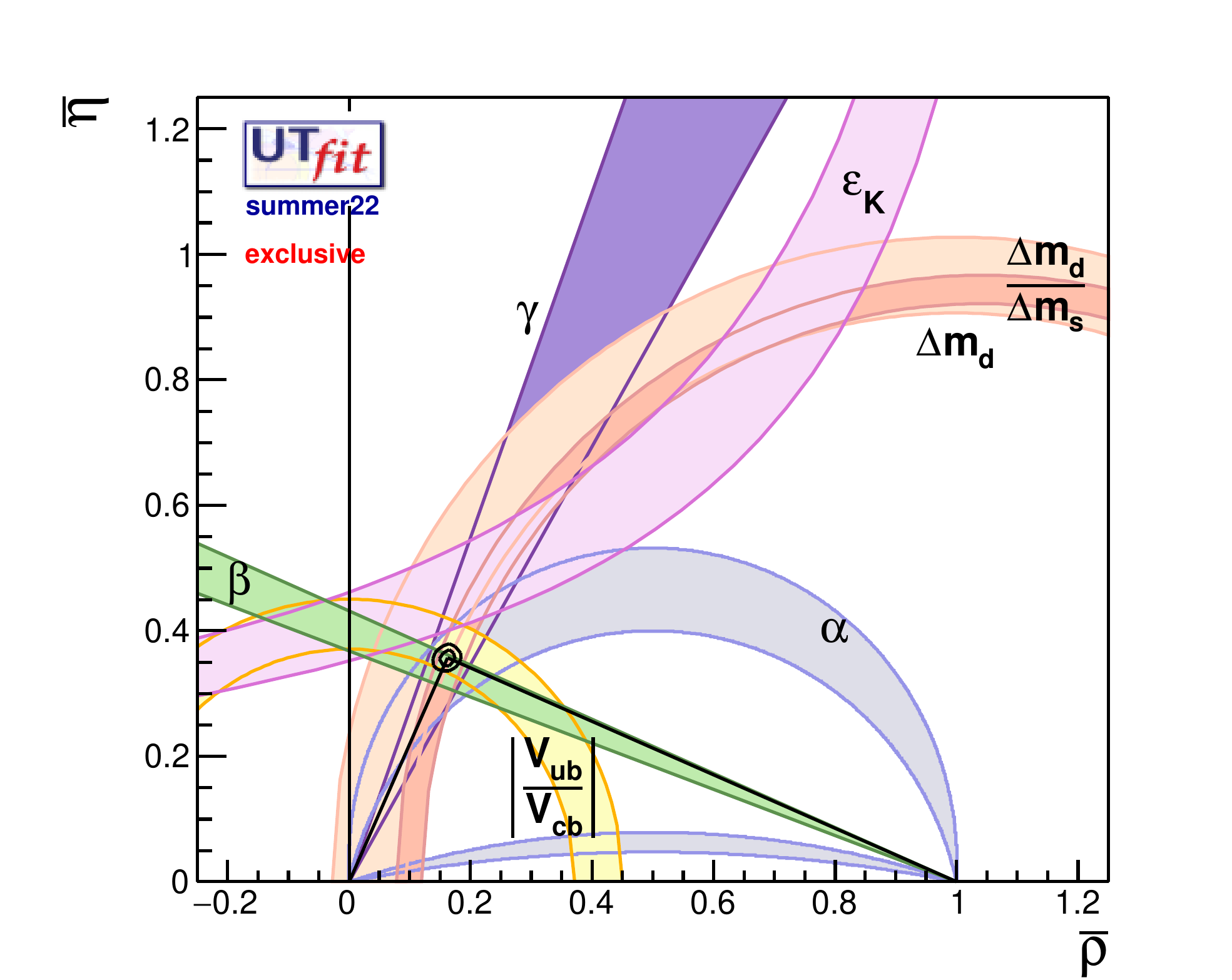} &
    \hspace*{-0.9cm}
    \includegraphics[width=0.39\linewidth,trim=0 -1.1cm 0 0]{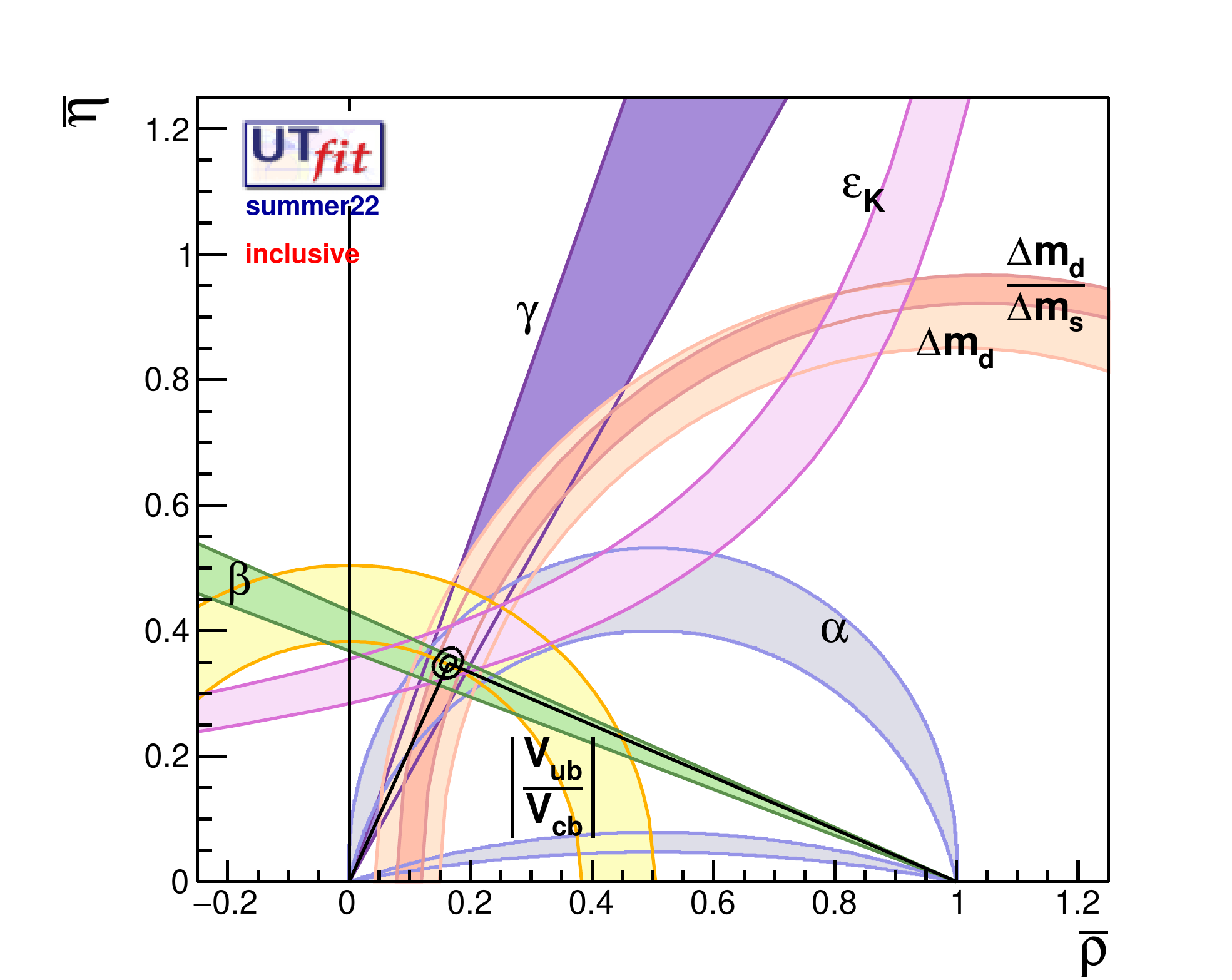}
    \hspace*{-0.9cm}
  \end{tabular}
  \vspace*{-0.3cm}
  \caption{{\it Left Panel: $|V_{cb}|$ vs $|V_{ub}|$ plane showing
    the values reported in Table~\ref{tab:fullSM}.
    We include  in the figure  the  ratio $|V_{cb}|/|V_{ub}|$ from ref.\,\cite{Aoki:2021kgd}  shown as a diagonal (blue)  band; 
    Central Panel: $\bar \rho$-$\bar \eta$ plane with the SM global fit results
    using only exclusive inputs for both $V_{ub}$ and $V_{cb}$; Right Panel:
   SM global fit results using only inclusive inputs. In  the central and right panels,  $\varepsilon_K=\vert \varepsilon\vert$ where  $\varepsilon$ 
   is defined in eq.\,(\ref{eq:eve}).}}
  \label{fig:vubvcb}
\end{figure}

The values of the CKM matrix elements $\vert V_{ub}\vert$ and $\vert V_{cb}\vert$  from exclusive  and inclusive decays,  eqs.\,(\ref {eq:vcbex}) and (\ref{eq:eq9}) and table\,\ref{table:vcb}
and  eq.\,(\ref{eq:eq9}) respectively,   and the ratio $\vert V_{ub}\vert/\vert V_{cb}\vert$ in eq.\,(\ref{eq:raubcb}) are  shown in  the left plot of fig.~\ref{fig:vubvcb}.    This figure highlights the  inclusive-vs-exclusive tensions already  discussed
by the \utfit~collaboration since $2006$~\cite{UTfit:2006vpt}.
In the same figure the allowed  two-dimensional (2D) region,   eq.(\ref {eq:usedVubcb}),  calculated with a 2D procedure
inspired by the skeptical method of Ref.~\cite{DAgostini:1999niu} with $\sigma=1$ and denoted as \utfit\ \,average, is   shown in the plot.
The allowed region, as  determined by the global fit and   denoted as global SM  \utfit\ \, is also given there. 
The global fit strongly prefers  a  value  of  $\vert V_{cb}\vert$  close to its   value  from inclusive decays  and a value of $\vert V_{ub}\vert$ close to its exclusive value. This results find a further support by the values denoted as  \utfit\ predictions shown in   the fourth column of table\,\ref{tab:fullSM}. 

\section{CP violation in  the $K^0$-$\overline{K^0}$ system: $\varepsilon$ and $\varepsilon^\prime/\varepsilon$}
\label{sec:eps}
In this section we discuss an  update in  the theoretical evaluation  of  $\varepsilon$  and the inclusion, for the first time in the UTA  analysis,  
of  the ratio  $\varepsilon^\prime/\varepsilon$\,\cite{RBC:2020kdj}. Although the latter still suffers from large uncertainties, its determination is a very important step in our understanding of  CP violation in the SM.

\subsection{An  updated evaluation of $\varepsilon$ }
 The parameter $\varepsilon$, which describes indirect CP-violation in the $K^0 - \overline{K}^0$ system, represents one of the most interesting observables in Flavor Physics.  It plays an important role in the UTA  both within the SM  and beyond,  see\,\cite{Ciuchini:2000de}-\cite{Alpigiani:2017lpj} and  references therein.  As the phenomenon of $K^0 - \overline{K}^0$ mixing is a loop process further suppressed by the GIM mechanism, $\varepsilon$ turns out to be a powerful constraint on New Physics models, for which it is important to have experimental and theoretical uncertainties well under control.

The standard formula for the evaluation of  $\varepsilon$  is  the following
\be
\varepsilon = e^{i \, \phi_\varepsilon} \sin \phi_\varepsilon \, \left( \frac{\mathrm{Im}M_{12}}{\Delta m_K} + \xi_0 \right)\, , \label{eq:eve}
\ee
where $\xi_0 = \mathrm{Im }A_0 / \mathrm{Re} A_0$.   At the leading order in the expansion  in inverse powers of the charm mass, $m_c$, $\mathrm{Im} M_{12}$  is  given by 
\bea \mathrm{Im} M_{12}= 
\langle \overline{K}^0 \vert  H^{ \mathrm{eff}} \vert K^0\rangle &=& \frac{G_F^2}{16\, \pi^2} \, M_W^2 \, \left[ 
 \mathrm{Im}\left(\lambda_c^2\right) \, \left( \eta_1 S_0(x_c) - \eta_3 S_0(x_c,x_t) \right) + \right. \nn \\
&& \left. +  \mathrm{Im}\left(\lambda_t^2\right) \, \left( \eta_2 S_0(x_t) - \eta_3 S_0(x_c,x_t) \right) \right] \, \frac{8}{3} \, f_K^2 \,  M_K^2 \, \hat B_K \ ,
\label{eq:leading}
\eea
where $\lambda_i = V_{id} V^*_{is}$; the   $S_0$'s   are the  Inami-Lim  functions computed by matching full and effective amplitudes for external states of four quarks with zero momentum\,\cite{Inami:1980fz};  the $\eta_i$ are  the  corrections to the Wilson coefficient of the four-fermion operator  $Q=\bar{s}\gamma^{\mu}_L d\ \bar{s}\gamma^{\mu}_Ld$ (in squared parentheses  in Eq.\,(\ref {eq:leading})), computed at short distance in perturbative  QCD  in order to match the Standard Model to the effective Hamiltonian  and $\hat B_K $ is the renormalisation group invariant bag parameter of the four-fermion operator. 

We wish to note that the UTfit Collaboration always computed the Wilson coefficient by evaluating all the $\eta_i$ simultaneously event by event  so that the reduction of the uncertainty noted in\,\cite{Brod:2019rzc} was always automatic and active in our calculations.

\vskip 0.5 cm 
On the one hand  $\varepsilon$ is experimentally measured with a $0.4$\% accuracy, 
$\varepsilon^{\rm exp}_K=\vert \varepsilon^{\rm exp}\vert =2.228 (11) \times  10^{-3}$\,\cite{ParticleDataGroup:2022pth}, on the other hand  the theoretical accuracy of the SM prediction is approaching a few percent level, mainly thanks to the improvement of the lattice determination of the relevant bag parameter $\hat B_K$\,\cite{Aoki:2021kgd},
\bea   N_f=2+1  \qquad  \hat B_K  = 0.7625 \, (97) \, ,  \qquad \qquad  N_f=2+1 +1 \qquad  \hat B_K  = 0.717 \, (18)(16) \, . \eea 
Following our general approach  we average  the above  two numbers  using the PDG method\,\cite{ParticleDataGroup:2022pth} and use 
\bea    \hat B_K  = 0.756 \, (16)  \,. \eea
Given the improved accuracy, the $\xi_0$ term and the deviation of the phase $\phi_\varepsilon$ from $45^\circ$ appearing in  Eq.\,(\ref{eq:eve}) 
are not negligible\,\cite{Buras:2008nn}.  

For what concerns $ \mathrm{Im} M_{12}$,  the  calculation of the amplitude,  including long distance contributions,  is possible from first principles in lattice QCD\,\cite{Christ:2010gi,Christ:2015pwa}. This was  also attempted numerically\,\cite{Christ:2015phf} but, for the time being, the difficulty in making the calculation at the physical point in the light  quark masses and the extrapolation to the continuum limit, leave too large uncertainties to compete with the standard approach of Eq.\,(\ref{eq:leading}).

Within the standard approach, the dominant long-distance corrections  of  $O(1/m_c^2)$ to $\mathrm{Im} M_{12}$, $\delta_{BGI}$ below, due to the exchange of two pions, were  evaluated in\,\cite{Buras:2010pza}.  The inclusion of this correction and  of the more accurate values of   $\xi_0$ and  $\phi_\varepsilon$  reduces  by  $6$\%  the predicted central value of $\varepsilon$.  Given the increasing precision of the theoretical ingredients entering $\varepsilon$, it is  then  becoming important to include all terms expected to contribute to the theoretical evaluation of $\varepsilon$ at the percent level.  Very recently, in ref.\,\cite{Ciuchini:2021zgf},  other  power corrections due to the finite value of  the charm quark mass, denoted as $\delta_{m_c}$ below, coming from dimension-8 operators in the effective Hamiltonian were evaluated.   We have also included the electroweak corrections, computed in ref.\,\cite{Brod:2022har},   to the charm-top contribution to the coefficient function of the operator $Q$ defined above, denoted as $\delta_{\rm EW}$.   By consistency,  the electroweak corrections to the renormalisation of the operator $Q$ should be included but this  calculation is not available yet. 

Up to higher order terms, we may then write 
\bea \mathrm{Im} M_{12}=  \langle \bar  K^0 \vert  H^{ \mathrm{eff}} \vert K^0\rangle \, \left(1 +\delta_{\rm EW}+\delta_{BGI}+\delta_{m_c}\right)\, , \label{eq:theps}\eea
where $\delta_{BGI}=  0.02$ and   the  $\delta_{m_c}$ correction increases the theoretical prediction for $\varepsilon_K$ by 1\%.   The electroweak correction is $\delta_{\rm EW}\sim 0.15\%$.  Both $\delta_{\rm EW}$ and $\delta_{m_c}$ are   small corrections which are unable to remove the small tension,  corresponding to a pull of $-1.56$,  between the  UTA  theoretical prediction for $\varepsilon$ 
\begin{equation} \varepsilon_K=\vert  \varepsilon\vert  =  2.00 (15) \times 10^{-3}\, ,\end{equation}
and the experimental result.   From the UTA analysis within  the SM,  the comparison of  the experimental value of $\varepsilon_K$ with the theoretical prediction in Eq.\,(\ref{eq:theps}) allows the extraction of  a value of   $\hat B_K$ that can be compared to the result of lattice calculations.  This value is given in 
subsection\,\ref{susec:consLQCD}, where the corresponding pull is also presented. 

Another improvement producing a 2\% reduction of the central value is the inclusion of the available NNLO QCD corrections to the Wilson coefficients of the $\Delta S=2$ effective Hamiltonian\,\cite{Brod:2010mj,Brod:2011ty}.  We have not included these corrections, however, since the relevant matrix element, $\hat B_K$,  computed on the lattice  is matched to the $\overline{\mathrm{MS}}$ coefficient at the NLO only.  In this respect,  the perturbative calculation of the NNLO matching of $\hat B_K$  from the non-perturbative  RI-MOM/SMOM schemes used in   lattice calculations to  the $\overline{\mathrm{MS}}$ scheme would be welcome.

 \subsection{New: the lattice determination of $\varepsilon^\prime/\varepsilon$}
Since this is the first time that $\varepsilon^\prime/\varepsilon$ is included in the UTA, in this subsection we give some details of  its calculation. Our theoretical prediction  has been  obtained  by using the operator matrix elements  computed on the lattice by the RBC-UKQCD collaboration\,\cite{RBC:2020kdj} and the Wilson coefficients computed with the parameters used in the present work.  The calculation requires several steps: i) the evaluation  of the matrix elements of the bare lattice four fermion  operators in lattice QCD; ii) the matching of the matrix elements  of the bare operators to those of the operators  renormalised non-perturbatively in some version of the  RI-MOM scheme\,\cite{Martinelli:1994ty}, which in ref.\,\cite{RBC:2020kdj} was either in  the \smomq or in the \smomgamma schemes\,\cite{Sturm:2009kb}; iii) the matching of the  renormalised operators in the SMOM schemes to the operators  renormalised in the $\msb$ scheme in which the Wilson coefficient functions have been computed at the NLO\,\cite{Ciuchini:1993vr, Buras:1991jm, Buras:1992tc, Buchalla:1995vs}; iv) the combination of the operators in the  $\msb$ scheme and the Wilson coefficients computed at the NLO to compute  the relevant amplitudes.  In our analysis  we take the matrix elements in the SMOM scheme from ref.\,\cite{RBC:2020kdj} and perform the steps iii) and iv) using the parameters extracted in our UTA run. 

The expression of $\varepsilon^\prime/\varepsilon$ is given by 
\begin{equation}
    \mathrm{Re}\left(\frac{\varepsilon^\prime}{\varepsilon}\right)=-\frac{\omega\sin(\delta_2-\delta_0-\phi_\varepsilon)}{\sqrt{2}\vert\varepsilon\vert}\left(\frac{ \mathrm{ Im}A_2}{ \mathrm{Re} A_2}-\frac{ \mathrm{Im }A_0}{ \mathrm{Re} A_0}\left(1-\hat  \Omega_{\rm eff}\right)\right)\, , 
\end{equation}
where the isospin breaking effects (both from the mass difference $m_d-m_u$ and from electromagnetic corrections)  are encoded in the parameter $\hat \Omega_{\rm eff}$.  Since at present there is no lattice calculation of  this parameter, in this UTA analysis,   from   the estimate of ref.\,\cite{Cirigliano:2019cpi} namely $\hat \Omega_{\rm eff}=17.0^{+9.1}_{-9.0} \cdot 10^{-2}$, by taking a symmetric error,   we have used  $\hat \Omega_{\rm eff}=17.05(9.05) \cdot 10^{-2}$.  In the calculation of $\varepsilon^\prime/\varepsilon$  we use the experimental value of the parameters  $\omega$, $\delta_2$, $\delta_0$, $\phi_\varepsilon$ and  $\varepsilon_K=\vert\varepsilon\vert$.    For comparison with the experimental data, the theoretical  values of $\delta_{0,2}(s)$ in the fourth column of the table  have  been taken  from ref.\,\cite{Blum:2015ywa} whereas the value of $\varepsilon_K$ is extracted from this \utfit\ \, analysis. $ \mathrm{Re}\left(A_{2}\right)$ is extracted from the $K^+ \to \pi^+ \pi^0$ decay width;   $\mathrm{Re}\left(A_{0}\right)$ is computed  using the $K^0 \to \pi^+ \pi^- $ and $K^0 \to \pi^0 \pi^0$  decay widths.  To get $A_0$ from the corresponding  decay amplitudes, one takes  $x = |A_2|/|A_0| \sim  \mathrm{Re}\left(A_{2}\right)/ \mathrm{Re}\left(A_{0}\right)$ as a small parameter (its value is about $0.05$) and expands the amplitudes to leading order in this parameter.  The  values of   $\mathrm{Re} A_{0,2}$ extracted as explained above are  given  in table\,\ref{tab:fullSM},   the values of $\mathrm{Im} A_{0,2}$ are computed using the matrix elements of the RBC/UKQCD Collaboration\,\cite{RBC:2020kdj}.
The evaluation of  $\varepsilon^\prime/\varepsilon$  is much harder  due to the presence of large cancellations among the contribution of the different operators, particularly between the matrix elements of $Q_6$ and $Q_8$ defined below. 

The general expression of  the amplitude  relative to a given isospin channel (in the case of the $K\to \pi\pi$, either $I=0$ or $I=2$)   in the SM is given by
\begin{equation}
    A_{0,2} = \frac{G_F}{\sqrt{2}}V_{us}^*V_{ud}\sum_{i=1}^{10}\left(z_i^{\overline{\text{MS}}}(\mu)+\tau y_i^{\overline{\text{MS}}}(\mu) \right)M_i^{\overline{\text{MS}}}(\mu)_{I=0,2}\, ,  \label{eq:A02}
\end{equation}
where the Wilson coefficients $z_i,y_i$,   the matrix elements of the relevant renormalised operators, $M_i^{\overline{\text{MS}}}(\mu)_{I=0,2} = \langle \pi\pi | Q_i(\mu) |K\rangle_{I=0,2} $,     and $\tau = -{V_{ts}^*V_{td}}/{V_{us}^*V_{ud}}$ will be discussed in the following.
 \subsubsection{Operators, bases and matrix elements of bare lattice operators}
 The bare lattice operators which have been evaluated by the authors of ref.\,\cite{RBC:2020kdj} are the following:
\begin{itemize}
\item \textbf{Current-Current Operators}:
    \bea  Q_1 = (\bar{s}_i\gamma^\mu P_L u_j)(\bar{u}_j\gamma_\mu P_L d_i)\, ,  \qquad \qquad Q_2 = (\bar{s}\gamma^\mu P_L u)(\bar{u}\gamma_\mu P_L d)
\, ;\eea
    \item \textbf{QCD-Penguins Operators}
 \bea
        Q_3 &=& (\bar{s}\gamma^\mu P_L d)\sum_q (\bar{q}\gamma_\mu P_L q) \, ,  \qquad \qquad Q_4 = (\bar{s}_i\gamma^\mu P_L d_j)\sum_q (\bar{q}_j \gamma_\mu P_L q_i)\nonumber\\
        Q_5 &=& (\bar{s}\gamma^\mu P_L d)\sum_q (\bar{q}\gamma_\mu P_R q) \, ,  \qquad \qquad Q_6 = (\bar{s}_i \gamma^\mu d_j)\sum_q (\bar{q}_j\gamma_\mu P_R q_i)\, ;\eea
    \item \textbf{Electroweak-Penguins Operators}
    \bea
        Q_7 &=& \frac{3}{2}(\bar{s}\gamma^\mu P_L d)\sum_q e_q(\bar{q}\gamma_\mu P_R q)\, ,  \qquad \qquad Q_8 = \frac{3}{2}(\bar{s}_i\gamma^\mu P_L d_j)\sum_q e_q (\bar{q}_j \gamma_\mu P_R q_i)\nonumber\\
        Q_9 &=& \frac{3}{2}(\bar{s}\gamma^\mu P_L d)\sum_q e_q(\bar{q}\gamma_\mu P_L q) \, ,  \qquad \qquad Q_{10} = \frac{3}{2}(\bar{s}_i \gamma^\mu d_j)\sum_q e_q(\bar{q}_j\gamma_\mu P_L q_i)\, , 
    \eea
\end{itemize}
where $P_{L/R} = (1\mp\gamma_5) /2$ are the chiral projectors and $e_q$ is the quark charge in units of $e$.
This basis  is used  in  lattice calculations.  For renormalization, however,  the chiral  basis is better suited for the task since, in the usual $10$-operator basis, the operators are not linearly independent. In fact, by Fierz transforming the operators $Q_1$, $Q_2$ and $Q_3$, we define
\bea
    \tilde{Q}_1 = (\bar{s}\gamma^\mu P_L d)(\bar{u}\gamma_\mu P_L u)\, , \qquad \qquad
    \tilde{Q}_2 = (\bar{s}_i \gamma^\mu P_L d_j)(\bar{u}_j\gamma_\mu P_L u_i)\, , \qquad \qquad 
   \tilde{Q}_3 = \sum_q (\bar{s}_i\gamma^\mu P_L q_j)(\bar{q}_j\gamma_\mu P_L d_i)\nonumber
\eea
we can eliminate operators $Q_4$, $Q_9$ and $Q_{10}$ using the relations
\bea
    Q_4 = \tilde{Q}_2+\tilde{Q}_3-Q_1\, , \qquad \qquad
    Q_9 = \frac{3}{2}\tilde{Q}_1-\frac{1}{2}Q_3\, , \qquad \qquad
    Q_{10} = \frac{1}{2}(Q_1-\tilde{Q}_3)+\tilde{Q}_2\, . \nonumber
\eea
The remaining seven operators can then be recombined according to irreducible representations of the chiral flavour-symmetry group $SU(3)_L \otimes SU(3)_R$. The details of the decomposition can be found in\,\cite{Lehner:2011fz}. The chiral operator basis, which we will indicate by primed operators, is thus given by 
\begin{equation}
\begin{array}{cc}
    (27,1)& Q^\prime_1 = 3\tilde{Q}_1+2Q_2-Q_3,\\[10pt]
    (8,1)& Q^\prime_2 = \frac{1}{5}(2\tilde{Q}_1-2Q_2+Q_3),\\[10pt]
    (8,1)& Q^\prime_3 = \frac{1}{5}(-3\tilde{Q}_1+3Q_2+Q_3),\\[10pt]
    (8,1)& Q^\prime_{5,6} = Q_{5,6},\\[5pt]
    (8,8)& Q^\prime_{7,8} = Q_{7,8}
\end{array}    
\label{eq:irrep}
\end{equation}
where $(L,R)$ denotes the respective irreducible representations of $SU(3)_L\otimes SU(3)_R$.  We  recall that the bare lattice matrix elements are  converted to the chiral basis by a specific minimization procedure  based on the fact that the Fierz identities are not obeyed exactly by the bare lattice matrix elements, for details see ref.\,\cite{RBC:2020kdj}.  The resulting matrix elements   converted back to the $10$ operator basis and renormalised in  \smomq  are given in table\,\ref{tab:renorm_operators} (second column).

The operator $(27,1)$ renormalises multiplicatively and contributes to the $I=2$ channel only, the operators $Q_{7,8}$  mix  only among themselves  and thus all the $(8,1)$ operators.  
\begin{table}[]
    \centering
    \begin{tabular}{|c|c|c|c|}
    \hline  
        $\mathrm{i}$ & $M_i^{(\slashed{q}, \slashed{q})}(\mu_0)_{I=0}\, \left(\mathrm{GeV}^{3}\right)$ &  $\mathrm{i}$ &
           $M_i^{(\slashed{q}, \slashed{q})}(\mu_2)_{I=2}\, \left(\mathrm{GeV}^{3}\right)$ \\
         \hline 1 & $0.060(39)$& (27,1) &0.0506 (29)\\
        2 & $-0.125(19)$& $-$& $-$\\
        3 & $0.142(17)$ & $-$&$-$\\
        5 & $-0.351(62)$& $-$&$-$ \\
        6 & $-1.306(90)$ & $-$&$-$\\
        7 & $0.775(23)$&(8,8) &1.003 (0.037)\\ 
        8 & $3.312(63)$&(8,8)mx & 4.43 (18)\\ \hline
    \end{tabular}
    \caption{{\it Physical, extrapolated to the infinite-volume,  matrix elements in the \smomq    scheme  at $\mu_0=4$\,GeV and $\mu_2=3$\,GeV}}
    \label{tab:renorm_operators}
\end{table}

For the calculation of $A_2$ the authors of ref.\,\cite{Blum:2015ywa}  used the following operator basis 
\bea  Q_{(27,1)}&=& (\bar s_a \gamma^\mu (1-\gamma_5) d_a) (\bar u_b \gamma_\mu (1-\gamma_5) u_b-\bar d_b \gamma_\mu (1-\gamma_5) d_b)+(\bar s_a \gamma^\mu (1-\gamma_5) u_a) (\bar u_b \gamma_\mu (1-\gamma_5) d_b)  \, ,\nonumber \\
Q_{(8,8)}&=& (\bar s_a \gamma^\mu (1-\gamma_5) d_a) (\bar u_b \gamma_\mu (1+\gamma_5) u_b-\bar d_b \gamma_\mu (1+\gamma_5) d_b)+(\bar s_a \gamma^\mu (1-\gamma_5) u_a) (\bar u_b \gamma_\mu (1+\gamma_5) d_b)  \, ,\label{eq:op32} \\
Q_{(8,8)mx}&=& (\bar s_a \gamma^\mu (1-\gamma_5) d_b) (\bar u_b \gamma_\mu (1+\gamma_5) u_a-\bar d_b \gamma_\mu (1+\gamma_5) d_a)+(\bar s_a \gamma^\mu (1-\gamma_5) u_b) (\bar u_b \gamma_\mu (1+\gamma_5) d_a)  \, ,\nonumber 
 \eea
where $a,b$ are summed colour indices.  The values of their matrix elements, renormalised in \smomq,   are given in the fourth column of  table\,\ref{tab:renorm_operators}.
\vskip 0.2cm

\subsubsection{Non-perturbative renormalization of lattice matrix elements}
The matrix elements of the bare operators  computed on lattice   are matched non-perturbatively to a RI-MOM scheme to reduce the uncertainties related to lattice QCD perturbation theory\,\cite{Martinelli:1994ty}.  Since the Wilson coefficients are usually given in the $\msbar$ scheme, we have  then to translate the matrix elements to this scheme. This is done by converting the bare lattice matrix elements to the RI-SMOM scheme and then matching at NLO in perturbation theory   the result to $\msbar$.   

 The authors of refs.\,\cite{RBC:2020kdj} and   \cite{Blum:2015ywa} computed the matrix elements  either in   the \smomq or  in the \smomgamma  schemes but      used only the results  in  \smomq.  According to   their choice,  in the following, we give only the information  necessary  for the generation of the UTA events using the matrix elements computed in this  scheme. In refs.\,\cite{RBC:2020kdj} and  \cite{Blum:2015ywa} the calculations were presented,  however, at   two different renormalisation  scales: for $A_2$  they used $\mu_2=3$\,GeV whereas for $A_0$  the renormalisation scale was  
 $\mu_0=4$\,GeV.    
 
The conversion from   the bare lattice matrix elements to those of the  RI-SMOM scheme is operated by a  matrix  $Z_{ij}^{\text{RI}\leftarrow\text{lat}}(\mu_I \text{ GeV})$     according to the relation
\begin{equation}
    M_i^{\prime\,\text{RI}}(\mu_I\text{ GeV})_{I=0,2} =  \sum_j  \, Z_{ij}^{\text{RI}\leftarrow\text{lat}}(\mu_I \text{ GeV})\left(a^{-3}F_I M_j^{\prime\,\text{lat}}\right)_{I=0,2}
\end{equation}
where $a^{-1}$ is the inverse of the lattice spacing and $F_I$ is the Lellouch-L\"uscher factor accounting for leading finite-volume corrections to the lattice matrix elements in the  isospin $I=0,2$ channels.  We do not need this matrix in our analysis, it can be found in the original publication quoted above.

Once we have the operators in the $7$-operator basis, we can perform the non-perturbative renormalization  using a $7\times7$ renormalization matrix.
In the case of $A_0$,  the matrix elements in  $7$-operator basis at renormalization scale of $\mu = 4$ GeV in the \smomq  scheme, and the corresponding  covariance  matrix,  $\Sigma_{ij} = \rho_{ij}\sigma_i\sigma_j$,  are given in table\,\ref{tab:renorm_operators} and \ref{tab:covariance}  respectively;  $\rho_{ij}$ is the correlation matrix and $\sigma_i$ are the  errors associated to the matrix elements.   In the case of  $A_2$  it was not possible to obtain the correlation matrix from the authors. To be conservative in the evaluation of the uncertainties we considered three cases:  a) no correlation among the three operators of eq.\,(\ref{eq:op32}); b)  maximal correlation,  namely $\rho_{ij}=1$, and c) the same correlation for the operators mediating $\Delta  I=3/2$ transitions as the one computed for $\Delta  I=1/2$ transitions in  ref.\,\cite{RBC:2020kdj}.  In the latter case the correlation matrix  can be easily derived using  the covariance matrix given  in Table\,\ref{tab:covariance}. The difference between the results obtained  with a)-c)  is  tiny with respect to the overall uncertainty and  was  absorbed in the overall uncertainty.
\begin{table}[]
    \centering
 \begin{tabular}{|c|c|c|c|c|c|c|c|}
    \hline 
 $\Sigma_{ij}$&   $Q_1$ & $Q_2$ & $Q_3$ & $Q_5$ & $Q_6$ & $Q_7$ & $Q_8$  \\  \hline
   $Q_1$ &     $0.001516$ & $5.385 \times 10^{-5}$ & $-9.167 \times 10^{-5}$ & $0.0001252$ & $-0.0003965$ & $0.0004930$ & $0.0007192$ \\
   $Q_2$    &   $5.385 \times 10^{-5}$ & $0.0003563$ & $-4.099 \times 10^{-5}$ & $0.0007596$ & $0.0002981$ & $2.914 \times 10^{-5}$ & $-0.0002118$ \\
  $Q_3$ &      $-9.167 \times 10^{-5}$ & $-4.099 \times 10^{-5}$ & $0.0002808$ & $0.0003784$ & $0.0004679$ & $-4.656 \times 10^{-5}$ & $0.0001516$ \\
   $Q_5$   &  $0.0001252$ & $0.0007596$ & $0.0003784$ & $0.003904$ & $0.001679$ & $-8.000 \times 10^{-5}$ & $-0.0004013$ \\
     $Q_6$ &    $-0.0003965$ & $0.0002981$ & $0.0004679$ & $0.001679$ & $0.008188$ & $-0.0003817$ & $-0.002110$ \\
   $Q_7$ &      $0.0004930$ & $2.914 \times 10^{-5}$ & $-4.656 \times 10^{-5}$ & $-8.000 \times 10^{-5}$ & $-0.0003817$ & $0.0005395$ & $0.0009460$ \\
  $Q_8$&      $0.0007192$ & $-0.0002118$ & $0.0001516$ & $-0.0004013$ & $-0.002110$ & $0.0009460$ & $0.003937$ \\
    \hline
    \end{tabular}
    \caption{\it  The $7\times 7$ covariance matrix between the renormalized, infinite-volume matrix elements in the \smomq scheme in the chiral basis.}
    \label{tab:covariance}
\end{table}

At this point one   matches the  \smomq renormalized matrix elements  to the $\msbar$ scheme. This is done with another renormalization matrix 
$Z_{ij}^{\overline{\text{MS}}\leftarrow \text{RI}}(\mu)$ that is   found in ref.\,\cite{Lehner:2011fz}
  \begin{equation}
    Z_{ij}^{\overline{\text{MS}}\leftarrow \text{RI}}(\mu) = \delta_{ij}+\frac{\alpha_s(\mu)}{4\pi} \Delta r_{ij}^{\overline{\text{MS}}\leftarrow \text{RI}}\, . 
    \label{eq:renorm_ms}
\end{equation}

The non-zero matrix elements for the $\Delta r_{ij}^{\overline{\text{MS}}\leftarrow \text{RI}}$ matrix of eq.\,(\ref{eq:renorm_ms}) are, in the case of the matching between the \smomq and $\msbar$ schemes  given in Table\,\ref{tab:Drij}, where the gauge parameter is denoted by $\xi_G$ ($\xi_G=0,1$ correspond to the Landau and Feynman  gauges, respectively). In our case the appropriate value is $\xi_G=0$.
\begin{table}[]
    \centering
 \begin{tabular}{|c|c|}
 \hline 
    (i,j)&$\Delta r_{ij}^{\overline{\text{MS}}\leftarrow\text{RI}}$\\ &
   \\ \hline
    (1,1)&$\xi_G\left(\frac{C_0}{N_c}-C_0-\frac{4\log(2)}{N_c}+4\log(2)\right)-\frac{12\log(2)}{N_c}+12\log(2)+\frac{9}{N_c}-9$\\
    (2,2) & $\xi_G\left(\frac{4 C_{0} N_{c}}{5}+\frac{C_{0}}{N_{c}}-\frac{6 C_{0}}{5}-\frac{4 \log (2)}{N_{c}}-\frac{4 N_{c}}{5}+\frac{6}{5}\right) -\frac{12 \log (2)}{N_{c}}+\frac{8 N_{c}}{5}+\frac{9}{N_{c}}-\frac{12}{5}$ \\ 
    (2,3) & $\xi_G\left(\frac{6 C_{0} N_{c}}{5}-\frac{9 C_{0}}{5}+4 \log (2)-\frac{6 N_{c}}{5}+\frac{4}{5}\right)+12 \log (2)+\frac{12 N_{c}}{5}-\frac{53}{5}$ \\ 
    (3,2) & $\xi_G\left(-\frac{6 C_{0} N_{c}}{5}+\frac{4 C_{0}}{5}+4 \log (2)+\frac{6 N_{c}}{5}-\frac{9}{5}\right)+12 \log (2)-\frac{12 N_{c}}{5}+\frac{2}{3 N_{c}}-\frac{263}{45}$\\ 
    (3,3) &$ \xi_G\left(-\frac{9 C_{0} N_{c}}{5}+\frac{C_{0}}{N_{c}}+\frac{6 C_{0}}{5}-\frac{4 \log (2)}{N_{c}}+\frac{9 N_{c}}{5}-\frac{6}{5}\right)-\frac{12 \log (2)}{N_{c}}-\frac{18 N_{c}}{5}+\frac{85}{9 N_{c}}+\frac{26}{15} $\\ 
    (3,5) &$ \frac{2}{9 N_{c}}$ \\ 
    (3,6) & $-\frac{2}{9}$ \\ 
    (5,5) & $\xi_G\left(\frac{C_{0}}{2 N_{c}}-\frac{2 \log (2)}{N_{c}}-\frac{1}{2 N_{c}}\right)+\frac{3 C_{0}}{2 N_{c}}-\frac{2 \log (2)}{N_{c}} -\frac{2}{N_{c}}$\\
    (5,6) & $\xi_G\left(-\frac{C_{0}}{2}+2 \log (2)+\frac{1}{2}\right)-\frac{3 C_{0}}{2}+2 \log (2) +2$ \\ 
    (6,2) & $\frac{5}{N_{c}}-\frac{10}{3}$ \\ 
    (6,3) & $\frac{10}{3 N_{c}}-5 $\\ 
    (6,5) &$ \left(2 \log (2)-\frac{1}{2}\right) \xi_G+2 \log (2)+\frac{5}{3 N_{c}}-2 $\\ 
    (6,6) & $\xi_G\left(-\frac{C_{0} N_{c}}{2}+\frac{C_{0}}{2 N_{c}}-\frac{2 \log (2)}{N_{c}}+N_{c}-\frac{1}{2 N_{c}}\right)-\frac{3 C_{0} N_{c}}{2}+\frac{3 C_{0}}{2 N_{c}}-\frac{2 \log (2)}{N_{c}}+4 N_{c}-\frac{2}{N_{c}}-\frac{5}{3}$ \\
    (7,7) &$ \xi_G\left(\frac{C_{0}}{2 N_{c}}-\frac{2 \log (2)}{N_{c}}-\frac{1}{2 N_{c}}\right)+\frac{3 C_{0}}{2 N_{c}}-\frac{2 \log (2)}{N_{c}}-\frac{2}{N_{c}} $\\
    (7,8) & $\xi_G\left(-\frac{C_{0}}{2}+2 \log (2)+\frac{1}{2}\right)-\frac{3 C_{0}}{2}+2 \log (2)+2 $\\
    (8,7) &$ \left(2 \log (2)-\frac{1}{2}\right) \xi_G+2 \log (2)-2$ \\
    (8,8) & $\xi_G\left(-\frac{C_{0} N_{c}}{2}+\frac{C_{0}}{2 N_{c}}-\frac{2 \log (2)}{N_{c}}+N_{c}-\frac{1}{2 N_{c}}\right) -\frac{3 C_{0} N_{c}}{2}+\frac{3 C_{0}}{2 N_{c}}-\frac{2 \log (2)}{N_{c}}+4 N_{c}-\frac{2}{N_{c}}$\\ &\\
    \hline
    \end{tabular}
    \caption{\it  Non-zero matrix elements for the $\Delta r_{ij}^{\overline{\text{MS}}\leftarrow \text{RI}}$ matrix of eq.\,(\ref{eq:renorm_ms}) are shown in the case of the matching between the \smomq and $\msbar$ schemes. $C_0 =2/3 \Psi^\prime(1/3)-(2\pi/3)^2\sim 2.34391$, where $\Psi$ is the digamma function.}
    \label{tab:Drij}
\end{table}
\begin{figure}
\hspace*{-0.6cm}
  \centering
    \includegraphics[width=0.6\linewidth]{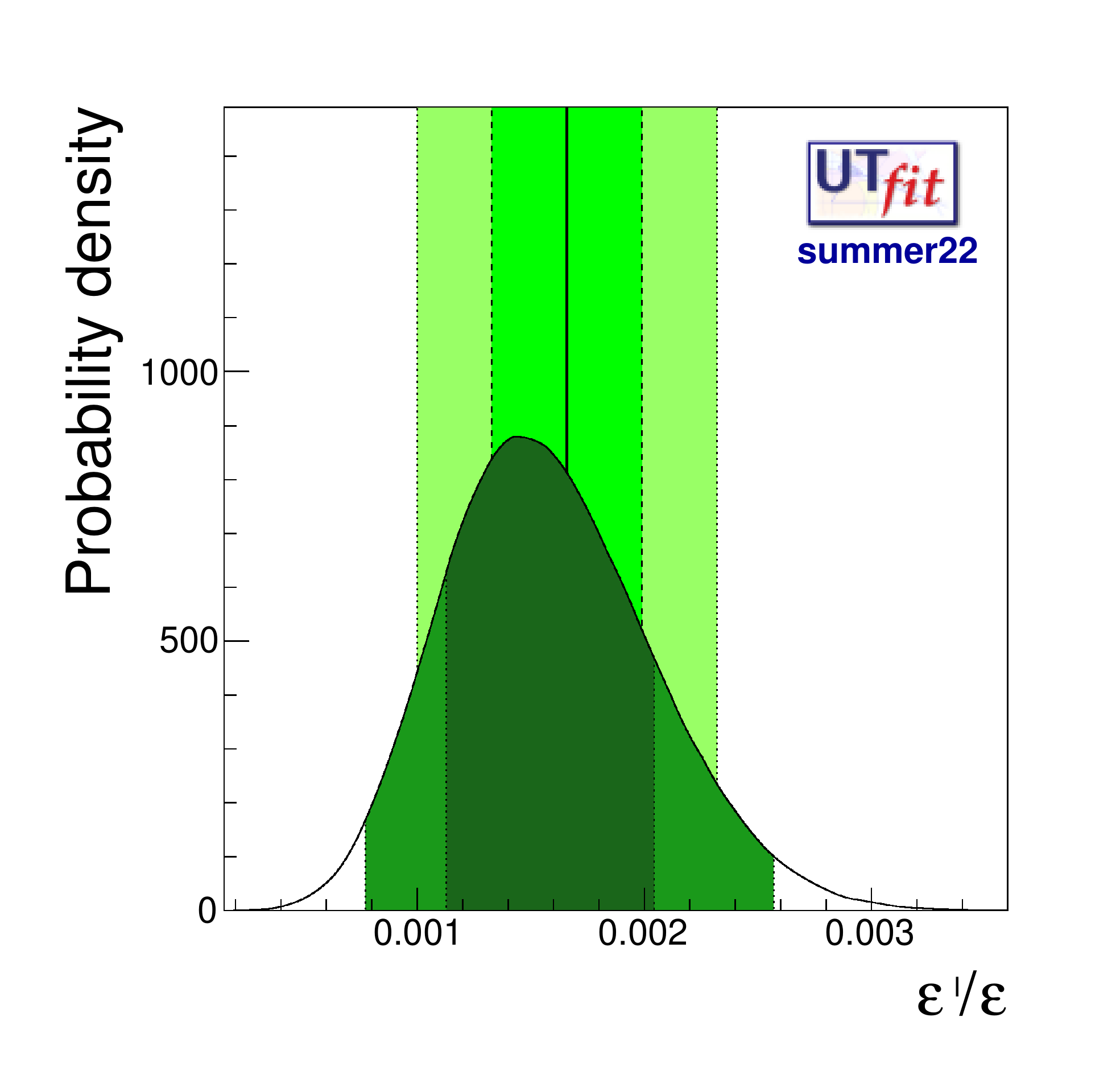}
  \vspace*{-0.3cm}
  \caption{{\it   The prediction  of $\varepsilon^\prime/ \varepsilon$  obtained within  this UT analysis. The vertical band represents the experimental measurement and uncertainty of this quantity.}}
  \label{fig:epse}
\end{figure}

\vskip 0.5cm 
\subsubsection{Wilson coefficients and final result}

 After the conversion of the matrix elements  in the $\msb$ scheme, we can compute the $A_0$ and $A_2$ amplitudes
using eq.\,(\ref{eq:A02}).  The expression of the Wilson coefficients $z_i,y_i$   can be found at NLO 
in the references\,\cite{Ciuchini:1993vr, Buras:1991jm, Buras:1992tc, Buchalla:1995vs}.  
 
 From the UTA  we  find
\beq  \tau = -\frac{V^*_{ts}V_{td}}{ V^*_{us}V_{ud}}=0.001482 (36) - i \,  0.000644(16) \, . \eeq
The values of the generated amplitudes $\mathrm{Re}\left(A_{0}\right)$ and $\mathrm{Re}\left(A_{2}\right)$  are given in table\,\ref{tab:fullSM}.We also find $\mathrm{Im }\left(A_0\right)= -6.75 (86) \times 10^{-11}$\,GeV and $\mathrm{Im }\left(A_2\right)= -8.4 (1.2) \times 10^{-13}$\,GeV.
For the calculation of  $\varepsilon^\prime/ \varepsilon$, however,  assuming the validity of the SM,  the real part of these amplitudes are taken from the experiments in order to  reduce the final theoretical  uncertainty.  For this quantity we get 
\beq \varepsilon^\prime/ \varepsilon  = 15.2(4.7) \cdot 10^{-4}\,  . \eeq
This number can  be compared with the RBC/UKQCD result, given without error,  which includes the isospin breaking corrections of ref.\,\cite{Cirigliano:2019cpi},  $\varepsilon^\prime/ \varepsilon  = 16.7 \cdot 10^{-4}$  (RBC/UKQCD quotes $21.7(8.4) \cdot 10^{-4}$  without isospin breaking corrections),  and with the experimental value  $\varepsilon^\prime/ \varepsilon  = 16.6(3.3)  \cdot 10^{-4}$.   The predicted  distribution of  $\varepsilon^\prime/ \varepsilon$  is shown in Fig.\,\ref{fig:epse}. Within still large theoretical uncertainties the SM predictions and experimental results  are in very good agreement and there is no sign of NP. The novelty  here  is the insertion of  the determination of $\varepsilon^\prime/ \varepsilon$ in the full UT analysis.

\subsection{The Unitarity Triangle angles}

\begin{figure}[!ht]
\hspace*{-0.6cm}
  \centering
  \begin{tabular}{ccc}
    \includegraphics[width=0.41\linewidth]{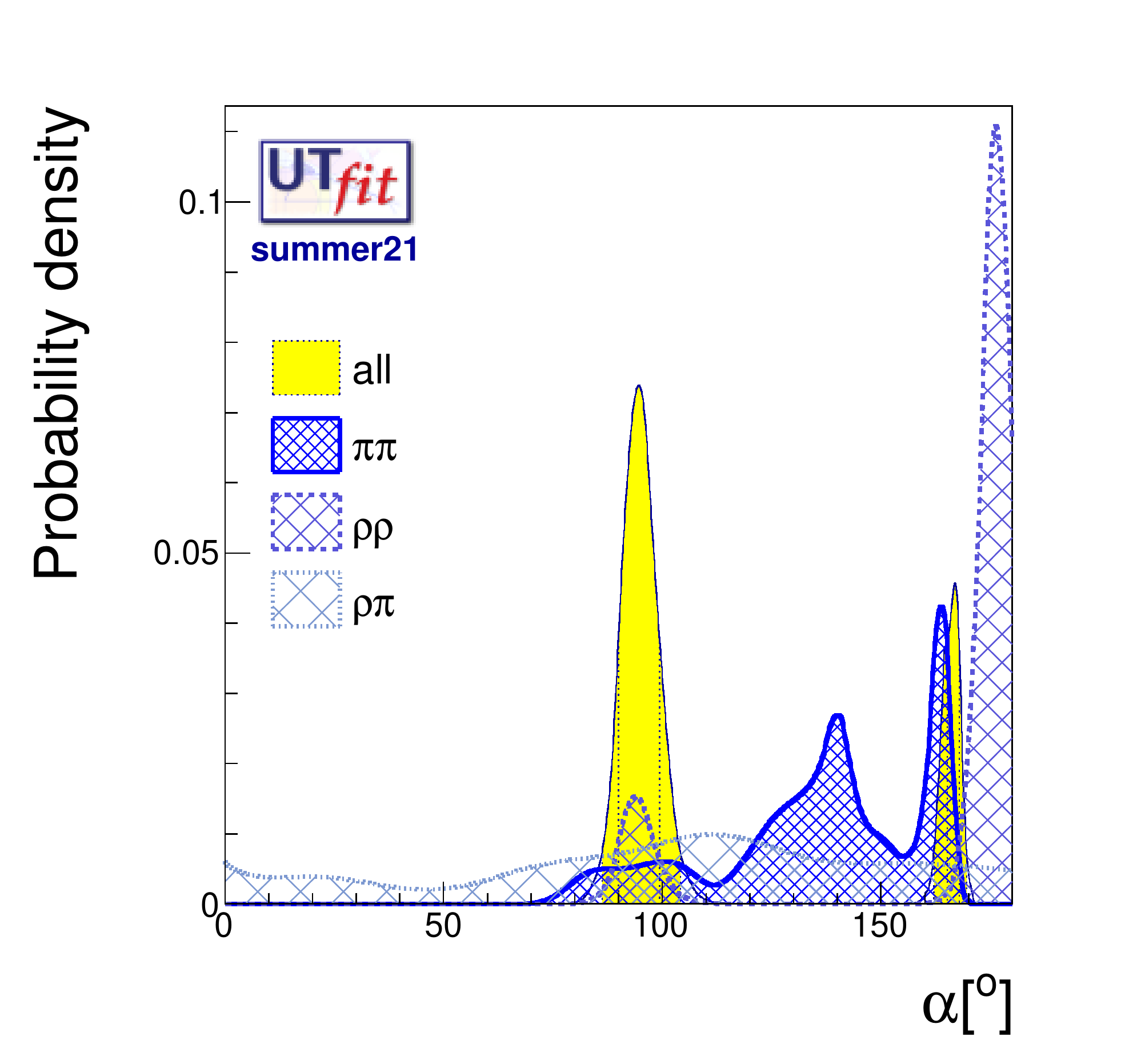} &
    \hspace*{-0.5cm}
    \includegraphics[width=0.36\linewidth]{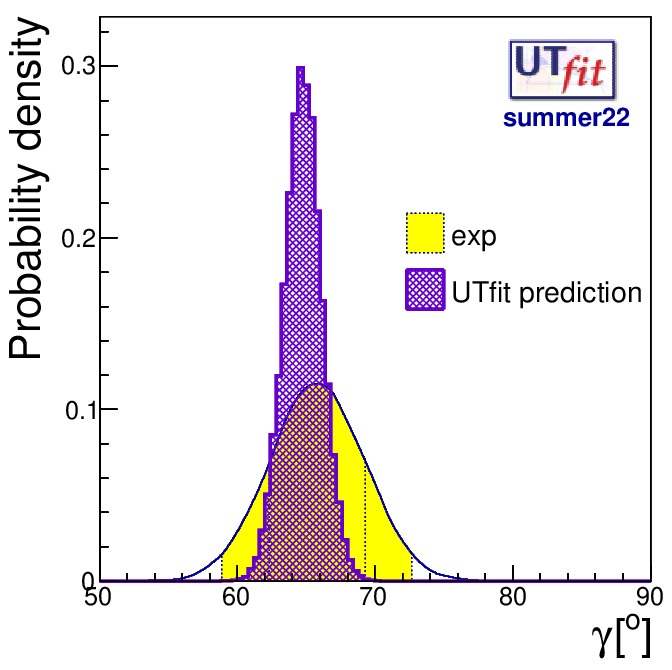} &
    \hspace*{-0.1cm}
  \end{tabular}
  \vspace*{-0.4cm}
  \caption{{\it  Left: global fit input distribution for the angle $\alpha$
      (in solid yellow histogram) with the three separate distributions
      coming from the three contributing final states $\pi\pi$, $\rho\rho$ and
      $\rho\pi$;  Right: global fit input distribution for the angle $\gamma$
      (in solid yellow histogram) obtained by the HFLAV~\cite{HFLAV:2019otj} average
      compared with the global \utfit\ prediction for the same angle.
      }}
  \label{fig:angles}
\end{figure}

For what concerns the values of the Unitarity Triangle angles, we used  the following  inputs:
\begin{itemize}
\item  $\beta$ (or $\phi_1$): the value of sin $\beta$ is taken from the latest HFLAV average\,\cite{HFLAV:2019otj}  with the   most updated inputs,   
 which gives $\sin 2 \beta = 0.688 (20)$.
We then add a correction factor of  $-0.01(1)$,  although strictly speaking this applies  exclusively  to the $J/\psi \,K^0$  channel,    as data-driven theory uncertainty obtained with the method described in ref.\,\cite{Ciuchini:2005mg};
\item $\alpha$ (or $\phi_2$): the value of the angle $\alpha$ is obtained by UTfit isospin analyses of the three contributing final
states $\pi\pi$,   $\rho\rho$ and $\rho\pi$. The various probability distributions are shown in Fig.\,\ref{fig:angles}  (left panel) together with
the combined one that is used as input to our global fit;
\item  $\gamma$ (or $\phi_3$):  the value of the angle $\gamma$ is taken from the latest HFLAV average\,\cite{HFLAV:2019otj} and the corresponding
probability distribution is shown in Fig.\,\ref{fig:angles} (right panel) together with the prediction from the global fit.
\end{itemize}

The full list of measurements used as inputs in the global fit is given in the first and second
columns of table\,\ref{tab:fullSM}.  $\varepsilon$, $\varepsilon^\prime/\varepsilon$, $\vert V_{ub}\vert$ and $\vert V_{cb}\vert$ have been discussed in the previous sections.

\section{Standard Model Unitarity Triangle Analysis}
\label{sec:SMUTA}
\begin{figure}[!ht]
  \centering
  \begin{tabular}{cc}
    \includegraphics[width=0.40\linewidth]{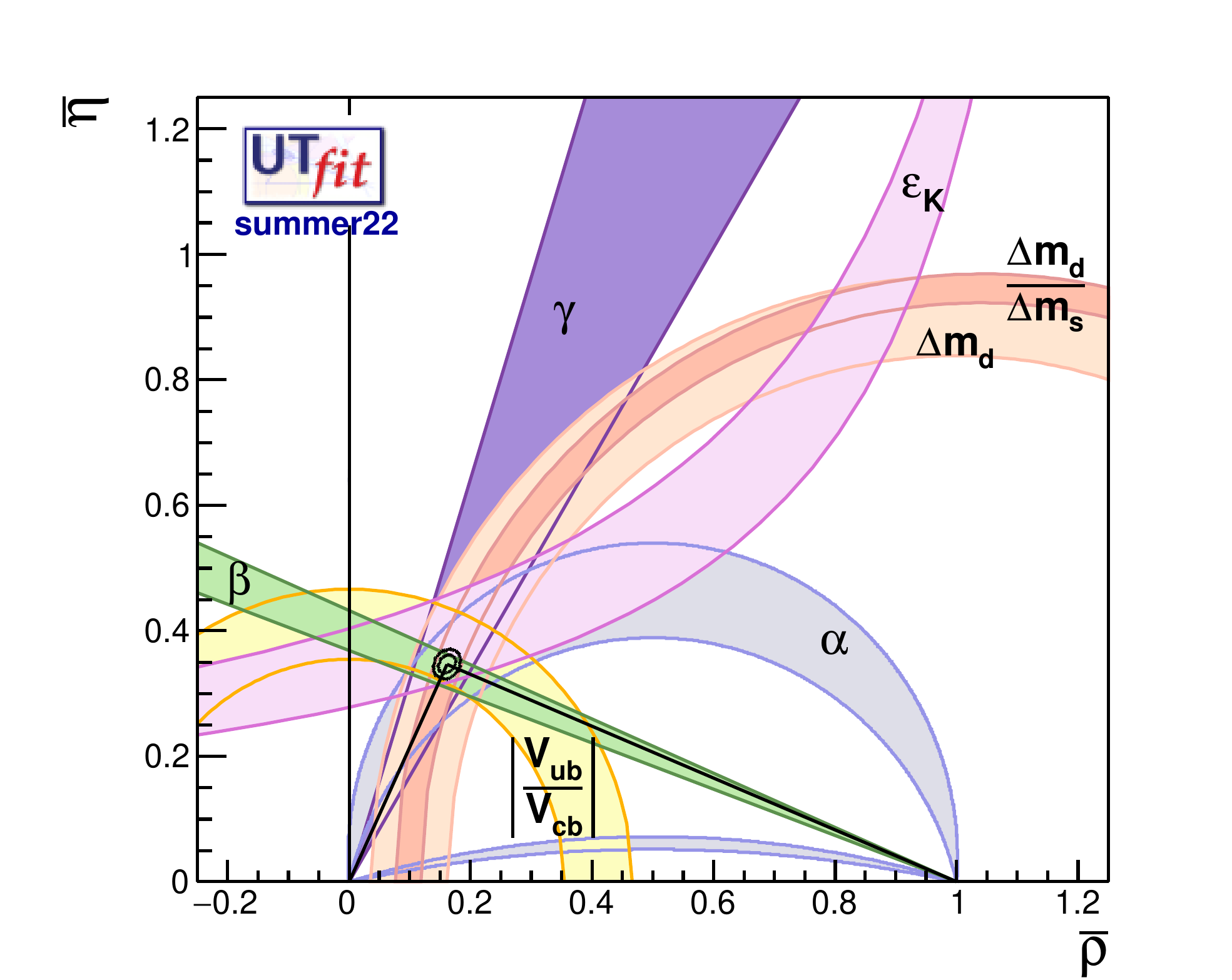} &
    \includegraphics[width=0.40\linewidth]{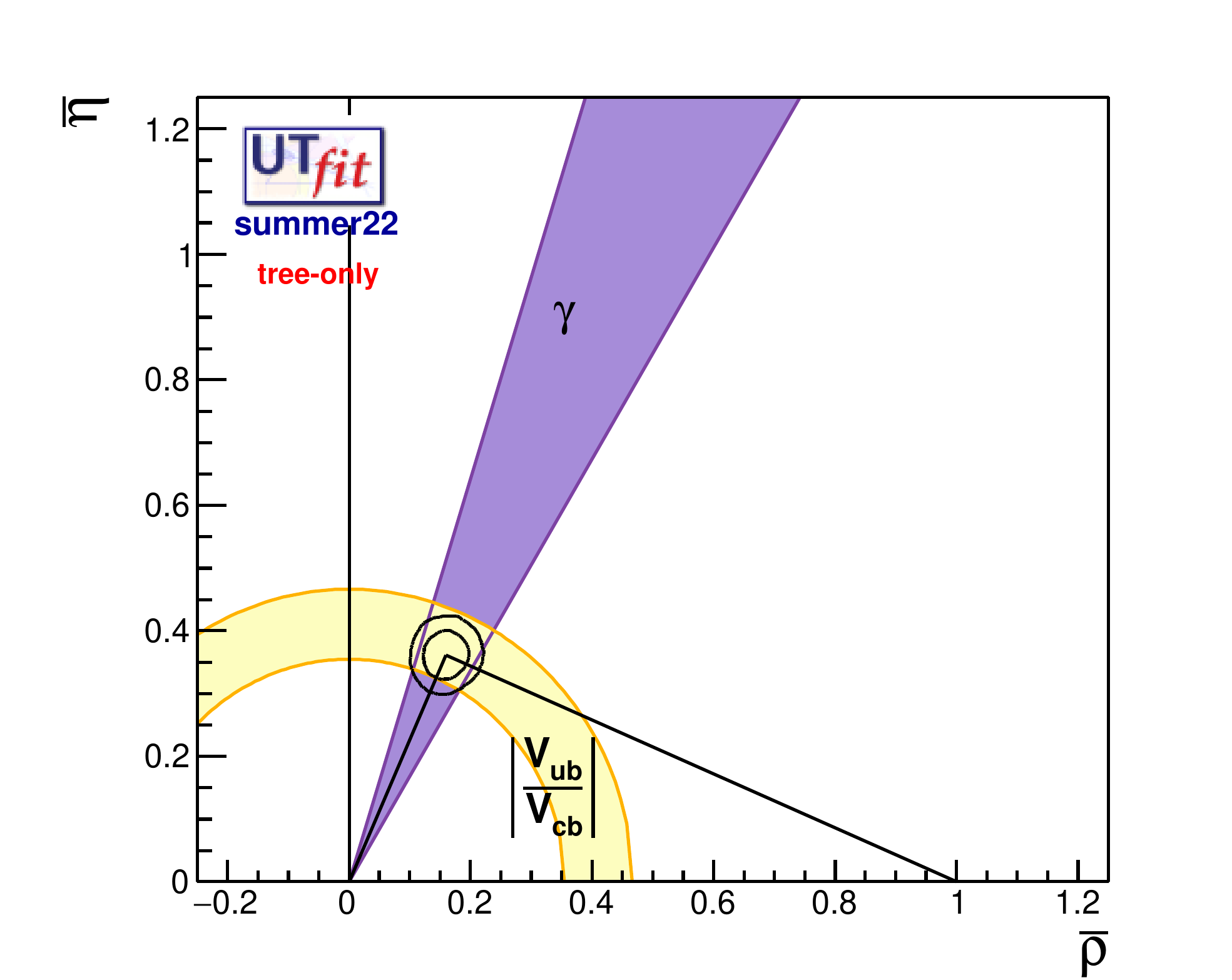} 
  \end{tabular}
  \begin{tabular}{cc}
    \includegraphics[width=0.40\linewidth]{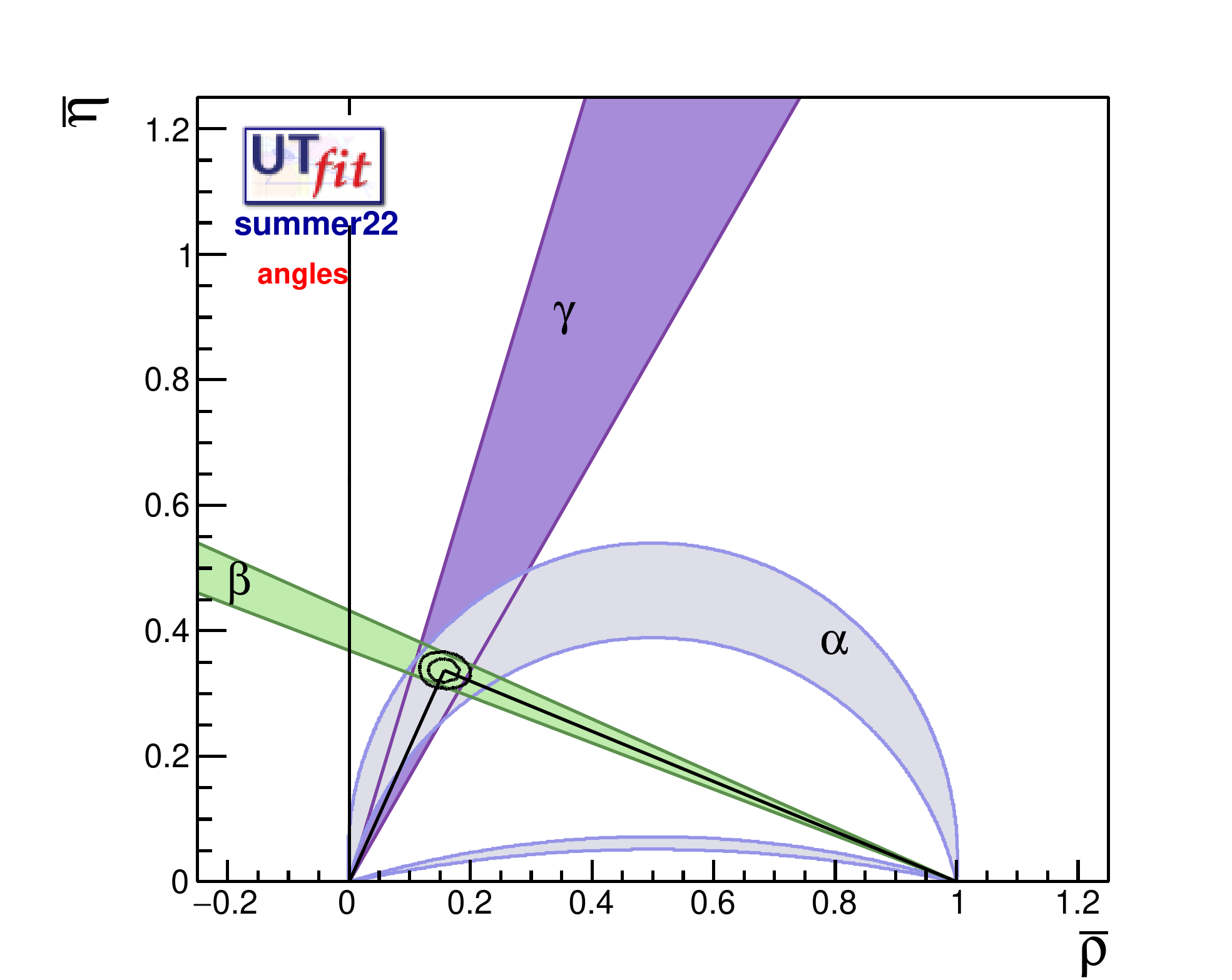} &
    \includegraphics[width=0.40\linewidth]{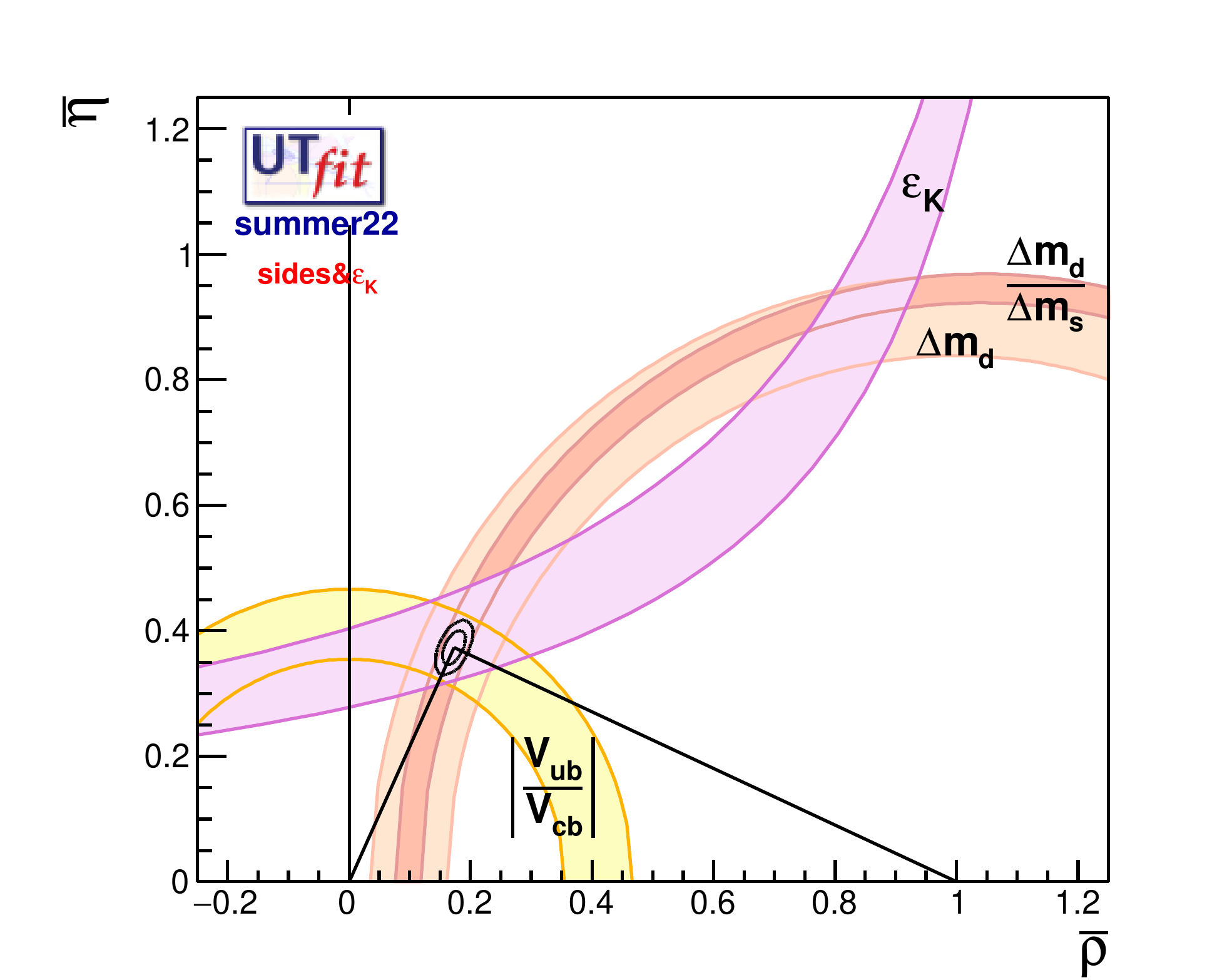}
  \end{tabular}
  \vspace*{-0.3cm}
  \caption{{\it $\bar \rho$-$\bar \eta$ planes with the SM global fit results in various
    configurations. The black contours display the 68\% and 95\% probability
    regions selected by the given global fit. The 95\% probability
    regions selected are also shown for each constraint considered.
    {\it{Top-Left}}: full SM fit;
    {\it{Top-Right}}: fit using as inputs the ``tree-only'' constraints; 
    {\it{Bottom-Left}}: fit using as inputs only the angle measurements; 
    {\it{Bottom-Right}}: fit using as inputs only the side measurements and the mixing parameter $\varepsilon_K$ in the kaon system.
    }}
  \label{fig:sm21}
\end{figure}
\begin{table}[ht]
\centering
\begin{tabular}{|c |c| c|}
\hline
fit configuration & $\bar\rho$ & $\bar\eta$ \\ [0.5ex]
\hline
full SM fit & $0.161 (10)$ & $0.347 (10)$ \\
tree-only fit & $\pm 0.158 (26)$ & $\pm  0.362 (27)$ \\
angle-only fit & $0.156 (17)$ & $0.334 (12)$ \\
no-angles fit & $0.157 (17)$ & $0.337 (12)$ \\
\hline
\end{tabular}
  \caption{\it Results for the $\bar\rho$ and $\bar\eta$ values as extracted from the
    various fit configurations. The Universal Unitarity Triangle (UUT) fit includes
    the three angles inputs and the semileptonic ratio $|V_{ub}/V_{cb}|$~\cite{Buras:2000dm}.}
\label{tab:rhoeta}
\end{table}
\begin{figure}[]
\hspace*{-1.0cm}
  \centering
  \begin{tabular}{ccc}
    \includegraphics[width=0.38\linewidth]{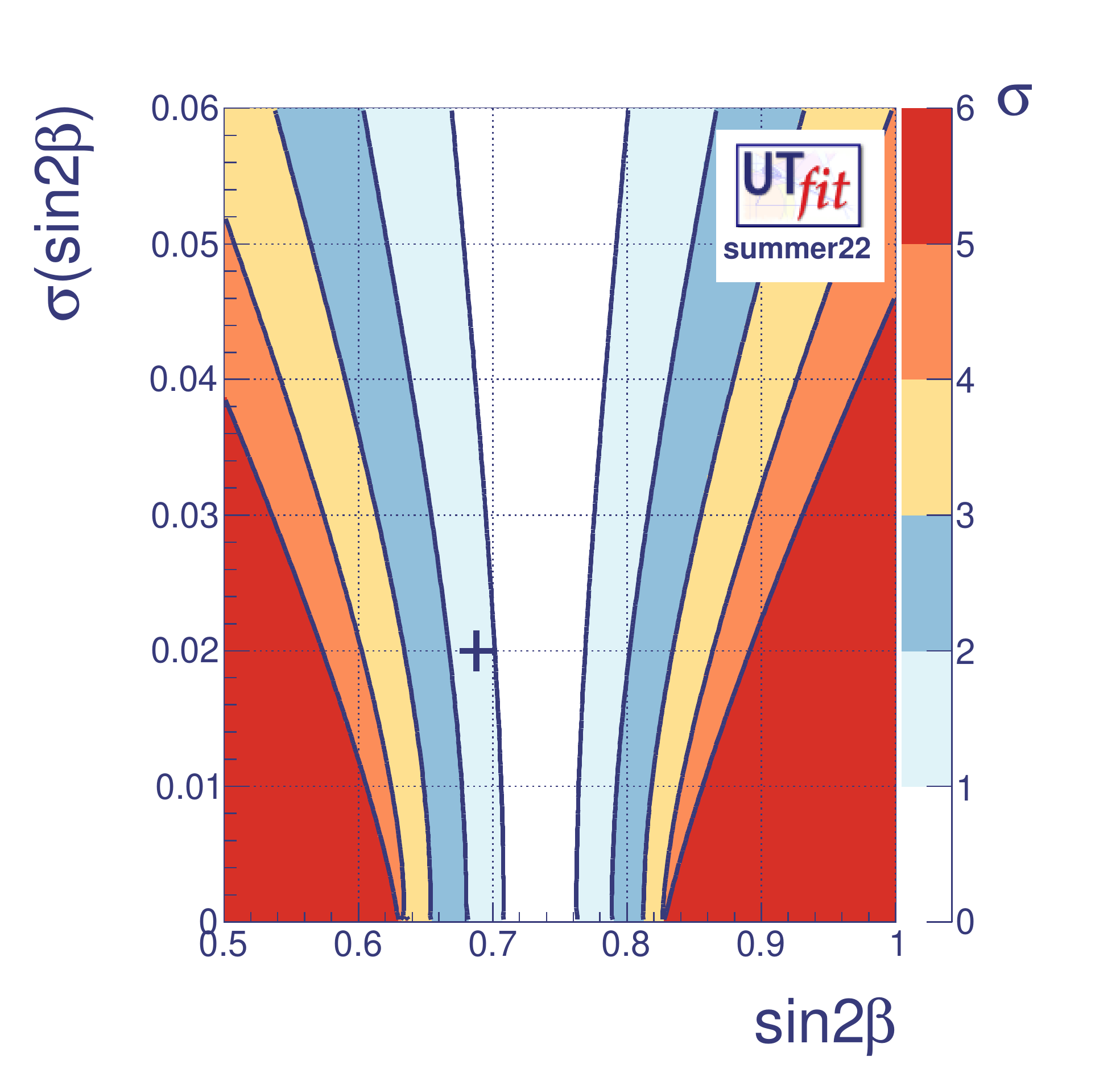} & 
    \hspace*{-0.6cm}
    \includegraphics[width=0.38\linewidth]{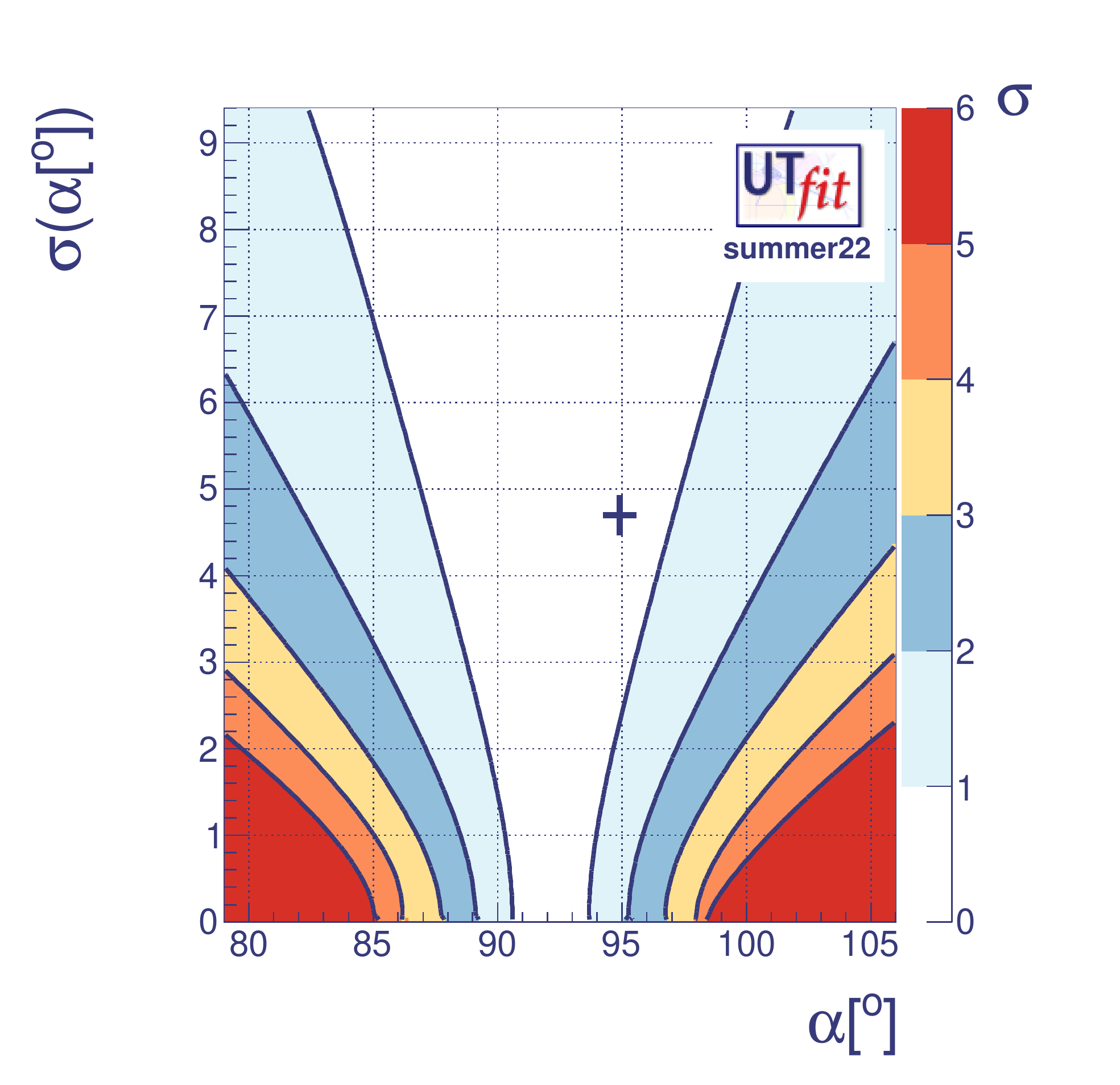} &
    \hspace*{-0.3cm}
    \includegraphics[width=0.38\linewidth]{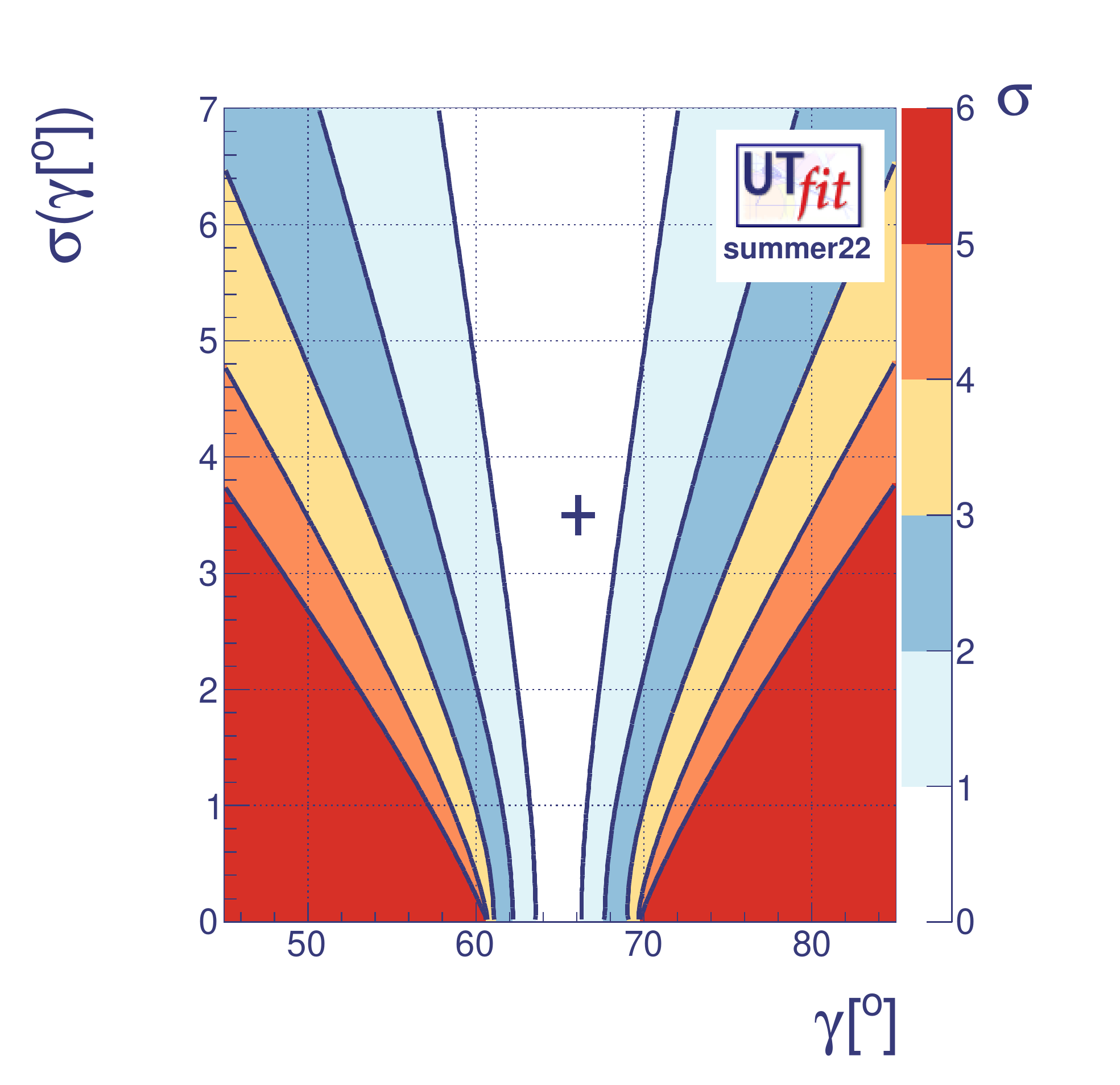} 
    \hspace*{-0.2cm}
\end{tabular}
  \begin{tabular}{cc}
    \includegraphics[width=0.38\linewidth]{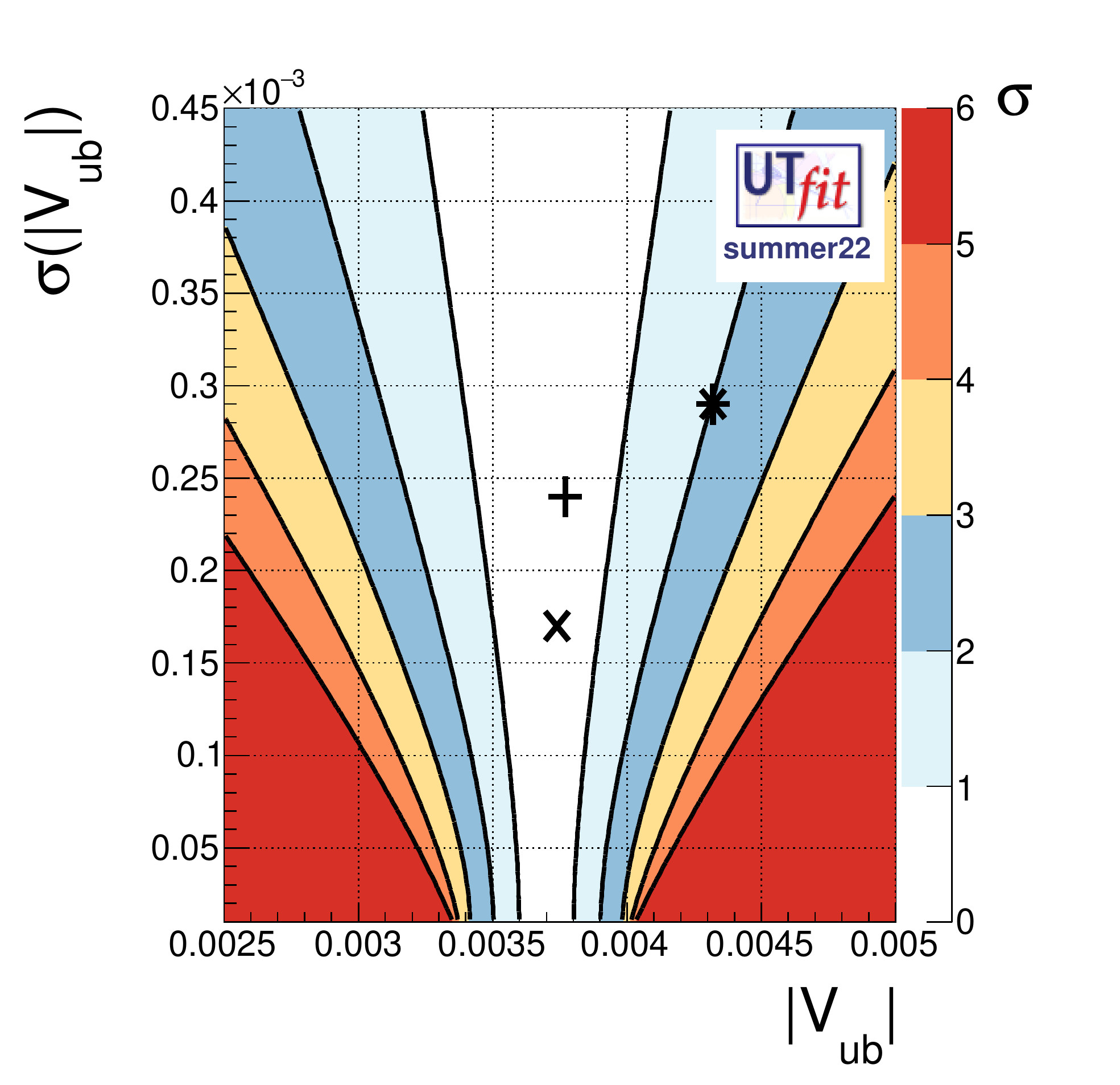} & 
    \hspace*{-0.6cm}
    \includegraphics[width=0.38\linewidth]{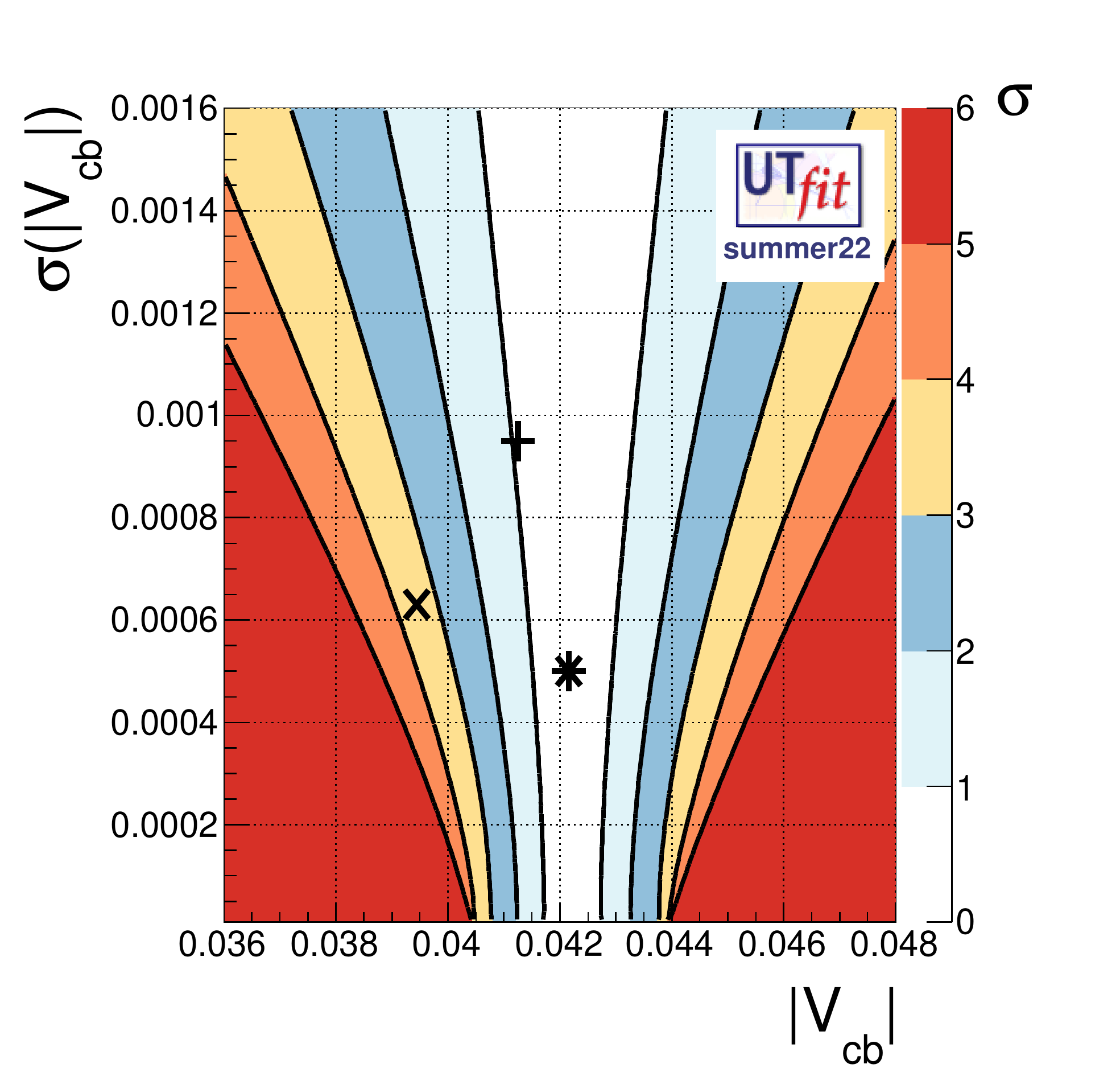} 
    \hspace*{-0.3cm}
  \end{tabular}
  \vspace*{-0.4cm}
  \caption{{\it  Pull plots (see text) for $\sin{2\beta}$ {\it{(top-left)}}, $\alpha$ {\it{(top-centre)}},
    $\gamma$ {\it{(top-right)}}, $|V_{ub}|$ {\it{(bottom-left)}} and $|V_{cb}|$ {\it{(bottom-right)}} inputs. The  crosses represent the input values reported in Table\,\ref{tab:fullSM}. In the case of $|V_{ub}|$  and $|V_{cb}|$ the {\rm x} and the {\rm *} represent the values extracted from exclusive and inclusive  semileptonic decays respectively.}}
  \label{fig:pulls}
\end{figure}
\begin{figure}
  \centering
    \includegraphics[width=0.50\linewidth]{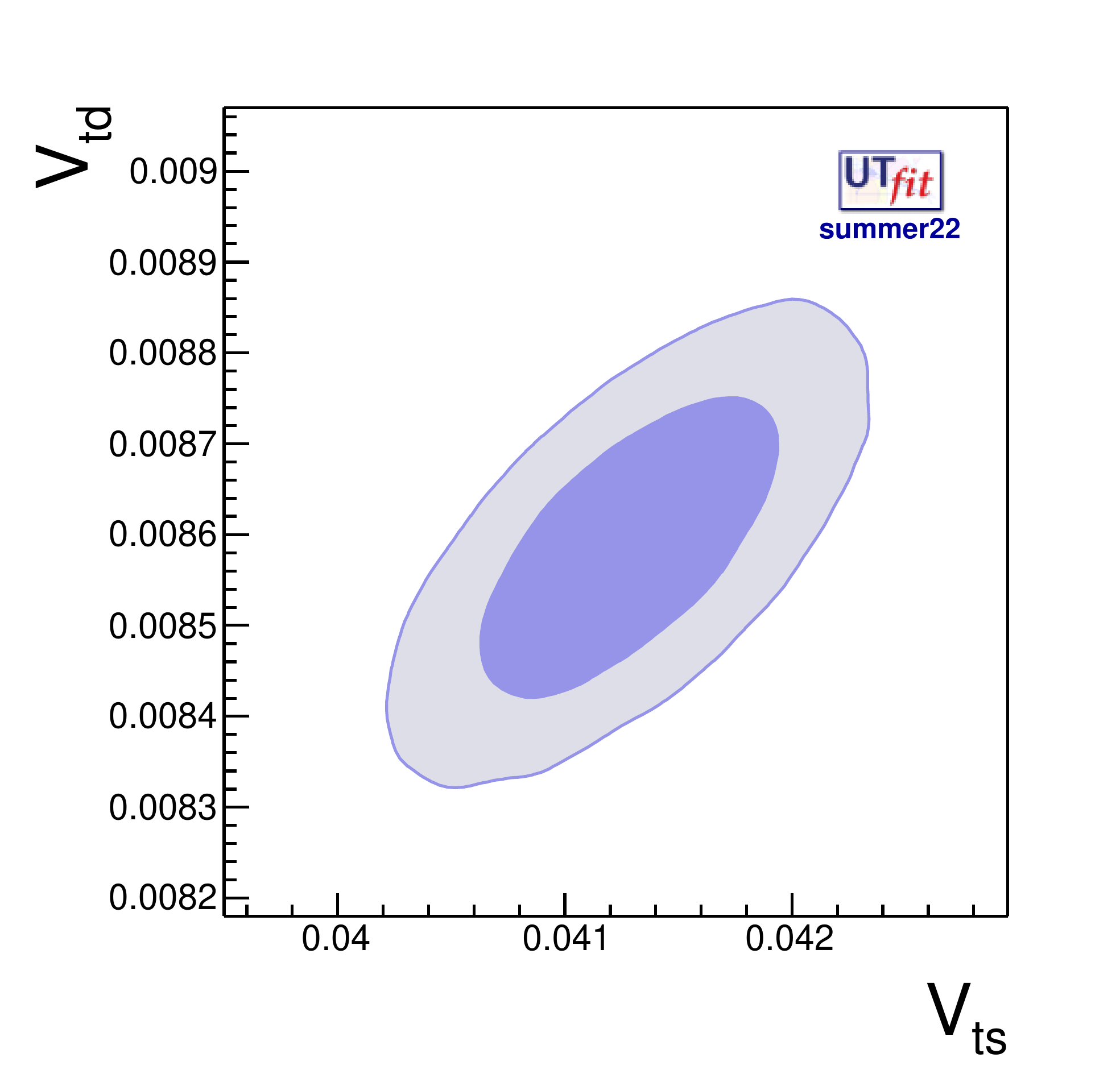}  
  \caption{{\it  Allowed region in the $|V_{td}|$-$|V_{ts}|$  plane.}}
  \label{fig:VtsvsVtd}
\end{figure}
\begin{figure}
  \centering
    \includegraphics[width=0.50\linewidth]{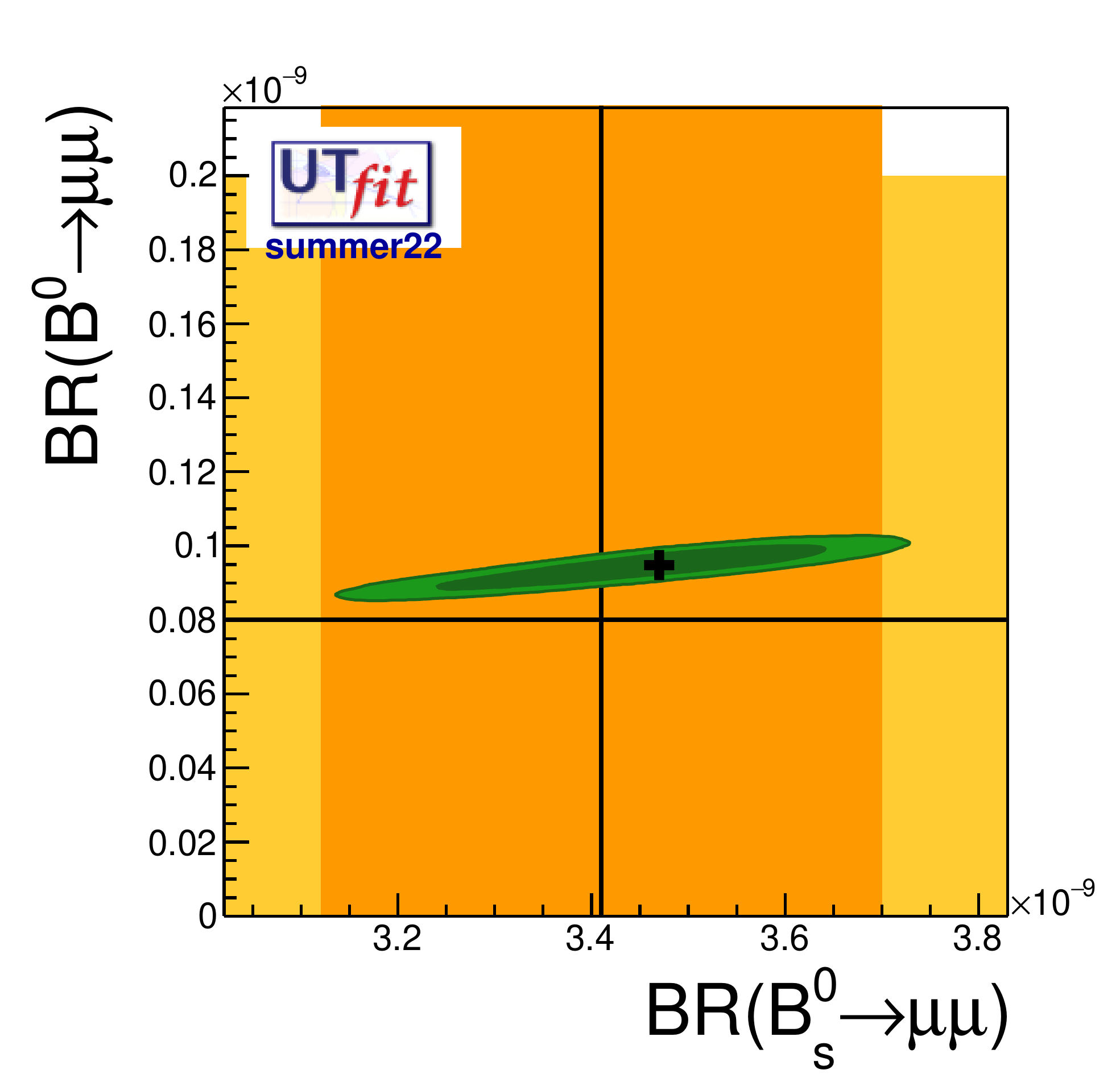}  
  \caption{{\it  Allowed region in the $BR(B^0_s\to \mu \mu)$-$BR(B^0\to \mu \mu)$  plane.  The vertical (orange) and  horizontal (yellow)  bands correspond to  the present experimental results ($1\sigma$ regions).}}
  \label{fig:Bmumu}
\end{figure}

The results of the global SM fit are given as two-dimensional probability distributions
in the plane  of the CKM parameters $\bar\rho$ and $\bar\eta$ and shown in Fig.~\ref{fig:sm21}.
The numerical results are in Table~\ref{tab:rhoeta}.   Besides the global fit  shown in the top-left panel, we have studied various configurations which
provide us further physical information:  
\begin{enumerate} 
\item By fitting  the  ``tree-only'' constraints, i.e. processes for which a contribution  from new physics  is with the highest probability absent,     we test the possibility that all the sources of  CP violation come from physics beyond the SM. The results shown in the top-right panel, which  have a two-fold sign ambiguity in  the $\bar \rho$-$\bar \eta$ values, show that the SM alone contributes to  the largest part of the observed CP violation at low energy;  
\item We  analysed the results   that can be obtained by using only the information coming from the measured angles, ``angle-only" fit, bottom-left panel;
\item We  analysed the results   that can be obtained  from the triangle sides fit and $\varepsilon$, ``sides+ $\varepsilon_K$"  fit,  bottom-right panel. 
 \end{enumerate}
 We observe that there is not a particular bias  from either case forcing the global fit to fill a specific region of the plane, all fits prefer essentially the same region in the $\bar \rho$-$\bar \eta$ plane.
We want also to give the CKM matrix in all its glory
\beq   V_{\rm CKM} = \left( \begin{array}{ ccc}
  0.97431(19)    &  0.22517(81) & 0.003715(93)\, e^{-i (65.1(1.3))^o} \\
  -0.22503(83) \, e^{+i (0.0351(1))^o}   &  0.97345(20)\, e^{-i (0.00187(5))^o} & 0.0420(5)  \\
  0.00859 (11)\, e^{-i (22.4(7))^o} & -0.04128 (46)  \, e^{+i (1.05(3))^o} & 0.999111(20)
 \end{array} \right) \, . 
\eeq

From the global fit we also  obtain
\beq  \lambda = 0.22519(83)\, ,  \qquad \qquad  {\cal A }=  0.828 (11) \, . \eeq
 Several other quantities that we have analysed in our fit can be found in Table\,\ref{tab:othersI} and\,\ref{tab:othersII} in the Appendix.

\subsection{Pull plots and allowed  regions}
\label{susec:pp}

For a given quantity $x$, the compatibility between its UTA  prediction $\bar x$,  given in Table\,\ref{tab:fullSM},  and its  direct measurement $\hat x$ is obtained integrating the probability density function (pdf) $p(\bar x - \hat x)$, in the region for which it acquires values smaller than $p(0)$,  namely the region for which the pdf value is smaller than that of the case $\bar x = \hat x$,  i.e., when the measurement matches the prediction. This two-sided $p$-value is then converted to the equivalent number of standard deviations for a Gaussian distribution. When  $ \bar x - \hat x$ is distributed according to a Gaussian p.d.f, this quantity coincides with the usual pull, i.e. with the ratio between $\vert \bar x - \hat x\vert $ and its standard deviation. The advantage of this approach is that no approximation is made on the shape of pdf's. 

The so-called ``pull plots``  are then constructed assuming a measured value and an experimental error for each point of the plane, with the procedure described above (assuming that the measurement has a Gaussian pdf).   In Fig.~\ref{fig:pulls}  these plots are used to assess the agreement
of a given measurement with the indirect determination from the fit using all the other inputs.
The coloured areas represent the level of agreement between the predicted values and the
measurements at better than $1$, $2$, \dots $n\, \sigma$.
The markers (crosses)  have the coordinates $(x,y)=($central value, error$)$ of the direct measurements
considered, see Table\,\ref{tab:fullSM}.  In the case of $|V_{ub}|$  and $|V_{cb}|$ the {\rm x} and the {\rm *} represent the values extracted from exclusive and inclusive  semileptonic decays respectively. These plots allow to visualise the tensions between each input and the rest of
them as in the pull column of Table~\ref{tab:fullSM}. It is clear that inputs as $\alpha$
and $\gamma$ show very good  agreement with the rest of the fit, while $\sin{2\beta}$,  $|V_{ub}|$ and $|V_{cb}|$
present various degrees of tension either directly or with respect to the different   exclusive or inclusive determinations.

Overall, the global fit proves a remarkable internal consistency with a better than $7\%$ precision in the determination of the fundamental CKM parameters $\bar\rho$ and $\bar\eta$. New physics effects, if present,  would give rather small contributions and for this reason it is necessary to improve both the  precision of the experiments and the accuracy of the theoretical calculations. 

We  also find   useful  to provide some further information coming form our global UT analysis: in Fig.\,\ref{fig:VtsvsVtd} we show  the  constraint on the third row of the CKM matrix given by   the allowed region, mainly determined by neutral $B$ meson mixing,  in the $|V_{td}|$-$|V_{ts}|$  plane;   
in Fig.\,\ref{fig:Bmumu} we show the  allowed region in the $BR(B^0_s\to \mu \mu)$-$BR(B^0\to \mu \mu)$  plane compared with the  experimental measurements given by the vertical (orange) and  horizontal (yellow)  bands corresponding to the present experimental constraints.

\subsection{Constraints on the lattice parameters in the Standard Model}
\label{susec:consLQCD}
Assuming the validity of the Standard Model, the constraints in the  $\bar\rho$-$\bar\eta$ plane 
allow the "experimental" determination of several hadronic
quantities which, in the previous fits,  were  taken from lattice QCD calculations. This approach has
 the advantage  that we can extract from the combined experimental measurements
the value of  $\hat B_K$,  of the $B^0$ mixing amplitudes, $f_{B_{s,d}} \hat B^{1/2}_{s,d}$ or $f_{B_{s}} \hat B^{1/2}_{s}$,  and of $\xi=f_{B_{s}} \hat B^{1/2}_{s}/f_{B_{d}} \hat B^{1/2}_{d}$. 
We have considered the following possibilities (the derived  lattice quantities and their uncertainties, obtained by combining the lattice inputs of Table  \ref{tab:fullLattice} are denoted by  "latt.*" ):

\begin{enumerate}
\item we remove from the lattice inputs $\hat B_K$ and we compare the \utfit \, value with the lattice result
\bea \hat B_K({\rm {\bf UT}}fit)=  0.840 (59) \, , \qquad   \hat B_K({\rm latt.})=0.756 (16)\,  ; \label{eq:uno} \eea
\item we only use $\hat B_{s}$ and $\hat B_{s}/\hat B_{d}$ and derive
\bea  f_{B_{d}} ({\rm {\bf UT}}fit) &=& 190.9 (7.2)\,{\rm MeV}  \, , \qquad  f_{B_{d}}({\rm latt.*})= 190.5(1.3)\,{\rm MeV}\, ; \nonumber \\ f_{B_{s}} ({\rm {\bf UT}}fit)&=& 229.4 (7.2)\,{\rm MeV}  \, , \qquad f_{B_{s}} ({\rm latt.*})= 230.1 (1.2)\,{\rm MeV}\, ; \label{eq:due} \\ \xi({\rm {\bf UT}}fit)&=&  1.204(27) \, , \qquad \qquad  \,\,\,\xi({\rm latt.*})=1.208(59)\, ;  \nonumber \eea
\item we use only the ratios  $f_{B_{s}} /f_{B_{d}}$ and $\hat B_{s}/\hat B_{d}$ but not $f_{B_{s}}$ and $\hat B_{s}$
\bea  f_{B_{d}} \hat B^{1/2}_{d}({\rm {\bf UT}}fit)&=& 216.9 (5.3)\,{\rm MeV}  \, , \qquad  f_{B_{d}} \hat B^{1/2}_{d}({\rm latt.*})= 214.2 (5.6)\,{\rm MeV}\, ; \nonumber \\  f_{B_{s}} \hat B^{1/2}_{s}({\rm {\bf UT}}fit)&=& 264.4 (6.0)\,{\rm MeV}  \, , \qquad  f_{B_{s}} \hat B^{1/2}_{s}({\rm latt.*})= 260.7 (6.1)\,{\rm MeV}\, ; \label{eq:tre} \\  \xi({\rm {\bf UT}}fit))&=&  1.219(12) \, , \qquad  \qquad  \,\,\,\xi({\rm latt.*})=1.208(51)\,  ;  \nonumber \eea
\item we only use $\hat B_K$ but not any of the other inputs of table\,\ref{tab:fullLattice}
\bea  f_{B_{d}} \hat B^{1/2}_{d}({\rm {\bf UT}}fit)&=& 210.5 (3.6)\,{\rm MeV}  \, , \qquad  f_{B_{d}} \hat B^{1/2}_{d}({\rm latt.*})= 214.2 (5.6)\,{\rm MeV}\, ; \nonumber \\  f_{B_{s}} \hat B^{1/2}_{s}({\rm {\bf UT}}fit)&=& 259.0 (3.4)\,{\rm MeV}  \, , \qquad  f_{B_{s}} \hat B^{1/2}_{s}({\rm latt.*})= 260.7 (6.1)\,{\rm MeV}\, ; \label{eq:quattro} \\  \xi({\rm {\bf UT}}fit)&=&  1.230(23) \, , \qquad  \qquad  \,\,\,\xi({\rm latt.*})=1.217(14)\,  .  \nonumber \eea
\end{enumerate}

We observe that the case 3. has simply slightly larger uncertainties than the case 4. and that the \utfit \,  predictions of the hadronic parameters are fully compatible with the lattice calculations.
For further information, we also repeated the case 1.  with $\vert V_{cb}\vert$ taken from Eq.\,(\ref{eq:vcbtotDM}) instead than Eq.\,(\ref{eq:vcbtot}) obtaining  
$B_K({\rm {\bf UT}}fit)=  0.831 (57) $  in substantial agreement  with the result in Eq.\,(\ref{eq:uno}).  This remains true for the other parameter condidered in all the other cases (2.-3.-4.).

\section{Conclusions}
\label{sec:concl}
Our main conclusions are the following:
\begin{itemize}
\item  The SM analysis shows a very good overall consistency;
\item The exclusive vs inclusive saga  is not concluded yet although there are signals that it could be quickly resolved.  We stress that, as in the past\,\cite{Alpigiani:2017lpj},  the unitarity triangle analysis, namely the analysis without including  the experimental measurements from semileptonic decays,  favours a large value of  $\vert V_{cb}\vert$,  close to the inclusive determination, and a smalle value of  $\vert V_{ub}\vert$,  close to the exclusive determination;
\item For $\vert V_{cb}\vert$,   on the theoretical side,  there  are signals that a more accurate determination of the 
form factors obtained from new and more accurate lattice calculations and   the DM approach\,\cite{Martinelli:2021myh},  combined with a more careful treatment of the experimental data,  could increase the central value and determine more realistically the uncertainty, thus reducing substantially the tension between the inclusive and exclusive values of this quantity.  The difference with respect to previous analyses is not only due 
to the use of the DM approach but also to a critical examination  of the experimental correlation matrix of the $B  \to D^*$ differential decay rates and to the theoretical determination of the momentum  dependence   of the form factors independently of the  experiments  and before fitting 
the data\,\cite{Martinelli:2021frl, Martinelli:2022adr, Martinelli:2022xir, Martinelli:2021myh}. 
\item Similar analyses are needed for semileptonic $B\to \pi$ decays, although in that case the tension is smaller since the uncertainties in both the inclusive and exclusive determinations of $\vert V_{ub}\vert $ are larger. In ref.\,\cite{Martinelli:2022tte} for example, using the DM approach,  the results are  
$\vert V_{ub} \vert \times 10^3 = 3.62 (47)$ from $B\to \pi$ decays and  $\vert V_{ub} \vert \times 10^3 = 3.77 (48)$  from $B_s \to K $ decays,  compatible with the latest inclusive determination  $\vert V_{ub}\vert \times 10^3 = 4.13 (26)$  from PDG\,\cite{ParticleDataGroup:2022pth};
\item On the experimental side, regarding the differences between exclusive and inclusive semileptonic  $B$ decays,  we really  need more  measurements  
 from LHCb and (mostly) Belle II;
\item Most of the quantities studied in this work refer to the quark down sector for which  the CKM paradigm  has been (and still is) a great success in predicting weak processes  (mainly to the physics of strange and beauty particles). Only  recently  we experimentally   started to investigate if there are signals of  New Physics in  the up sector singled by discrepancies between measurements and theoretical predictions. The discovery of CP violation in neutral D meson system  has opened a new sector of investigation and will be the subject of our study in a future publication.  
\end{itemize}

\acknowledgments
We thank Antonio Di Domenico, C. Kelly and C. Sachrajda for very useful  discussions. The work of M.V. is supported by the Simons Foundation under the Simons Bridge for Postdoctoral Fellowships at SCGP and YITP, award number 815892.
DD is thankful to HSE basic research program.  This work was partly supported  by the Italian Ministry of Research (MIUR) under grant PRIN 20172LNEEZL.V. acknowledge the hospitality of the Scuola Normale Superiore and of the INFN, Sezione di Pisa where part of this work was developed.
\newpage
\appendix
\section*{Appendix}
\label{sec:app}
\vskip 1 cm
\begin{table}[ht]
 \begin{tabular}{c|c|c} 
 \hline 
 Observable & Full Fit & Full Fit (95\%) \\ 
 \hline 
$\lambda$ & $ 0.22519 \pm 0.00083 $ & $[ 0.22359 , 0.22686 ]$  \\ 
${\cal A}$ & $ 0.828 \pm 0.011 $ & $[ 0.807 , 0.851 ]$  \\ 
$\bar{\rho}$ & $ 0.1609 \pm 0.0095 $ & $[ 0.1430 , 0.1794 ]$  \\ 
$\bar{\eta}$ & $ 0.347 \pm 0.010 $ & $[ 0.327 , 0.367 ]$  \\ 
$\beta$ & $ 22.46 \pm 0.68 $ & $[ 21.13 , 23.78 ]$  \\ 
$\alpha$ & $ 92.4 \pm 1.4 $ & $[ 89.9 , 95.4 ]$  \\ 
$\beta_s$ & $ -0.0368 \pm 0.0010 $ & $[ -0.0388 , -0.0347 ]$  \\ 
$\sin\theta_{12}$ & $ 0.22519 \pm 0.00083 $ & $[ 0.22359 , 0.22686 ]$  \\ 
$\sin\theta_{23}$ & $ 0.04200 \pm 0.00047 $ & $[ 0.04109 , 0.04290 ]$  \\ 
$\sin\theta_{13}$ & $ 0.003714 \pm 0.000092 $ & $[ 0.003528 , 0.003898 ]$  \\ 
$\delta[^\circ]$ & $ 1.137 \pm 0.022 $ & $[ 1.092 , 1.180 ]$  \\ 
$J_{CP}\times 10^5$ & $ 3.102 \pm 0.080 $ & $[ 2.946 , 3.264 ]$  \\ 
$R_{t}$ & $ 0.9082 \pm 0.0084 $ & $[ 0.8908 , 0.9244 ]$  \\ 
$R_{b}$ & $ 0.383 \pm 0.011 $ & $[ 0.362 , 0.404 ]$  \\ 
$\vert V_{td}/V_{ts}\vert$ & $ 0.2080 \pm 0.0020 $ & $[ 0.2041 , 0.2119 ]$  \\ 
BR$(B_d \to \mu \mu)\times 10^{10}$ & $ 0.949 \pm 0.037 $ & $[ 0.871 , 1.009 ]$  \\ 
BR$(B_s \to \mu \mu)\times 10^{9}$ & $ 3.25 \pm 0.12 $ & $[ 3.01 , 3.44 ]$  \\ 
$\vert V_{ud}\vert$ & $ 0.97431 \pm 0.00019 $ & $[ 0.97392 , 0.97468 ]$  \\ 
$\vert V_{us}\vert$ & $ 0.22517 \pm 0.00081 $ & $[ 0.22352 , 0.22682 ]$  \\ 
$\vert V_{ub}\vert$ & $ 0.003715 \pm 0.000093 $ & $[ 0.003532 , 0.003898 ]$  \\ 
$\vert V_{cd}\vert$ & $ 0.22503 \pm 0.00083 $ & $[ 0.22343 , 0.22669 ]$  \\ 
$\vert V_{cs}\vert$ & $ 0.97345 \pm 0.00020 $ & $[ 0.97305 , 0.97381 ]$  \\ 
$\vert V_{cb}\vert$ & $ 0.04200 \pm 0.00047 $ & $[ 0.04109 , 0.04290 ]$  \\ 
$\vert V_{td}\vert$ & $ 0.00859 \pm 0.00011 $ & $[ 0.00837 , 0.00880 ]$  \\ 
$\vert V_{ts}\vert$ & $ 0.04128 \pm 0.00046 $ & $[ 0.04038 , 0.04217 ]$  \\ 
$\vert V_{tb}\vert$ & $ 0.999111 \pm 0.000020 $ & $[ 0.999072 , 0.999149 ]$  \\ 
\end{tabular} 
\caption{\it Extra outputs of interest obtained from our UTA analysis I.}
\label{tab:othersI}
 \end{table} 
\begin{table} 
 \begin{tabular}{c|c|c} 
 \hline 
 Observable & Full Fit & Full Fit (95\%) \\ 
 \hline 
Re$(\lambda^{t}_{sd})$ & $ -0.0003252 \pm 0.0000080 $ & $[ -0.0003413 , -0.0003098 ]$  \\ 
Re$(\lambda^{c}_{sd})$ & $ -0.21908 \pm 0.00076 $ & $[ -0.22058 , -0.21757 ]$  \\ 
Re$(\lambda^{u}_{sd})$ & $ 0.21943 \pm 0.00077 $ & $[ 0.21789 , 0.22090 ]$  \\ 
Re$(\lambda^{t}_{bd})$ & $ 0.00794 \pm 0.00013 $ & $[ 0.00769 , 0.00818 ]$  \\ 
Re$(\lambda^{c}_{bd})$ & $ -0.00945 \pm 0.00011 $ & $[ -0.00967 , -0.00924 ]$  \\ 
Re$(\lambda^{u}_{bd})$ & $ 0.001522 \pm 0.000090 $ & $[ 0.001344 , 0.001699 ]$  \\ 
Re$(\lambda^{t}_{bs})$ & $ -0.04123 \pm 0.00046 $ & $[ -0.04216 , -0.04037 ]$  \\ 
Re$(\lambda^{c}_{bs})$ & $ 0.04089 \pm 0.00045 $ & $[ 0.04001 , 0.04177 ]$  \\ 
Re$(\lambda^{u}_{bs})$ & $ -0.04123 \pm 0.00046 $ & $[ -0.04216 , -0.04037 ]$  \\ 
Im$(\lambda^{t}_{sd})\times 10^5$ & $ 14.13 \pm 0.37 $ & $[ 13.42 , 14.85 ]$  \\ 
Im$(\lambda^{c}_{sd})\times 10^5$ & $ -14.13 \pm 0.37 $ & $[ -14.85 , -13.42 ]$  \\ 
Im$(\lambda^{t}_{bd})\times 10^5$ & $ -327.6 \pm 8.1 $ & $[ -343.2 , -311.3 ]$  \\ 
Im$(\lambda^{c}_{bd})\times 10^5$ & $ -0.578 \pm 0.018 $ & $[ -0.615 , -0.545 ]$  \\ 
Im$(\lambda^{u}_{bd})\times 10^5$ & $ 328.1 \pm 8.3 $ & $[ 311.6 , 344.0 ]$  \\ 
Im$(\lambda^{t}_{bs})\times 10^5$ & $ -75.7 \pm 1.9 $ & $[ -79.4 , -71.8 ]$  \\ 
Im$(\lambda^{c}_{bs})\times 10^5$ & $ -0.1336 \pm 0.0041 $ & $[ -0.1420 , -0.1258 ]$  \\ 
Im$(\lambda^{u}_{bs})\times 10^5$ & $ -75.7 \pm 1.9 $ & $[ -79.4 , -71.8 ]$  \\ 
$\vert\lambda^{t}_{sd}\vert$ & $ 0.0003545 \pm 0.0000075 $ & $[ 0.0003398 , 0.0003698 ]$  \\ 
$\vert\lambda^{c}_{sd}\vert$ & $ 0.21908 \pm 0.00076 $ & $[ 0.21757 , 0.22058 ]$  \\ 
$\vert\lambda^{u}_{sd}\vert$ & $ 0.21943 \pm 0.00077 $ & $[ 0.21789 , 0.22090 ]$  \\ 
$\vert\lambda^{t}_{bd}\vert$ & $ 0.00858 \pm 0.00011 $ & $[ 0.00837 , 0.00880 ]$  \\ 
$\vert\lambda^{c}_{bd}\vert$ & $ 0.00945 \pm 0.00011 $ & $[ 0.00924 , 0.00967 ]$  \\ 
$\vert\lambda^{u}_{bd}\vert$ & $ 0.003618 \pm 0.000091 $ & $[ 0.003436 , 0.003794 ]$  \\ 
$\vert\lambda^{t}_{bs}\vert$ & $ 0.04125 \pm 0.00045 $ & $[ 0.04037 , 0.04213 ]$  \\ 
$\vert\lambda^{c}_{bs}\vert$ & $ 0.04089 \pm 0.00045 $ & $[ 0.04001 , 0.04177 ]$  \\ 
$\vert\lambda^{u}_{bs}\vert$ & $ 0.04125 \pm 0.00045 $ & $[ 0.04037 , 0.04213 ]$  \\ 
Im$(\tau)$ & $ -0.000644 \pm 0.000016 $ & $[ -0.000678 , -0.000612 ]$  \\ 
Re$(\tau)$ & $ 0.001482 \pm 0.000036 $ & $[ 0.001410 , 0.001557 ]$  \\ 
\end{tabular} 
\caption{\it Extra outputs of interest obtained from   our UTA analysis II. $\lambda^q_{ij} = V_{qi} V_{qj}^*$.}
\label{tab:othersII}
 \end{table} 
\newpage

\bibliographystyle{JHEP}
\bibliography{rendicontiLincei}

\providecommand{\href}[2]{#2}\begingroup\raggedright\begin{thebibliography}{10}

\bibitem{Ciuchini:2000de}
M.~Ciuchini, G.~D'Agostini, E.~Franco, V.~Lubicz, G.~Martinelli, F.~Parodi
  et~al., \emph{{2000 CKM triangle analysis: A Critical review with updated
  experimental inputs and theoretical parameters}},
  \href{http://dx.doi.org/10.1088/1126-6708/2001/07/013}{\emph{JHEP} {\bf 07}
  (2001) 013}, [\href{http://arxiv.org/abs/hep-ph/0012308}{{\tt
  hep-ph/0012308}}].

\bibitem{Alpigiani:2017lpj}
C.~Alpigiani et~al., \emph{{Unitarity Triangle Analysis in the Standard Model
  and Beyond}},  in \emph{{5th Large Hadron Collider Physics Conference}}, 10,
  2017.
\newblock \href{http://arxiv.org/abs/1710.09644}{{\tt 1710.09644}}.

\bibitem{Bona:2022zhn}
M.~Bona et~al., \emph{{Unitarity Triangle global fits beyond the Standard
  Model: UTfit 2021 NP update}},
  \href{http://dx.doi.org/10.22323/1.398.0500}{\emph{PoS} {\bf EPS-HEP2021}
  (2022) 500}.

\bibitem{Bona:2022xnf}
M.~Bona et~al., \emph{{Unitarity Triangle global fits testing the Standard
  Model: UT$fit$ 2021 SM update}},
  \href{http://dx.doi.org/10.22323/1.398.0512}{\emph{PoS} {\bf EPS-HEP2021}
  (2022) 512}.

\bibitem{Buras:2022qip}
A.~J. Buras, \emph{{Standard Model Predictions for Rare K and B Decays without
  New Physics Infection}},  \href{http://arxiv.org/abs/2209.03968}{{\tt
  2209.03968}}.

\bibitem{Cabibbo:1963yz}
N.~Cabibbo, \emph{{Unitary Symmetry and Leptonic Decays}},
  \href{http://dx.doi.org/10.1103/PhysRevLett.10.531}{\emph{Phys. Rev. Lett.}
  {\bf 10} (1963) 531--533}.

\bibitem{Kobayashi:1973fv}
M.~Kobayashi and T.~Maskawa, \emph{{CP Violation in the Renormalizable Theory
  of Weak Interaction}},
  \href{http://dx.doi.org/10.1143/PTP.49.652}{\emph{Prog. Theor. Phys.} {\bf
  49} (1973) 652--657}.

\bibitem{Wolfenstein:1983yz}
L.~Wolfenstein, \emph{{Parametrization of the Kobayashi-Maskawa Matrix}},
  \href{http://dx.doi.org/10.1103/PhysRevLett.51.1945}{\emph{Phys. Rev. Lett.}
  {\bf 51} (1983) 1945}.

\bibitem{Buras:1994ec}
A.~J. Buras, M.~E. Lautenbacher and G.~Ostermaier, \emph{{Waiting for the top
  quark mass, ${K}^+ \to \pi^+ \nu \bar{\nu}$, ${B}_s^0 - \bar{B}_s^0$ mixing
  and CP asymmetries in B decays}},
  \href{http://dx.doi.org/10.1103/PhysRevD.50.3433}{\emph{Phys. Rev. D} {\bf
  50} (1994) 3433--3446}, [\href{http://arxiv.org/abs/hep-ph/9403384}{{\tt
  hep-ph/9403384}}].

\bibitem{Glashow:1970gm}
S.~L. Glashow, J.~Iliopoulos and L.~Maiani, \emph{{Weak Interactions with
  Lepton-Hadron Symmetry}},
  \href{http://dx.doi.org/10.1103/PhysRevD.2.1285}{\emph{Phys. Rev. D} {\bf 2}
  (1970) 1285--1292}.

\bibitem{Gambino:2013rza}
P.~Gambino and C.~Schwanda, \emph{{Inclusive semileptonic fits, heavy quark
  masses, and $V_{cb}$}},
  \href{http://dx.doi.org/10.1103/PhysRevD.89.014022}{\emph{Phys. Rev. D} {\bf
  89} (2014) 014022}, [\href{http://arxiv.org/abs/1307.4551}{{\tt 1307.4551}}].

\bibitem{Alberti:2014yda}
A.~Alberti, P.~Gambino, K.~J. Healey and S.~Nandi, \emph{{Precision
  Determination of the Cabibbo-Kobayashi-Maskawa Element $V_{cb}$}},
  \href{http://dx.doi.org/10.1103/PhysRevLett.114.061802}{\emph{Phys. Rev.
  Lett.} {\bf 114} (2015) 061802}, [\href{http://arxiv.org/abs/1411.6560}{{\tt
  1411.6560}}].

\bibitem{Gambino:2016jkc}
P.~Gambino, K.~J. Healey and S.~Turczyk, \emph{{Taming the higher power
  corrections in semileptonic B decays}},
  \href{http://dx.doi.org/10.1016/j.physletb.2016.10.023}{\emph{Phys. Lett. B}
  {\bf 763} (2016) 60--65}, [\href{http://arxiv.org/abs/1606.06174}{{\tt
  1606.06174}}].

\bibitem{BaBar:2007nwi}
{\scshape BaBar} collaboration, B.~Aubert et~al., \emph{{Measurement of the
  Decay $B^{-} \to D^{*0} e^{-} \bar{\nu}_e$}},
  \href{http://dx.doi.org/10.1103/PhysRevLett.100.231803}{\emph{Phys. Rev.
  Lett.} {\bf 100} (2008) 231803}, [\href{http://arxiv.org/abs/0712.3493}{{\tt
  0712.3493}}].

\bibitem{BaBar:2007cke}
{\scshape BaBar} collaboration, B.~Aubert et~al., \emph{{Determination of the
  form-factors for the decay $B^0 \to D^{*-} \ell^{+} \nu_{l}$ and of the CKM
  matrix element $|V_{cb}|$}},
  \href{http://dx.doi.org/10.1103/PhysRevD.77.032002}{\emph{Phys. Rev. D} {\bf
  77} (2008) 032002}, [\href{http://arxiv.org/abs/0705.4008}{{\tt 0705.4008}}].

\bibitem{BaBar:2008zui}
{\scshape BaBar} collaboration, B.~Aubert et~al., \emph{{Measurements of the
  Semileptonic Decays $\bar{{B}} \to {D} \ell \bar{\nu}$ and $\bar{{B}} \to
  {D^*} \ell \bar{\nu}$ Using a Global Fit to ${D} {X} \ell \bar{\nu}$ Final
  States}}, \href{http://dx.doi.org/10.1103/PhysRevD.79.012002}{\emph{Phys.
  Rev. D} {\bf 79} (2009) 012002}, [\href{http://arxiv.org/abs/0809.0828}{{\tt
  0809.0828}}].

\bibitem{BaBar:2009zxk}
{\scshape BaBar} collaboration, B.~Aubert et~al., \emph{{Measurement of $\vert
  V_{cb} \vert$ and the Form-Factor Slope in $\bar{{B}} \to {D} \ell^-
  \bar{\nu}_{\ell}$ Decays in Events Tagged by a Fully Reconstructed {B}
  Meson}}, \href{http://dx.doi.org/10.1103/PhysRevLett.104.011802}{\emph{Phys.
  Rev. Lett.} {\bf 104} (2010) 011802},
  [\href{http://arxiv.org/abs/0904.4063}{{\tt 0904.4063}}].

\bibitem{Belle:2010qug}
{\scshape Belle} collaboration, W.~Dungel et~al., \emph{{Measurement of the
  form factors of the decay ${B^0} \to {D^{*-}} \ell^+ \nu_{\ell}$ and
  determination of the CKM matrix element $\vert V_{cb} \vert$}},
  \href{http://dx.doi.org/10.1103/PhysRevD.82.112007}{\emph{Phys. Rev. D} {\bf
  82} (2010) 112007}, [\href{http://arxiv.org/abs/1010.5620}{{\tt 1010.5620}}].

\bibitem{Belle:2015pkj}
{\scshape Belle} collaboration, R.~Glattauer et~al., \emph{{Measurement of the
  decay $B\to D\ell\nu_\ell$ in fully reconstructed events and determination of
  the Cabibbo-Kobayashi-Maskawa matrix element $|V_{cb}|$}},
  \href{http://dx.doi.org/10.1103/PhysRevD.93.032006}{\emph{Phys. Rev. D} {\bf
  93} (2016) 032006}, [\href{http://arxiv.org/abs/1510.03657}{{\tt
  1510.03657}}].

\bibitem{Belle:2017rcc}
{\scshape Belle} collaboration, A.~Abdesselam et~al., \emph{{Precise
  determination of the CKM matrix element $\left| V_{cb}\right|$ with $\bar B^0
  \to D^{*\,+} \, \ell^- \, \bar \nu_\ell$ decays with hadronic tagging at
  Belle}},  \href{http://arxiv.org/abs/1702.01521}{{\tt 1702.01521}}.

\bibitem{Belle:2018ezy}
{\scshape Belle} collaboration, E.~Waheed et~al., \emph{{Measurement of the CKM
  matrix element $|V_{cb}|$ from $B^0\to D^{*-}\ell^ {+} \nu_\ell$ at Belle}},
  \href{http://dx.doi.org/10.1103/PhysRevD.100.052007}{\emph{Phys. Rev. D} {\bf
  100} (2019) 052007}, [\href{http://arxiv.org/abs/1809.03290}{{\tt
  1809.03290}}].

\bibitem{HFLAV:2019otj}
{\scshape HFLAV} collaboration, Y.~S. Amhis et~al., \emph{{Averages of
  b-hadron, c-hadron, and $\tau $-lepton properties as of 2018}},
  \href{http://dx.doi.org/10.1140/epjc/s10052-020-8156-7}{\emph{Eur. Phys. J.
  C} {\bf 81} (2021) 226}, [\href{http://arxiv.org/abs/1909.12524}{{\tt
  1909.12524}}].

\bibitem{BaBar:2012obs}
{\scshape BaBar} collaboration, J.~P. Lees et~al., \emph{{Evidence for an
  excess of $\bar{B} \to D^{(*)} \tau^-\bar{\nu}_\tau$ decays}},
  \href{http://dx.doi.org/10.1103/PhysRevLett.109.101802}{\emph{Phys. Rev.
  Lett.} {\bf 109} (2012) 101802}, [\href{http://arxiv.org/abs/1205.5442}{{\tt
  1205.5442}}].

\bibitem{BaBar:2013mob}
{\scshape BaBar} collaboration, J.~P. Lees et~al., \emph{{Measurement of an
  Excess of $\bar{B} \to D^{(*)}\tau^- \bar{\nu}_\tau$ Decays and Implications
  for Charged Higgs Bosons}},
  \href{http://dx.doi.org/10.1103/PhysRevD.88.072012}{\emph{Phys. Rev. D} {\bf
  88} (2013) 072012}, [\href{http://arxiv.org/abs/1303.0571}{{\tt 1303.0571}}].

\bibitem{LHCb:2015gmp}
{\scshape LHCb} collaboration, R.~Aaij et~al., \emph{{Measurement of the ratio
  of branching fractions $\mathcal{B}(\bar{B}^0 \to
  D^{*+}\tau^{-}\bar{\nu}_{\tau})/\mathcal{B}(\bar{B}^0 \to
  D^{*+}\mu^{-}\bar{\nu}_{\mu})$}},
  \href{http://dx.doi.org/10.1103/PhysRevLett.115.111803}{\emph{Phys. Rev.
  Lett.} {\bf 115} (2015) 111803}, [\href{http://arxiv.org/abs/1506.08614}{{\tt
  1506.08614}}].

\bibitem{Belle:2015qfa}
{\scshape Belle} collaboration, M.~Huschle et~al., \emph{{Measurement of the
  branching ratio of $\bar{B} \to D^{(\ast)} \tau^- \bar{\nu}_\tau$ relative to
  $\bar{B} \to D^{(\ast)} \ell^- \bar{\nu}_\ell$ decays with hadronic tagging
  at Belle}}, \href{http://dx.doi.org/10.1103/PhysRevD.92.072014}{\emph{Phys.
  Rev. D} {\bf 92} (2015) 072014}, [\href{http://arxiv.org/abs/1507.03233}{{\tt
  1507.03233}}].

\bibitem{Belle:2016ure}
{\scshape Belle} collaboration, Y.~Sato et~al., \emph{{Measurement of the
  branching ratio of $\bar{B}^0 \rightarrow D^{*+} \tau^- \bar{\nu}_{\tau}$
  relative to $\bar{B}^0 \rightarrow D^{*+} \ell^- \bar{\nu}_{\ell}$ decays
  with a semileptonic tagging method}},
  \href{http://dx.doi.org/10.1103/PhysRevD.94.072007}{\emph{Phys. Rev. D} {\bf
  94} (2016) 072007}, [\href{http://arxiv.org/abs/1607.07923}{{\tt
  1607.07923}}].

\bibitem{Belle:2016dyj}
{\scshape Belle} collaboration, S.~Hirose et~al., \emph{{Measurement of the
  $\tau$ lepton polarization and $R(D^*)$ in the decay $\bar{B} \to D^* \tau^-
  \bar{\nu}_\tau$}},
  \href{http://dx.doi.org/10.1103/PhysRevLett.118.211801}{\emph{Phys. Rev.
  Lett.} {\bf 118} (2017) 211801}, [\href{http://arxiv.org/abs/1612.00529}{{\tt
  1612.00529}}].

\bibitem{LHCb:2017smo}
{\scshape LHCb} collaboration, R.~Aaij et~al., \emph{{Measurement of the ratio
  of the $B^0 \to D^{*-} \tau^+ \nu_{\tau}$ and $B^0 \to D^{*-} \mu^+
  \nu_{\mu}$ branching fractions using three-prong $\tau$-lepton decays}},
  \href{http://dx.doi.org/10.1103/PhysRevLett.120.171802}{\emph{Phys. Rev.
  Lett.} {\bf 120} (2018) 171802}, [\href{http://arxiv.org/abs/1708.08856}{{\tt
  1708.08856}}].

\bibitem{Belle:2017ilt}
{\scshape Belle} collaboration, S.~Hirose et~al., \emph{{Measurement of the
  $\tau$ lepton polarization and $R(D^*)$ in the decay $\bar{B} \rightarrow D^*
  \tau^- \bar{\nu}_\tau$ with one-prong hadronic $\tau$ decays at Belle}},
  \href{http://dx.doi.org/10.1103/PhysRevD.97.012004}{\emph{Phys. Rev. D} {\bf
  97} (2018) 012004}, [\href{http://arxiv.org/abs/1709.00129}{{\tt
  1709.00129}}].

\bibitem{LHCb:2017rln}
{\scshape LHCb} collaboration, R.~Aaij et~al., \emph{{Test of Lepton Flavor
  Universality by the measurement of the $B^0 \to D^{*-} \tau^+ \nu_{\tau}$
  branching fraction using three-prong $\tau$ decays}},
  \href{http://dx.doi.org/10.1103/PhysRevD.97.072013}{\emph{Phys. Rev. D} {\bf
  97} (2018) 072013}, [\href{http://arxiv.org/abs/1711.02505}{{\tt
  1711.02505}}].

\bibitem{LHCb:2021trn}
{\scshape LHCb} collaboration, R.~Aaij et~al., \emph{{Test of lepton
  universality in beauty-quark decays}},
  \href{http://dx.doi.org/10.1038/s41567-021-01478-8}{\emph{Nature Phys.} {\bf
  18} (2022) 277--282}, [\href{http://arxiv.org/abs/2103.11769}{{\tt
  2103.11769}}].

\bibitem{LHCb:2021lvy}
{\scshape LHCb} collaboration, R.~Aaij et~al., \emph{{Tests of lepton
  universality using $B^0\to K^0_S \ell^+ \ell^-$ and $B^+\to K^{*+} \ell^+
  \ell^-$ decays}},
  \href{http://dx.doi.org/10.1103/PhysRevLett.128.191802}{\emph{Phys. Rev.
  Lett.} {\bf 128} (2022) 191802}, [\href{http://arxiv.org/abs/2110.09501}{{\tt
  2110.09501}}].

\bibitem{LHCb:2017avl}
{\scshape LHCb} collaboration, R.~Aaij et~al., \emph{{Test of lepton
  universality with $B^{0} \rightarrow K^{*0}\ell^{+}\ell^{-}$ decays}},
  \href{http://dx.doi.org/10.1007/JHEP08(2017)055}{\emph{JHEP} {\bf 08} (2017)
  055}, [\href{http://arxiv.org/abs/1705.05802}{{\tt 1705.05802}}].

\bibitem{Belle:2019oag}
{\scshape Belle} collaboration, A.~Abdesselam et~al., \emph{{Test of
  Lepton-Flavor Universality in ${B\to K^\ast\ell^+\ell^-}$ Decays at Belle}},
  \href{http://dx.doi.org/10.1103/PhysRevLett.126.161801}{\emph{Phys. Rev.
  Lett.} {\bf 126} (2021) 161801}, [\href{http://arxiv.org/abs/1904.02440}{{\tt
  1904.02440}}].

\bibitem{RBC:2020kdj}
{\scshape RBC, UKQCD} collaboration, R.~Abbott et~al., \emph{{Direct CP
  violation and the $\Delta I=1/2$ rule in $K\to\pi\pi$ decay from the standard
  model}}, \href{http://dx.doi.org/10.1103/PhysRevD.102.054509}{\emph{Phys.
  Rev. D} {\bf 102} (2020) 054509},
  [\href{http://arxiv.org/abs/2004.09440}{{\tt 2004.09440}}].

\bibitem{Ciuchini:2021zgf}
M.~Ciuchini, E.~Franco, V.~Lubicz, G.~Martinelli, L.~Silvestrini and
  C.~Tarantino, \emph{{Power corrections to the CP-violation parameter
  \ensuremath{\varepsilon}$_{K}$}},
  \href{http://dx.doi.org/10.1007/JHEP02(2022)181}{\emph{JHEP} {\bf 02} (2022)
  181}, [\href{http://arxiv.org/abs/2111.05153}{{\tt 2111.05153}}].

\bibitem{ParticleDataGroup:2022pth}
{\scshape Particle Data Group} collaboration, R.~L. Workman et~al.,
  \emph{{Review of Particle Physics}},
  \href{http://dx.doi.org/10.1093/ptep/ptac097}{\emph{PTEP} {\bf 2022} (2022)
  083C01}.

\bibitem{Aoki:2021kgd}
{\scshape Flavour Lattice Averaging Group (FLAG)} collaboration, Y.~Aoki
  et~al., \emph{{FLAG Review 2021}},
  \href{http://dx.doi.org/10.1140/epjc/s10052-022-10536-1}{\emph{Eur. Phys. J.
  C} {\bf 82} (2022) 869}, [\href{http://arxiv.org/abs/2111.09849}{{\tt
  2111.09849}}].

\bibitem{Bordone:2021oof}
M.~Bordone, B.~Capdevila and P.~Gambino, \emph{{Three loop calculations and
  inclusive $V_{cb}$}},
  \href{http://dx.doi.org/10.1016/j.physletb.2021.136679}{\emph{Phys. Lett. B}
  {\bf 822} (2021) 136679}, [\href{http://arxiv.org/abs/2107.00604}{{\tt
  2107.00604}}].

\bibitem{Blum:2015ywa}
T.~Blum et~al., \emph{{$K \rightarrow \pi\pi$ $\Delta I=3/2$ decay amplitude in
  the continuum limit}},
  \href{http://dx.doi.org/10.1103/PhysRevD.91.074502}{\emph{Phys. Rev. D} {\bf
  91} (2015) 074502}, [\href{http://arxiv.org/abs/1502.00263}{{\tt
  1502.00263}}].

\bibitem{deBlas:2022hdk}
J.~de~Blas, M.~Pierini, L.~Reina and L.~Silvestrini, \emph{{Impact of the
  recent measurements of the top-quark and W-boson masses on electroweak
  precision fits}},  \href{http://arxiv.org/abs/2204.04204}{{\tt 2204.04204}}.

\bibitem{Moulson:2017ive}
M.~Moulson, \emph{{Experimental determination of $V_{us}$ from kaon decays}},
  \href{http://dx.doi.org/10.22323/1.291.0033}{\emph{PoS} {\bf CKM2016} (2017)
  033}, [\href{http://arxiv.org/abs/1704.04104}{{\tt 1704.04104}}].

\bibitem{Marciano:2005ec}
W.~J. Marciano and A.~Sirlin, \emph{{Improved calculation of electroweak
  radiative corrections and the value of $V_{ud}$}},
  \href{http://dx.doi.org/10.1103/PhysRevLett.96.032002}{\emph{Phys. Rev.
  Lett.} {\bf 96} (2006) 032002},
  [\href{http://arxiv.org/abs/hep-ph/0510099}{{\tt hep-ph/0510099}}].

\bibitem{Seng:2018yzq}
C.-Y. Seng, M.~Gorchtein, H.~H. Patel and M.~J. Ramsey-Musolf, \emph{{Reduced
  Hadronic Uncertainty in the Determination of $V_{ud}$}},
  \href{http://dx.doi.org/10.1103/PhysRevLett.121.241804}{\emph{Phys. Rev.
  Lett.} {\bf 121} (2018) 241804}, [\href{http://arxiv.org/abs/1807.10197}{{\tt
  1807.10197}}].

\bibitem{Seng:2018qru}
C.~Y. Seng, M.~Gorchtein and M.~J. Ramsey-Musolf, \emph{{Dispersive evaluation
  of the inner radiative correction in neutron and nuclear $\beta$ decay}},
  \href{http://dx.doi.org/10.1103/PhysRevD.100.013001}{\emph{Phys. Rev. D} {\bf
  100} (2019) 013001}, [\href{http://arxiv.org/abs/1812.03352}{{\tt
  1812.03352}}].

\bibitem{Czarnecki:2019mwq}
A.~Czarnecki, W.~J. Marciano and A.~Sirlin, \emph{{Radiative Corrections to
  Neutron and Nuclear Beta Decays Revisited}},
  \href{http://dx.doi.org/10.1103/PhysRevD.100.073008}{\emph{Phys. Rev. D} {\bf
  100} (2019) 073008}, [\href{http://arxiv.org/abs/1907.06737}{{\tt
  1907.06737}}].

\bibitem{Hardy:2020qwl}
J.~C. Hardy and I.~S. Towner, \emph{{Superallowed $0^+ \to 0^+$ nuclear $\beta$
  decays: 2020 critical survey, with implications for V$_{ud}$ and CKM
  unitarity}}, \href{http://dx.doi.org/10.1103/PhysRevC.102.045501}{\emph{Phys.
  Rev. C} {\bf 102} (2020) 045501}.

\bibitem{Giusti:2017dwk}
D.~Giusti, V.~Lubicz, G.~Martinelli, C.~T. Sachrajda, F.~Sanfilippo, S.~Simula
  et~al., \emph{{First lattice calculation of the QED corrections to leptonic
  decay rates}},
  \href{http://dx.doi.org/10.1103/PhysRevLett.120.072001}{\emph{Phys. Rev.
  Lett.} {\bf 120} (2018) 072001}, [\href{http://arxiv.org/abs/1711.06537}{{\tt
  1711.06537}}].

\bibitem{DiCarlo:2019thl}
M.~Di~Carlo, D.~Giusti, V.~Lubicz, G.~Martinelli, C.~T. Sachrajda,
  F.~Sanfilippo et~al., \emph{{Light-meson leptonic decay rates in lattice
  QCD+QED}}, \href{http://dx.doi.org/10.1103/PhysRevD.100.034514}{\emph{Phys.
  Rev. D} {\bf 100} (2019) 034514},
  [\href{http://arxiv.org/abs/1904.08731}{{\tt 1904.08731}}].

\bibitem{DAgostini:1999niu}
G.~D'Agostini, \emph{{Sceptical combination of experimental results: General
  considerations and application to $\epsilon' / \epsilon$}},
  \href{http://arxiv.org/abs/hep-ex/9910036}{{\tt hep-ex/9910036}}.

\bibitem{Hudspith:2017vew}
R.~J. Hudspith, R.~Lewis, K.~Maltman and J.~Zanotti, \emph{{A resolution of the
  inclusive flavor-breaking $\tau$ $|V_{us}|$ puzzle}},
  \href{http://dx.doi.org/10.1016/j.physletb.2018.03.074}{\emph{Phys. Lett. B}
  {\bf 781} (2018) 206--212}, [\href{http://arxiv.org/abs/1702.01767}{{\tt
  1702.01767}}].

\bibitem{Maltman:2019xeh}
K.~Maltman et~al., \emph{{Current Status of inclusive hadronic $\tau$
  determinations of $\vert V_{us} \vert$}},
  \href{http://dx.doi.org/10.21468/SciPostPhysProc.1.006}{\emph{SciPost Phys.
  Proc.} {\bf 1} (2019) 006}.

\bibitem{Cirigliano:2022yyo}
V.~Cirigliano, A.~Crivellin, M.~Hoferichter and M.~Moulson, \emph{{Scrutinizing
  CKM unitarity with a new measurement of the $K_{\mu 3}/K_{\mu 2}$ branching
  fraction}},  \href{http://arxiv.org/abs/2208.11707}{{\tt 2208.11707}}.

\bibitem{Blanke:2018cya}
M.~Blanke and A.~J. Buras, \emph{{Emerging $\Delta M_{d}$ -anomaly from
  tree-level determinations of $|V_{cb}|$ and the angle $\gamma $}},
  \href{http://dx.doi.org/10.1140/epjc/s10052-019-6667-x}{\emph{Eur. Phys. J.
  C} {\bf 79} (2019) 159}, [\href{http://arxiv.org/abs/1812.06963}{{\tt
  1812.06963}}].

\bibitem{Buras:2021nns}
A.~J. Buras and E.~Venturini, \emph{{Searching for New Physics in Rare $K$ and
  $B$ Decays without $|V_{cb}|$ and $|V_{ub}|$ Uncertainties}},
  \href{http://dx.doi.org/10.5506/APhysPolB.53.6-A1}{\emph{Acta Phys. Polon. B}
  {\bf 53} (9, 2021) A1}, [\href{http://arxiv.org/abs/2109.11032}{{\tt
  2109.11032}}].

\bibitem{Bernlochner:2022ucr}
F.~Bernlochner, M.~Fael, K.~Olschewsky, E.~Persson, R.~van Tonder, K.~K. Vos
  et~al., \emph{{First extraction of inclusive V$_{cb}$ from q$^{2}$ moments}},
  \href{http://dx.doi.org/10.1007/JHEP10(2022)068}{\emph{JHEP} {\bf 10} (2022)
  068}, [\href{http://arxiv.org/abs/2205.10274}{{\tt 2205.10274}}].

\bibitem{FermilabLattice:2021cdg}
{\scshape Fermilab Lattice, MILC} collaboration, A.~Bazavov et~al.,
  \emph{{Semileptonic form factors for $B \to D^\ast\ell\nu$ at nonzero recoil
  from 2 + 1-flavor lattice QCD}},  \href{http://arxiv.org/abs/2105.14019}{{\tt
  2105.14019}}.

\bibitem{DiCarlo:2021dzg}
M.~Di~Carlo, G.~Martinelli, M.~Naviglio, F.~Sanfilippo, S.~Simula and
  L.~Vittorio, \emph{{Unitarity bounds for semileptonic decays in lattice
  QCD}}, \href{http://dx.doi.org/10.1103/PhysRevD.104.054502}{\emph{Phys. Rev.
  D} {\bf 104} (2021) 054502}, [\href{http://arxiv.org/abs/2105.02497}{{\tt
  2105.02497}}].

\bibitem{Martinelli:2021frl}
G.~Martinelli, S.~Simula and L.~Vittorio, \emph{{Constraints for the
  semileptonic B\textrightarrow{}D(*) form factors from lattice QCD simulations
  of two-point correlation functions}},
  \href{http://dx.doi.org/10.1103/PhysRevD.104.094512}{\emph{Phys. Rev. D} {\bf
  104} (2021) 094512}, [\href{http://arxiv.org/abs/2105.07851}{{\tt
  2105.07851}}].

\bibitem{Martinelli:2022adr}
G.~Martinelli, S.~Simula and L.~Vittorio, \emph{{$\vert V_{cb} \vert$ and
  $R(D^{(*)})$ using lattice QCD and unitarity}},
  \href{http://dx.doi.org/10.1103/PhysRevD.105.034503}{\emph{Phys. Rev. D} {\bf
  105} (2022) 034503}, [\href{http://arxiv.org/abs/2105.08674}{{\tt
  2105.08674}}].

\bibitem{Martinelli:2022xir}
G.~Martinelli, M.~Naviglio, S.~Simula and L.~Vittorio, \emph{{$\vert V_{cb}
  \vert$, lepton flavor universality and $SU(3)_F$ symmetry breaking in
  Bs\textrightarrow{}Ds(*)\ensuremath{\ell}\ensuremath{\nu}$_{\ensuremath{\ell}}$
  decays through unitarity and lattice QCD}},
  \href{http://dx.doi.org/10.1103/PhysRevD.106.093002}{\emph{Phys. Rev. D} {\bf
  106} (2022) 093002}, [\href{http://arxiv.org/abs/2204.05925}{{\tt
  2204.05925}}].

\bibitem{Martinelli:2021myh}
G.~Martinelli, S.~Simula and L.~Vittorio, \emph{{Exclusive determinations of
  $\vert V_{cb} \vert $ and $R(D^{*})$ through unitarity}},
  \href{http://dx.doi.org/10.1140/epjc/s10052-022-11050-0}{\emph{Eur. Phys. J.
  C} {\bf 82} (2022) 1083}, [\href{http://arxiv.org/abs/2109.15248}{{\tt
  2109.15248}}].

\bibitem{Gambino:2019sif}
P.~Gambino, M.~Jung and S.~Schacht, \emph{{The $V_{cb}$ puzzle: An update}},
  \href{http://dx.doi.org/10.1016/j.physletb.2019.06.039}{\emph{Phys. Lett. B}
  {\bf 795} (2019) 386--390}, [\href{http://arxiv.org/abs/1905.08209}{{\tt
  1905.08209}}].

\bibitem{Jaiswal:2020wer}
S.~Jaiswal, S.~Nandi and S.~K. Patra, \emph{{Updates on extraction of $\vert
  V_{cb}\vert$ and SM prediction of R(D*) in $B\to D^{*}\ell\nu_\ell$ decays}},
  \href{http://dx.doi.org/10.1007/JHEP06(2020)165}{\emph{JHEP} {\bf 06} (2020)
  165}, [\href{http://arxiv.org/abs/2002.05726}{{\tt 2002.05726}}].

\bibitem{Sirlin:1981ie}
A.~Sirlin, \emph{{Large $m_W$, $m_Z$ Behavior of the O($\alpha$) Corrections to
  Semileptonic Processes Mediated by W}},
  \href{http://dx.doi.org/10.1016/0550-3213(82)90303-0}{\emph{Nucl. Phys. B}
  {\bf 196} (1982) 83--92}.

\bibitem{Belle-II:2022imn}
{\scshape Belle-II} collaboration, K.~Adamczyk et~al., \emph{{Determination of
  $|V_{ub}|$ from untagged $B^0\to\pi^- \ell^+ \nu_{\ell}$ decays using
  2019-2021 Belle II data}},  \href{http://arxiv.org/abs/2210.04224}{{\tt
  2210.04224}}.

\bibitem{FermilabLattice:2015mwy}
{\scshape Fermilab Lattice, MILC} collaboration, J.~A. Bailey et~al.,
  \emph{{$|V_{ub}|$ from $B\to\pi\ell\nu$ decays and (2+1)-flavor lattice
  QCD}}, \href{http://dx.doi.org/10.1103/PhysRevD.92.014024}{\emph{Phys. Rev.
  D} {\bf 92} (2015) 014024}, [\href{http://arxiv.org/abs/1503.07839}{{\tt
  1503.07839}}].

\bibitem{Flynn:2015mha}
J.~M. Flynn, T.~Izubuchi, T.~Kawanai, C.~Lehner, A.~Soni, R.~S. Van~de Water
  et~al., \emph{{$B \to \pi \ell \nu$ and $B_s \to K \ell \nu$ form factors and
  $|V_{ub}|$ from 2+1-flavor lattice QCD with domain-wall light quarks and
  relativistic heavy quarks}},
  \href{http://dx.doi.org/10.1103/PhysRevD.91.074510}{\emph{Phys. Rev. D} {\bf
  91} (2015) 074510}, [\href{http://arxiv.org/abs/1501.05373}{{\tt
  1501.05373}}].

\bibitem{Colquhoun:2022atw}
{\scshape JLQCD} collaboration, B.~Colquhoun, S.~Hashimoto, T.~Kaneko and
  J.~Koponen, \emph{{Form factors of
  B\textrightarrow{}\ensuremath{\pi}\ensuremath{\ell}\ensuremath{\nu} and a
  determination of $\vert V_{ub} \vert$ with M\"obius domain-wall fermions}},
  \href{http://dx.doi.org/10.1103/PhysRevD.106.054502}{\emph{Phys. Rev. D} {\bf
  106} (2022) 054502}, [\href{http://arxiv.org/abs/2203.04938}{{\tt
  2203.04938}}].

\bibitem{UTfit:2006vpt}
{\scshape UTfit} collaboration, M.~Bona et~al., \emph{{The Unitarity Triangle
  Fit in the Standard Model and Hadronic Parameters from Lattice QCD: A
  Reappraisal after the Measurements of $\Delta m_s$ and $BR(B \to \tau
  \nu_{\tau})$}},
  \href{http://dx.doi.org/10.1088/1126-6708/2006/10/081}{\emph{JHEP} {\bf 10}
  (2006) 081}, [\href{http://arxiv.org/abs/hep-ph/0606167}{{\tt
  hep-ph/0606167}}].

\bibitem{Inami:1980fz}
T.~Inami and C.~S. Lim, \emph{{Effects of Superheavy Quarks and Leptons in
  Low-Energy Weak Processes $K_L \to \mu \bar{\mu}$, $K^+ \to \pi^+ \nu
  \bar{\nu}$ and $K^0 \leftrightarrow \bar{K}^0$}},
  \href{http://dx.doi.org/10.1143/PTP.65.297}{\emph{Prog. Theor. Phys.} {\bf
  65} (1981) 297}.

\bibitem{Brod:2019rzc}
J.~Brod, M.~Gorbahn and E.~Stamou, \emph{{Standard-Model Prediction of
  $\epsilon_K$ with Manifest Quark-Mixing Unitarity}},
  \href{http://dx.doi.org/10.1103/PhysRevLett.125.171803}{\emph{Phys. Rev.
  Lett.} {\bf 125} (2020) 171803}, [\href{http://arxiv.org/abs/1911.06822}{{\tt
  1911.06822}}].

\bibitem{Buras:2008nn}
A.~J. Buras and D.~Guadagnoli, \emph{{Correlations among new CP violating
  effects in $\Delta F = 2$ observables}},
  \href{http://dx.doi.org/10.1103/PhysRevD.78.033005}{\emph{Phys. Rev. D} {\bf
  78} (2008) 033005}, [\href{http://arxiv.org/abs/0805.3887}{{\tt 0805.3887}}].

\bibitem{Christ:2010gi}
{\scshape RBC, UKQCD} collaboration, N.~H. Christ, \emph{{Long-distance
  contributions to weak amplitudes}},  in \emph{{28th International Symposium
  on Lattice Field Theory}}, 12, 2010.
\newblock \href{http://arxiv.org/abs/1012.6034}{{\tt 1012.6034}}.

\bibitem{Christ:2015pwa}
N.~H. Christ, X.~Feng, G.~Martinelli and C.~T. Sachrajda, \emph{{Effects of
  finite volume on the $K_L$-$K_S$ mass difference}},
  \href{http://dx.doi.org/10.1103/PhysRevD.91.114510}{\emph{Phys. Rev. D} {\bf
  91} (2015) 114510}, [\href{http://arxiv.org/abs/1504.01170}{{\tt
  1504.01170}}].

\bibitem{Christ:2015phf}
N.~H. Christ and Z.~Bai, \emph{{Computing the long-distance contributions to
  $\varepsilon_K$}}, \href{http://dx.doi.org/10.22323/1.251.0342}{\emph{PoS}
  {\bf LATTICE2015} (2016) 342}.

\bibitem{Buras:2010pza}
A.~J. Buras, D.~Guadagnoli and G.~Isidori, \emph{{On $\epsilon_K$ Beyond Lowest
  Order in the Operator Product Expansion}},
  \href{http://dx.doi.org/10.1016/j.physletb.2010.04.017}{\emph{Phys. Lett. B}
  {\bf 688} (2010) 309--313}, [\href{http://arxiv.org/abs/1002.3612}{{\tt
  1002.3612}}].

\bibitem{Brod:2022har}
J.~Brod, S.~Kvedaraite, Z.~Polonsky and A.~Youssef, \emph{{Electroweak
  Corrections to the Charm-Top-Quark Contribution to $\epsilon_K$}},
  \href{http://arxiv.org/abs/2207.07669}{{\tt 2207.07669}}.

\bibitem{Brod:2010mj}
J.~Brod and M.~Gorbahn, \emph{{$\epsilon_K$ at Next-to-Next-to-Leading Order:
  The Charm-Top-Quark Contribution}},
  \href{http://dx.doi.org/10.1103/PhysRevD.82.094026}{\emph{Phys. Rev. D} {\bf
  82} (2010) 094026}, [\href{http://arxiv.org/abs/1007.0684}{{\tt 1007.0684}}].

\bibitem{Brod:2011ty}
J.~Brod and M.~Gorbahn, \emph{{Next-to-Next-to-Leading-Order Charm-Quark
  Contribution to the $CP$ Violation Parameter $\epsilon_K$ and $\Delta M_K$}},
  \href{http://dx.doi.org/10.1103/PhysRevLett.108.121801}{\emph{Phys. Rev.
  Lett.} {\bf 108} (2012) 121801}, [\href{http://arxiv.org/abs/1108.2036}{{\tt
  1108.2036}}].

\bibitem{Martinelli:1994ty}
G.~Martinelli, C.~Pittori, C.~T. Sachrajda, M.~Testa and A.~Vladikas, \emph{{A
  General method for nonperturbative renormalization of lattice operators}},
  \href{http://dx.doi.org/10.1016/0550-3213(95)00126-D}{\emph{Nucl. Phys. B}
  {\bf 445} (1995) 81--108}, [\href{http://arxiv.org/abs/hep-lat/9411010}{{\tt
  hep-lat/9411010}}].

\bibitem{Sturm:2009kb}
C.~Sturm, Y.~Aoki, N.~H. Christ, T.~Izubuchi, C.~T.~C. Sachrajda and A.~Soni,
  \emph{{Renormalization of quark bilinear operators in a momentum-subtraction
  scheme with a nonexceptional subtraction point}},
  \href{http://dx.doi.org/10.1103/PhysRevD.80.014501}{\emph{Phys. Rev. D} {\bf
  80} (2009) 014501}, [\href{http://arxiv.org/abs/0901.2599}{{\tt 0901.2599}}].

\bibitem{Ciuchini:1993vr}
M.~Ciuchini, E.~Franco, G.~Martinelli and L.~Reina, \emph{{The $\Delta S = 1$
  effective Hamiltonian including next-to-leading order QCD and QED
  corrections}},
  \href{http://dx.doi.org/10.1016/0550-3213(94)90118-X}{\emph{Nucl. Phys. B}
  {\bf 415} (1994) 403--462}, [\href{http://arxiv.org/abs/hep-ph/9304257}{{\tt
  hep-ph/9304257}}].

\bibitem{Buras:1991jm}
A.~J. Buras, M.~Jamin, M.~E. Lautenbacher and P.~H. Weisz, \emph{{Effective
  Hamiltonians for $\Delta S = 1$ and $\Delta B = 1$ nonleptonic decays beyond
  the leading logarithmic approximation}},
  \href{http://dx.doi.org/10.1016/0550-3213(92)90345-C}{\emph{Nucl. Phys. B}
  {\bf 370} (1992) 69--104}.

\bibitem{Buras:1992tc}
A.~J. Buras, M.~Jamin, M.~E. Lautenbacher and P.~H. Weisz, \emph{{Two loop
  anomalous dimension matrix for $\Delta S = 1$ weak nonleptonic decays I:
  $\mathcal{O}(\alpha_s^2)$}},
  \href{http://dx.doi.org/10.1016/0550-3213(93)90397-8}{\emph{Nucl. Phys. B}
  {\bf 400} (1993) 37--74}, [\href{http://arxiv.org/abs/hep-ph/9211304}{{\tt
  hep-ph/9211304}}].

\bibitem{Buchalla:1995vs}
G.~Buchalla, A.~J. Buras and M.~E. Lautenbacher, \emph{{Weak decays beyond
  leading logarithms}},
  \href{http://dx.doi.org/10.1103/RevModPhys.68.1125}{\emph{Rev. Mod. Phys.}
  {\bf 68} (1996) 1125--1144}, [\href{http://arxiv.org/abs/hep-ph/9512380}{{\tt
  hep-ph/9512380}}].

\bibitem{Cirigliano:2019cpi}
V.~Cirigliano, H.~Gisbert, A.~Pich and A.~Rodr\'\i{}guez-S\'anchez,
  \emph{{Isospin-violating contributions to $\epsilon'/\epsilon$}},
  \href{http://dx.doi.org/10.1007/JHEP02(2020)032}{\emph{JHEP} {\bf 02} (2020)
  032}, [\href{http://arxiv.org/abs/1911.01359}{{\tt 1911.01359}}].

\bibitem{Lehner:2011fz}
C.~Lehner and C.~Sturm, \emph{{Matching factors for $\Delta S=1$ four-quark
  operators in RI/SMOM schemes}},
  \href{http://dx.doi.org/10.1103/PhysRevD.84.014001}{\emph{Phys. Rev. D} {\bf
  84} (2011) 014001}, [\href{http://arxiv.org/abs/1104.4948}{{\tt 1104.4948}}].

\bibitem{Ciuchini:2005mg}
M.~Ciuchini, M.~Pierini and L.~Silvestrini, \emph{{The Effect of penguins in
  the $B_d \to J / \psi K^0$ CP asymmetry}},
  \href{http://dx.doi.org/10.1103/PhysRevLett.95.221804}{\emph{Phys. Rev.
  Lett.} {\bf 95} (2005) 221804},
  [\href{http://arxiv.org/abs/hep-ph/0507290}{{\tt hep-ph/0507290}}].

\bibitem{Buras:2000dm}
A.~J. Buras, P.~Gambino, M.~Gorbahn, S.~Jager and L.~Silvestrini,
  \emph{{Universal unitarity triangle and physics beyond the standard model}},
  \href{http://dx.doi.org/10.1016/S0370-2693(01)00061-2}{\emph{Phys. Lett. B}
  {\bf 500} (2001) 161--167}, [\href{http://arxiv.org/abs/hep-ph/0007085}{{\tt
  hep-ph/0007085}}].

\bibitem{Martinelli:2022tte}
G.~Martinelli, S.~Simula and L.~Vittorio, \emph{{Exclusive semileptonic B
  \textrightarrow{} \ensuremath{\pi}\ensuremath{\ell}\ensuremath{\nu}$_{\ell}$
  and B$_{s}$ \textrightarrow{} K\ensuremath{\ell}\ensuremath{\nu}$_{\ell}$
  decays through unitarity and lattice QCD}},
  \href{http://dx.doi.org/10.1007/JHEP08(2022)022}{\emph{JHEP} {\bf 08} (2022)
  022}, [\href{http://arxiv.org/abs/2202.10285}{{\tt 2202.10285}}].

\end{thebibliography}\endgroup


\providecommand{\href}[2]{#2}\begingroup\raggedright\endgroup

\end{document}